\documentclass[aps,rmp,reprint,amsmath,amssymb,graphicx,longbibliography,floatfix,preprintnumbers,eqsecnum]{revtex4-2}

\usepackage[english]{babel} 
\usepackage{microtype} 
\usepackage{xspace} 
\usepackage[T1]{fontenc} 
\usepackage{float}
\usepackage{dsfont}

\usepackage{soul}
\usepackage{setspace} 

\usepackage{xspace}
\usepackage[dvipsnames,table]{xcolor}
\usepackage{amsfonts,amsmath,amssymb,bm,mathtools,slashed} 

\usepackage{graphicx}
\usepackage{epstopdf}
\graphicspath{{./Figures/}{./figures/}}
\usepackage{tikz}
\usepackage[compat=1.1.0]{tikz-feynman}

\definecolor{rmp}{RGB}{41, 43, 133}
\newcommand{\LinkColor}[0]{rmp}
\usepackage{hyperref}
\hypersetup{
	colorlinks, 
	bookmarksopen, 
	bookmarksnumbered,
	pdftoolbar=false, 
	pdfmenubar=false, 
	pdffitwindow=false, 
	pdfstartview={FitH},	 
	citecolor=\LinkColor, 
	linkcolor=\LinkColor,	
	urlcolor=\LinkColor, 
	pdftitle={The Standard Model effective field theory at work}, 
	pdfsubject={SMEFT}, 
	pdfauthor={Felix Wilsch}, 
	pdfnewwindow=true, 
	linktoc=all
}

\makeatletter
\g@addto@macro\bfseries{\boldmath} 
\makeatother

\allowdisplaybreaks

\newcommand{\cmd}[1]{{\texttt{#1}}\xspace}
\newcommand{\dd}{\mathop{}\!\mathrm{d}}

\newcommand{\brackets}[1]{\left( #1 \right)}
\newcommand{\squarebrackets}[1]{\left[ #1 \right]}

\renewcommand{\Re}{\mathop{\mathrm{Re}}}
\renewcommand{\L}{\mathcal{L}}
\newcommand{\ord}[1]{\mathcal{O}\!\left( #1 \right)}

\newcommand{\vev}[0]{vacuum expectation value\xspace}

\newcommand{\SU}[1]{\mathrm{SU}(#1)}

\newcommand{\be}{\begin{equation}}
\newcommand{\ee}{\end{equation}}
\newcommand{\bea}{\begin{eqnarray}}
\newcommand{\eea}{\end{eqnarray}}
\newcommand{\ba}{\begin{array}}
\newcommand{\ea}{\end{array}}
\newcommand{\no}{\nonumber}
\newcommand{\cL}{\mathcal{L}}
\newcommand{\cO}{\mathcal{O}}
\newcommand{\cG}{\mathcal{G}}

\newcommand{\myBlue}[0]{RoyalBlue}

\begin{document}

\title{The Standard Model effective field theory at work}

\author{Gino Isidori}
\email{gino.isidori@physik.uzh.ch}
\author{Felix Wilsch}
\email{felix.wilsch@physik.uzh.ch}
\author{Daniel Wyler}
\email{wyler@physik.uzh.ch}
\affiliation{Physik-Institut, Universit\"at Z\"urich, CH-8057~Z\"urich, Switzerland}

\preprint{ZU-TH 14/23}


\begin{abstract}
The striking success of the Standard Model in explaining precision data and, at the same time, its lack of explanations for various fundamental phenomena, such as dark matter or the baryon asymmetry of the universe, suggests new physics at an energy scale much larger than the electroweak scale. 
In the absence of a short-range--long-range conspiracy, the Standard Model can be viewed as the leading term of an effective ‚remnant‘ theory (referred to as the SMEFT) of a more fundamental structure. 
Over the last years, many aspects of the SMEFT have been investigated and it has become a standard tool to analyze experimental results in an integral way. 
In this article, after briefly presenting the salient features of the Standard Model, we review the construction of the SMEFT. 
We discuss the range of its applicability and bounds on its coefficients imposed by general theoretical considerations. 
Since new-physics models are likely to exhibit exact or approximate accidental global symmetries, especially in the flavor sector, we also discuss their implications for the SMEFT.  
The main focus of our review is the phenomenological analysis of experimental results. 
We show explicitly how to use various effective field theories to study the phenomenology of theories beyond the Standard Model. 
We give a detailed description of the matching procedure and the use of the renormalization group equations, allowing to connect multiple effective theories valid at different energy scales. 
Explicit examples from low-energy experiments and from high-$p_T$ physics illustrate the workflow. 
We also comment on the non-linear realization of the electroweak symmetry breaking and its phenomenological implications. 
\end{abstract}

{
\maketitle
{
\hypersetup{linkcolor=black}
\begin{spacing}{0.97}
\tableofcontents{}
\end{spacing}
}
}

\newpage

\section{Introduction}
\begin{quotation}
\noindent {\it Maybe the main point of our analysis is that it demonstrates explicitly  how remarkable the standard electroweak theory is. 
\cite{Buchmuller:1985jz}}
\end{quotation}
The Standard Model~(SM) of particle physics, formulated some 50~years ago, and judiciously completed over the years, forms the basis of our understanding of the fundamental interactions. More precisely, the SM is the quantum field theory~(QFT) that describes how the basic matter constituents (quarks and leptons) interact at the microscopic level via weak, strong, and electromagnetic forces. While all data from earth-based laboratory experiments agrees with the SM predictions (possibly with a few exceptions that we will comment on later), there is some indirect evidence, derived from cosmological observations, that the model is not complete: It does not explain the baryon asymmetry of the universe, dark matter, and dark energy. 
These are all phenomena that could naturally find their explanation in the domain of particle physics or, more generally, within~QFT.
There are also theoretical concerns about the SM itself, such as the strong sensitivity of the Higgs mass term to high-energy modes in the renormalization procedure (the so-called ``hierarchy problem''), the absence of an explanation for the hierarchical structure of the fermion spectrum, and the lack of a bridge to quantum gravity. Last but not least, non-vanishing neutrino masses cannot be accounted for by the ``classical version'' of the Standard Model, containing only left-handed neutrinos and only renormalizable interactions.

In order to address these problems, a large number of new ``fundamental'' theories beyond the Standard Model~(BSM) were formulated over the last 40--50~years. In fact, the 1980s and, to a lesser extent, the 1990s saw a downright explosion of model building. While some of them addressed specific questions, others offered veritable extensions of the basics of the SM, such as supersymmetric models, or models with composite Higgs sectors, and/or composite quarks and leptons: concepts that might become important again in the future, possibly within a new context, such as string theory. These models have new particles and interactions, generally at energies (well) above the Fermi scale. They were designed to explain some of the facts that the SM cannot, such as the 
quantization of the electric charge, the hierarchical generational structure of quarks and leptons, the possible unification of interaction strengths, etc. Unfortunately, many of these models have been shown to be inconsistent with data 
or are not testable with present and near-future experimental facilities.

In order to look for these new-physics scenarios, most likely manifested in small discrepancies between the SM predictions and the observations, both theoretical and experimental progress is necessary. 
Over many years, and with increasing intensity and success in the new century, theoretical work on the Standard Model has improved enormously. Apart from devising new calculational tools, this progress has been made possible by developing and applying the concepts of effective field theory~(EFT) in several relevant areas. 
Roughly speaking, a quantum EFT is a quantum field theory which is not considered to be ``fundamental'', being valid only in a limited range of energies or distances, or even in specific kinematic configurations. The wide separation between the Fermi scale (or the $W$-boson mass,~$m_W$) and the masses of the $B$-mesons or the charmed particles has allowed to successfully use EFT and renormalization group techniques to calculate the expected (inclusive) decay rates of these mesons with astonishing accuracy. The formulation of new quantum EFTs like HQET (heavy quark effective theory) and SCET (soft collinear effective theory) have lead to accurate predictions also for exclusive decays. Equally, high-energy calculations, such as used for jet dynamics at the LHC, have benefited from EFT techniques. Also the oldest effective field theory of the SM, namely ChPT (chiral perturbation theory), has been extensively used to obtain precision results for low-energy meson dynamics. 
We expect this quest for ever higher precision, both on the theoretical and the experimental side, to continue, in the hope to find deviations from the SM for which there are well motivated reasons. 

In this perspective, it is very natural to consider the original formulation of the SM as the effective low-energy ``remnant'' of a more fundamental theory, whose new heavy degrees of freedom are removed 
in favor of generating additional effective contact interactions between the known SM fields. As argued by \textcite{Wilson:1983xri}, the true physics of the ``full'' theory below the cutoff scale can be recovered by including all possible interactions allowed by the particles and symmetries of the theory. 
The effective Lagrangian thus obtained consists of a string of local interaction terms (operators), each characterized by an appropriate coefficient (effective coupling/Wilson coefficient), organized in a series of increasing dimensionality, corresponding to the expected decreasing relevance. 
As usual for EFTs, this construction is not renormalizable in the usual strict sense, because it involves an infinite number of coupling constants. It is however renormalizable order by order in an energy/momentum expansion reflected in the operator expansion. Actually the independence from the renormalization scale of physical amplitudes can be exploited by the renormalization group flow of the operator coefficients, allowing to identify and resum the largest quantum corrections. 

Given the success of effective theories so far, this approach seems a good way to access the next layer of physics, as proposed by \textcite{Buchmuller:1985jz} even before the last building blocks of the SM where experimentally identified. 
In this review, we will trace its development and highlight some of the most recent results.
Our main scope is to illustrate how considering the SM as an EFT can help in identifying properties of new physics and single out future research directions. The EFT approach provides indeed not only a systematic way for analyzing experimental results, but also a precious tool to correlate different observables obtaining a deeper insights on where to look for the next layer. 

This review is organized as follows: in the rest of
this section we introduce the SM, briefly recalling also the motivations why we want to go beyond it, we review general aspects of EFT, and finally introduce the so-called Standard Model effective field theory~(SMEFT). A detailed analysis of the SMEFT, with special focus on the structure of operators of dimension six, is presented in 
Sec.~\ref{sect:SMEFT}. The role of global symmetries in the SMEFT, with a particular emphasis on exact and approximate flavor symmetries, is discussed in Sec.~\ref{sec:GlobalSymmetries}. Section~\ref{sec:HEFT} is devoted to a discussion of the differences between the SMEFT and the more general case of a non-linearly realized electroweak symmetry. In Sec.~\ref{sect:LEFT} we briefly review the low-energy ($E \ll M_W$) effective theory  of the SMEFT, in particular in comparison to the Standard Model.  Finally, in Sec.~\ref{sect:practical} we present two concrete examples of the SMEFT at work, i.e., of applications of the SMEFT to analyze concrete phenomenological problems. 
In Appendix~\ref{app:dim-reg} we discuss some technical details of dimensional regularization showing up in SMEFT computations.


\subsection{The Standard Model of particle physics}

Within the Standard Model\footnote{For a pedagogical introduction to the SM see, e.g., \cite{Grossman:2023sm,Donoghue:1992dd}.} the three fundamental 
forces are described via the principle of gauge 
invariance, requiring the theory to be 
invariant under the local symmetry group
\begin{align}
\mathcal{G}_\mathrm{SM} &= \SU{3}_c \times \SU{2}_L \times \mathrm{U}(1)_Y \, .
\end{align}
The quantum fields can be divided in three categories: $i)$~the gauge fields associated to the local gauge symmetry groups~$(G_\mu,W_\mu,B_\mu)$; $ii)$~the matter (fermion) fields~($\ell,e,q,u,d$); $iii)$~the Higgs boson doublet~$H$ responsible for the breaking of the electroweak subgroup of~$\mathcal{G}_\mathrm{SM}$ down to the QED group~$\mathrm{U}(1)_e$
\begin{align}
\SU{2}_L \times \mathrm{U}(1)_Y \longrightarrow \mathrm{U}(1)_e \, .
\end{align}
The field content of the SM is shown in Tab.~\ref{tab:SM_field-content} together with the transformation properties of each field under the different gauge groups and the hypercharge assignments.\footnote{In principle, one could extend the fermion content including right-handed neutrinos. However, these fields would be completely neutral under~$\mathcal{G}_\mathrm{SM}$. We prefer to define the SM as the theory of the chiral fermions with non-trivial transformation properties under~$\mathcal{G}_\mathrm{SM}$, that acquire mass via the Higgs mechanism. As such, right-handed neutrinos are not SM fields.}
The basic fermion family ($\ell,e,q,u,d$) is replicated three times.

\begin{table}
\centering 
\begin{tabular}{ | c | c c c c c | c | c c c | }
\hline
& $\ell$ & $e$ & $q$& $u$ & $d$ & $H$ & $G$ & $W$ & $B$ \\ \hline
$\SU{3}_c$ representation & $\mathbf{1}$ & $\mathbf{1}$ & $\mathbf{3}$ & $\mathbf{3}$ & $\mathbf{3}$ & $\mathbf{1}$ & $\mathbf{8}$ & $\mathbf{1}$ & $\mathbf{1}$ \\ \hline
$\SU{2}_L$ representation & $\mathbf{2}$ & $\mathbf{1}$ & $\mathbf{2}$ & $\mathbf{1}$ & $\mathbf{1}$ & $\mathbf{2}$ & $\mathbf{1}$ & $\mathbf{3}$ & $\mathbf{1}$ \\ \hline
$\mathrm{U}(1)_Y$ charge & $-\frac{1}{2}$ & $-1$ & $\frac{1}{6} $ & $\frac{2}{3}$ & $-\frac{1}{3}$ & $\frac{1}{2}$ & $0$ & $0$ & $0$ \\ \hline
\end{tabular}
\caption{Standard Model field content with the transformation properties of the fields under $\SU{3}_c \times \SU{2}_L$, and the hypercharge assignments. The fields are divided into fermions~$(\ell,e,q,u,d)$, the Higgs doublet~$(H)$, and gauge fields~$(G,W,B)$.
\label{tab:SM_field-content}
}
\end{table}

The SM Lagrangian 
is the most general renormalizable expression that can be constructed out of the fields in Tab.~\ref{tab:SM_field-content} that is invariant under~$\mathcal{G}_\mathrm{SM}$:
\begin{align}
& \L_\mathrm{SM} = -\frac{1}{4} G_{\mu\nu}^A G^{A\mu\nu} -\frac{1}{4} W_{\mu\nu}^I W^{I\mu\nu} -\frac{1}{4} B_{\mu\nu} B^{\mu\nu} 
\nonumber\\
&-  \frac{\theta_3g_3^2 }{32\pi^2}  G_{\mu\nu}^A \widetilde{G}^{A\mu\nu} - \frac{\theta_2g_2^2 }{32\pi^2}  W_{\mu\nu}^I \widetilde{W}^{I\mu\nu} - \frac{\theta_1g_1^2 }{32\pi^2} B_{\mu\nu} \widetilde{B}^{\mu\nu}  
\nonumber\\
&+ i \brackets{\overline{\ell}_p \slashed{D} \ell_p + \overline{e}_p \slashed{D} e_p + \overline{q}_p \slashed{D} q_p + \overline{u}_p \slashed{D} u_p + \overline{d}_p \slashed{D} d_p} 
\label{eq:SM_Lagrangian}\\
&+ \brackets{D_\mu H}^\dagger \brackets{D^\mu H} + m^2 H^\dagger H - \frac{\lambda}{2} \brackets{H^\dagger H}^2 
\nonumber\\
&- \brackets{[Y_e]_{pr}\, \overline{\ell}_p e_r H + [Y_u]_{pr}\, \overline{q}_p u_r \widetilde{H} + [Y_d]_{pr}\, \overline{q}_p d_r H + \mathrm{h.c.}} .
\nonumber
\end{align}

\subsubsection{The gauge sector}
The first three lines of Eq.~\eqref{eq:SM_Lagrangian} contain all gauge interaction in the SM. The gauge couplings associated to the gauge groups $\SU{3}_c$, $\SU{2}_L$, and~$\mathrm{U}(1)_Y$ are $g_3$, $g_2$, and~$g_1$. The indices $A=1,...,8$ and $I=1,2,3$ denote adjoint $\SU{3}_c$ or $\SU{2}_L$ gauge indices, respectively.
In the first line of Eq.~\eqref{eq:SM_Lagrangian} the field-strength tensors are defined by 
\begin{subequations}
\begin{align}
G_{\mu\nu}^A &= \partial_\mu G_\nu^A - \partial_\nu G_\mu^A + g_3 f^{ABC} G_\mu^B G_\nu^C \, , 
\\
W_{\mu\nu}^I &= \partial_\mu W_\nu^I - \partial_\nu W_\mu^I + g_2 \varepsilon^{IJK} W_\mu^J W_\nu^K \, , 
\label{eq:SM_fieldstrength_W}
\\
B_{\mu\nu} &= \partial_\mu B_\nu - \partial_\nu B_\mu \, ,
\end{align}
\end{subequations}
where $\smash{f^{ABC}}$ and $\smash{\varepsilon^{IJK}}$ are the totally anti-symmetric structure constants of $\smash{\SU{3}_c}$ and~$\smash{\SU{2}_L}$. They contain the kinetic terms for the gauge fields as well as all interactions among the gauge fields themselves. 

In the second line, the dual field-strength tensors are defined by $\smash{\widetilde{F}^{\mu\nu}=\frac{1}{2}\varepsilon^{\mu\nu\rho\sigma}F_{\rho\sigma}}$ for $\smash{F=G^A,W^I,B}$ with the totally anti-symmetric Levi-Civita tensor defined by~$\varepsilon^{0123}=-\varepsilon_{0123}=+1$. The Lagrangian terms containing dual field-strength tensors are proportional to total derivatives, meaning we can rewrite them as $\smash{G_{\mu\nu}^A \tilde{G}^{A,\mu\nu} = 2 \varepsilon^{\mu\nu\alpha\beta}\partial_\mu \big( G_\nu^A \partial_\alpha G_\beta^A + \frac{1}{3} g_3 f^{ABC} G_\nu^A G_\alpha^B G_\beta^C \big)}$. Therefore, they can only contribute to topological effects. For simplicity, 
we drop them from here on.

The third line comprises the kinetic terms of the fermion fields, as well as their gauge interactions. The latter are encoded in the gauge covariant derivative 
\begin{align}
D_\mu &= \partial_\mu - i g_3 T^A G_\mu^A - i g_2 t^I W_\mu^I - i g_1 \mathsf{y} B_\mu \, ,
\end{align}
where $T^A = \lambda^A / 2$ and $t^I = \tau^I / 2$  are the generators of the fundamental representation of $\SU{3}_c$ and $\SU{2}_L$, respectively, with the Gell-Mann matrices~$\lambda^A$ and the Pauli matrices~$\tau^I$. The hypercharge generator is denoted~$\mathsf{y}$.

\subsubsection{The Higgs sector}
The last two lines of Eq.~\eqref{eq:SM_Lagrangian} include the Higgs and Yukawa sector of the~SM 
written in a symmetric notation before electroweak symmetry  breaking. The complex Higgs doublet is denoted by~$H$ and we define $\widetilde{H}=i \tau_2 H^\ast$. Minimizing the scalar potential
\begin{equation}
V(H) = -m^2 H^\dagger H + \frac{\lambda}{2}\brackets{H^\dagger H}^2
\label{eq:Hpotential}
\end{equation}
 yields a non-vanishing \vev~(vev) for the Higgs field,
$v^2 = 2 \langle 0 | H^\dagger H| 0  \rangle$, whose tree-level 
expression reads $v^2= 2 m^2/\lambda$. Considering the breaking of the electroweak symmetry, it is convenient to re-write
the Higgs doublet as
\begin{align}
H &= \frac{1}{\sqrt{2}} 
	\begin{pmatrix}
		\varphi^2 + i \varphi^1 \\
		v + h - i \varphi^3
	\end{pmatrix} \,,
 \label{eq:Hdec}
\end{align}
where $h$ is the massive physical Higgs boson and~$\varphi^a$ denote the three Goldstone bosons, that, in the unitary gauge, are ``eaten'' by the massive gauge bosons. The tree-level mass of the physical Higgs is~$m^2_h = 2 m^2$.

The Yukawa couplings~$[Y_i]_{pr}$ for $i=e,u,d$ are complex ${3 \times 3}$~matrices in flavor space contracted to the fermion fields via the global flavor indices $p$ and~$r$, which run from 1~to~3.
After electroweak symmetry breaking the Yukawa interactions in Eq.~\eqref{eq:SM_Lagrangian} yield the fermion mass terms as well as the Yukawa interactions with the physical Higgs boson~$h$. The Yukawa matrices~$Y_{u,d}$ are the only source of flavor violation in the SM,
as the gauge interactions are all flavor diagonal. 
They are also the only source of CP~violation in the SM, apart from the topological terms associated to the dual field-strength tensors, which are shown in the second line of Eq.~\eqref{eq:SM_Lagrangian}.

\subsubsection{The success of the Standard Model}
\begin{figure}[t]
\begin{center}
\includegraphics[width=0.75\linewidth]{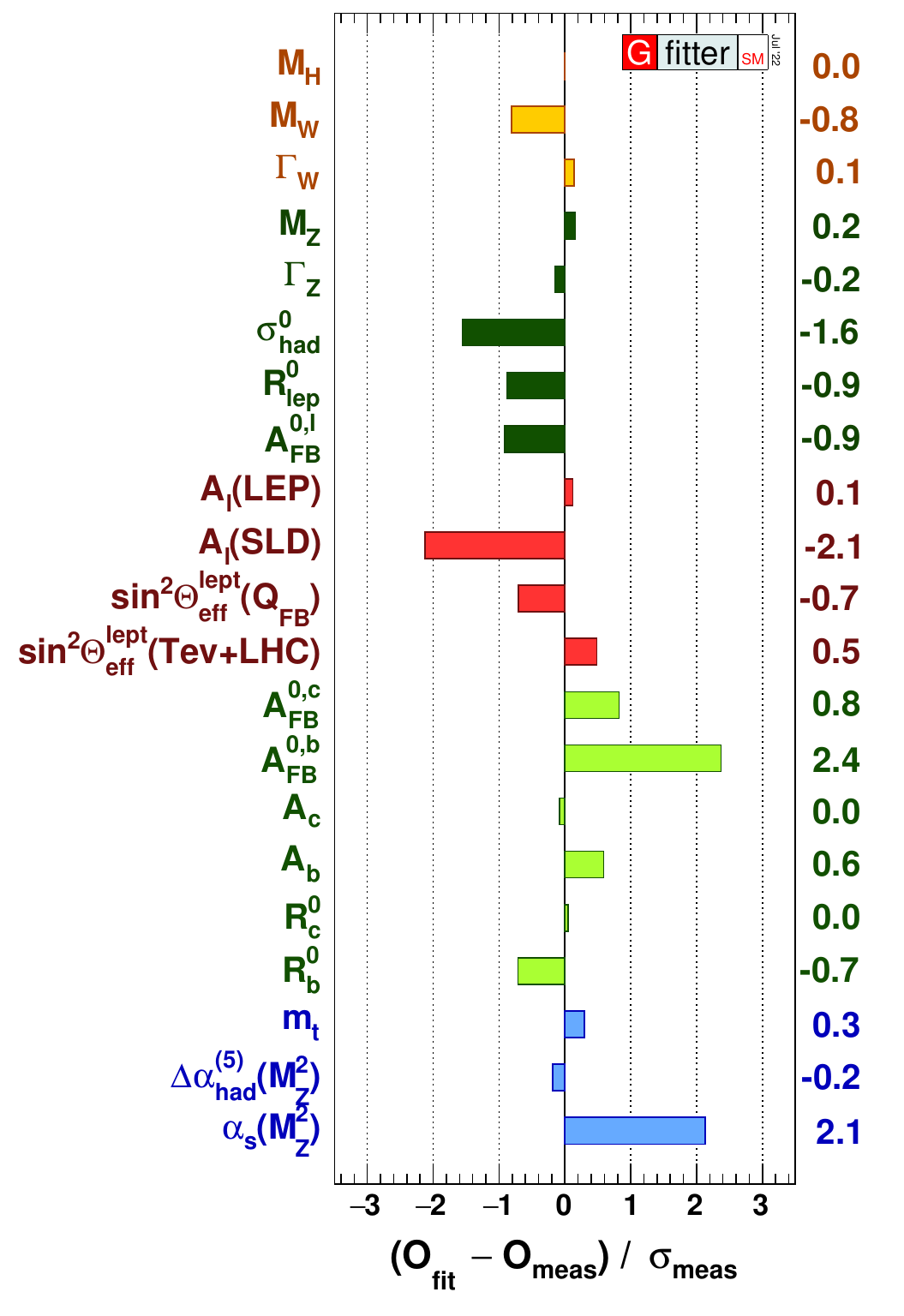}
\caption{{\em Pulls} of the electroweak observables as obtained by a global SM fit, namely differences between SM predictions and direct measurements, normalized to the experimental uncertainties. From \cite{Haller:2022eyb} see also \cite{Haller:2018nnx}.}
\label{fig:pulls2}
\end{center}
\end{figure}

With the discovery of the Higgs boson by the ATLAS \cite{ATLAS:2012yve} and CMS \cite{CMS:2012qbp} experiments at the Large Hadron Collider~(LHC) in 2012, the last missing piece of the Standard Model was observed. The measurement of the Higgs mass also made it possible to complete the determination of all the free parameters of the SM Lagrangian, but for the topological terms.
The overall agreement of the theoretical predictions of the SM with the plethora of available experimental data is remarkable. Especially in the electroweak sector the achieved precision is very high \cite{Haller:2018nnx,deBlas:2021wap}, as highlighted by the results in Fig.~\ref{fig:pulls2}. It is worth stressing that the results shown in this figure are only a small subset of the many tests successfully passed by the SM in the last few years, including also flavor-violating transitions of both quarks and leptons \cite{Isidori:2014rba,Bona:2022xnf}, and high-energy processes \cite{Boyd:2020qox}. In particular, no clear deviation from the SM predictions has been observed in the high-energy distributions analyzed so far by the ATLAS and CMS experiments, which collected an integrated luminosity of about $140\,\text{pb}^{-1}$ each in proton-proton collisions at an energy of $\sqrt{s}=13\,\text{TeV}$ at the~LHC.

\subsection{Motivations and hints for new physics}

Despite the outstanding agreement of the SM with experimental data, there are well known deficiencies that hint at a more fundamental theory. The most important is arguably the lack to incorporate gravity, the fourth known fundamental force of nature, into a coherent QFT framework  valid at arbitrary energy scales. 
As anticipated, the SM does not provide an explanation for cosmological observations such as the baryon asymmetry, dark matter, and dark energy. These phenomena do not necessarily need to find an explanation in the domain of particle physics. However, no convincing alternative explanations have been provided yet and, if interpreted in a QFT framework, they unavoidably point to the existence of new degrees of freedom beyond the SM ones. 

The clear experimental evidence of non-vanishing neutrino masses is also an unambiguous indication that the SM Lagrangian in (\ref{eq:SM_Lagrangian}) is not complete. As we shall discuss in Sec.~\ref{subsec:SMEFT_Operator-basis}, a natural solution to this problem is obtained when interpreting (\ref{eq:SM_Lagrangian}) as the  first  part --more precisely, the leading part containing operators of dimension up to four-- of a more general EFT Lagrangian.
A~serious consistency problem of the SM is also the instability of the Higgs quadratic term in (\ref{eq:Hpotential}) with respect to quantum corrections, the so-called electroweak hierarchy problem~\cite{Barbieri:2017uzd}.
While none of the problems mentioned above points to a well-defined energy scale for the breakdown of the SM, a solution of the 
electroweak hierarchy problem would necessarily require new physics not far from the Fermi scale ($v\approx 246$~GeV). More precisely, we should expect some new degrees of freedom in the few-TeV energy domain able to screen the quadratic sensitivity of the mass term in (\ref{eq:Hpotential}) to possible higher scales in the theory.
The fact that no clear evidence of new physics has been found yet at the LHC has led to consider explanations of this problem beyond the EFT framework~\cite{Giudice:2017pzm}.
However, it is worth stressing that the few-TeV energy domain is still largely unexplored and 
many solutions within the EFT domain are still possible.
This motivates a deeper study of the SM as the low-energy limit of a more complete theory with new degrees of freedom not far from the Fermi scale and thus potentially detectable in near-future experiments.

Beside these general considerations, there are a few specific hints of deviations from the SM predictions observed in precision measurements. 
None of these hints is statistically compelling yet. 
However, they provide a clear illustration of the type of deviations we can expect in the near future, and of the type of effects we can describe within the EFT approach to new physics. This is why we discuss two such hints in more detail below: we will use these results in Sec.~\ref{sect:practical} to illustrate, in practice, the power of the EFT  approach.

\subsubsection{Muon anomalous magnetic moment}
\label{sect:gm2intro}
A long-standing discrepancy between SM predictions and observations concerns the anomalous magnetic moment of the muon.
The magnetic moment of the muon,~$\boldsymbol{\mu}_\mu$, is defined as
\begin{align}
\boldsymbol{\mu}_\mu &= g_\mu \brackets{\frac{e}{2m_\mu}} \boldsymbol{s} \, ,
\end{align}
where $\boldsymbol{s}$ denotes the muon spin and $g_\mu$ is 
the so-called $g$-factor. The prediction from the Dirac equations is $g_\mu = 2$; however, in QFT this value is modified by quantum effects sensitive to heavy degrees of freedom. The interesting quantum effects are parametrized by the anomalous magnetic moment, $a_\mu = \frac{1}{2}\brackets{g_\mu - 2}$. According to the detailed analysis by~\textcite{Aoyama:2020ynm}, the current SM prediction is ${a_\mu^\mathrm{SM}=116 591 810(43) \times 10^{-11}}$. The E989 experiment at FNAL~\cite{Muong-2:2021ojo} recently measured a deviation from this value that, combined  with the previous BNL E821 experiment~\cite{Muong-2:2006rrc}, yields a $4.2\,\sigma$ discrepancy:
\begin{equation}
\Delta a_\mu = a_\mu^\mathrm{Exp} - a_\mu^\mathrm{SM} = \brackets{251 \pm 59} \times 10^{-11} \, .
\label{eq:gm2exp}
\end{equation}
The chance of a statistical fluctuation of this size is below $0.003\,\%$ making this an interesting hint of possible BSM dynamics. We will discuss the possible interpretation of this effect in terms of the SM effective field theory 
in Sec.~\ref{sec:g-2_SMEFT}. However, we warn the reader that there is an intense debate on the reliability of the error in the SM prediction entering 
(\ref{eq:gm2exp}). The main uncertainty is due to hadronic contributions to the photon vacuum-polarization amplitude. The latter is computed either via $\sigma(e^+e^- \to {\rm hadron})$ data and dispersion relations, or via lattice~QCD.  
Recent results from lattice QCD \cite{Borsanyi:2020mff} [see also \cite{FermilabLattice:2022izv,Ce:2022kxy,Alexandrou:2022amy}]
hint at a possibly smaller deviation from the SM than what was obtained in \cite{Aoyama:2020ynm} using dispersive techniques, see also \cite{Colangelo:2022vok}.
More recently, a new measurements of $\smash{\sigma(e^+e^- \to {\rm hadron})}$, presented in \cite{CMD-3:2023alj},
 also shows some discrepancies with previous experimental inputs used in the dispersive approach.

\subsubsection{Lepton universality violation}
\label{sec:LFUV-intro}
Deviations from the SM predictions have recently been reported in tests of lepton flavor universality in semileptonic $B$-meson decays. 
These tests are performed via
universality ratios, such as 
\begin{align}
\label{eq:RD-ratio}
R_{D^{(\ast)}} = \frac{\mathcal{B}\!\brackets{B \to D^{(\ast)} \tau \nu_\tau}}{\mathcal{B}\!\brackets{B \to D^{(\ast)} \ell \nu_\ell}} \,,
\end{align}
where $\ell \in \{\mu,e\}$,
probing the quark-level amplitude  ${b \to c \ell \nu}$,
and similar ratios in neutral-current processes of the type 
${b \to s \ell \ell}$. These ratios can be predicted with high accuracy within the SM due the cancellation of hadronic uncertainties. The latest results on 
$R_{D^{(\ast)}}$ indicate a $3.1\,\sigma$ deviation from the SM predictions \cite{HFLAV:2022pwe}.
We will discuss the possible interpretation of this effect in terms of the SM effective field theory 
in Sec.~\ref{sec:Drell-Yan}. 
Till recently, an even more significant deviation was reported by the LHCb experiment in universality ratios 
in ${b \to s \ell \ell}$ decays; however, this effect has not been confirmed by the latest analysis \cite{LHCb:2022qnv}.


\subsection{Effective field theories}

In physics we are interested in very different length or energy scales. Starting from the scale of the whole universe for cosmological studies all the way down to the scales of elementary particle physics at the LHC, the relevant energy scales indeed vary by many orders of magnitude. Each energy region usually requires its own physical theory to describe its phenomena. Remarkably, 
we often do not need to know in detail the laws at all energies if we want to describe processes at a given scale: it often suffices to set scales that are small or large compared to the process of interest to zero or to infinity, respectively, to get correct results. This is the basic principle of effective theory. We state it as principle; however, in a wide class of quantum field theories, and specifically when considering effective theories with an ultraviolet cutoff, this principle follows from the decoupling theorem 
~\cite{Appelquist:1974tg}.\footnote{Possible exceptions are discussed in~\cite{Donoghue:2009mn}.}

Computations in an effective theory are usually simpler than in the full theory and reproduce the complete results with a  degree of accuracy that can be systematically improved. A~common example is Newtonian mechanics, which is the effective theory of special relativity in the limit of small energies and small velocities. Relativistic (or post-Newtonian) corrections are included by an expansion in the small parameter $v^2/c^2$ to the desired accuracy. 
An excellent description about the essence of effective quantum field theories
is the review by \textcite{Georgi:1994qn}, the article by \textcite{Weinberg:2016kyd}, or \cite{Manohar:2018aog,Skiba:2010xn,Falkowski:2023hsg},
on which our discussion below is based. Recent and further information can be found in the {\em All Things EFT} lecture series \cite{AllThings}.

Quantum EFTs as we use them today grew out of the
attempts to simplify and systematize the calculations of low-energy pion observables, originally based on current algebra techniques. \textcite{Weinberg:1978kz}\footnote{For early work see \cite{Weinberg:1966fm,Dashen:1969ez}. See also~\cite{Weinberg:1980wa, Weinberg:2009bg}.} argued that adhering just to the relevant symmetry properties embodied in current algebra, it is possible to construct an effective Lagrangian of the pion fields able to reproduce all known results and greatly simplifying the treatment. This, together with the work of \textcite{Wilson:1969zs}, that clarified the concept of integrating out heavy states in QFT obtaining universal results, and the related decoupling theorem by~\textcite{Appelquist:1974tg},
put EFTs on a solid basis. Starting from this basis, the systematic construction of the
effective theory of low-energy QCD, namely ChPT, was developed by \textcite{Gasser:1984gg, Gasser:1983yg}. This theory, whose leading expansion parameter is $E/\Lambda_\mathrm{QCD}$, where $E$ is the energy of the process, has been applied with great success to describe with high precision a multitude of low-energy systems. For a recent review see~\cite{Ananthanarayan:2023gzw}.  

Another well-known example of an effective theory is Fermi's theory of weak interactions~\cite{Fermi:1934hr}, 
which is part of the EFT of the Standard Model, and actually the first quantum EFT considered in particle physics (although its recognition as quantum theory, valid also beyond lowest order in the loop expansion, arrived much later).  
While certain amplitudes of the Fermi theory diverge at high energies, thereby violating unitarity, this does not spoil the low-energy limit of the SM, and particularly the 
infrared~(IR) behavior of QCD and~QED, which are still correctly reproduced.

To elucidate in simple terms the basic concepts of quantum EFT,
let us consider a theory containing two types of fields $\phi_L$ and $\phi_H$. Let us further assume $m \ll M$, where $m$ denote the mass of the excitations of $\phi_L$, and $M$ the one associated to $\phi_H$. 
The generating functional of the sources~$J_L$ associated to the light fields, and the corresponding 
EFT Lagrangian, can be obtained by performing the path integral over the heavy fields
\begin{align}
Z[J_L] &= \! \int \!\!  \mathcal{D} \phi_L \exp\! \left[ \int \!\! \dd^4 x  \left(\L_\mathrm{EFT}\!\brackets{\phi_L} + \phi_L J_L\right) \right] 
\label{eq:EFT_integrate-out}
\\
&=\! \int \!\!  \mathcal{D} \phi_H   \mathcal{D} \phi_L \exp\!\left[ \int \!\! \dd^4 x \left( \L\!\brackets{\phi_L , \phi_H} + \phi_L J_L\right) \right]. 
\nonumber
\end{align}
This formal manipulation, usually referred to as \textit{integrating out} the heavy degrees of freedom, essentially amounts to averaging over all $\phi_H$ configurations. The $\L_\mathrm{EFT}\brackets{\phi_L}$ thus
obtained contains non-local operators built only out of the light fields. Using an operator product expansion we can then express $\L_\mathrm{EFT}$ as a generally infinite sum of higher-dimensional operators
\begin{align}
\L_\mathrm{EFT} &= \L_{d \leq 4} + \sum_{d=5}^\infty \frac{1}{M^{d-4}} \sum_{i=1}^{n_d} C_i^{(d)} Q_i^{(d)} \, ,
\label{eq:EFT_Lagrangian}
\end{align}
where $d$ is the (mass) dimension of the operator~$Q_i^{(d)}$, 
and $n_d$ is the number of independent operators at a given dimension~$d$, 
which is always finite. The effective couplings~$\smash{C_i^{(d)}}$, 
associated to each operator, are dubbed \textit{Wilson coefficients}.
This procedure of integrating out the heavy fields changes the ultraviolet~(UV) structure of the theory, but it ensures that the EFT is constructed in such a way as to reproduce the same low-energy behavior as the original theory.

As can be seen in Eq.~\eqref{eq:EFT_Lagrangian}, the higher-dimensional operators are suppressed by inverse powers of the mass scale~$M$ of the heavy fields. 
Computing physical observables using $\L_\mathrm{EFT}$ thus leads to an expansion in powers of $E/M$, where $E$ is the typical energy scale of the process of interest. 
The EFT description is valid if $E \sim m \ll M$, i.e., if the energies probed are far below the mass scale of the heavy states and only the light particles can be produced on-shell.
This energy region is exactly where the EFT offers a valid approximation of the underlying theory. 
It is then sufficient to truncate the sum over~$d$ in Eq.~\eqref{eq:EFT_Lagrangian} at some finite order depending on the required accuracy of the result, since  higher-dimensional operators contribute with higher powers of the suppression factor~$E/\Lambda$.
More details on the validity of the EFT approach can be found in Sec.~\ref{sect:dim8}.

Since the operators~$\smash{Q_i^{(d)}}$ in Eq.~\eqref{eq:EFT_Lagrangian} are of mass dimension~$d>4$, these terms are non-renormalizable in the traditional sense, that is all infinities cannot be absorbed in a finite number of coefficients. For example, a divergent Feynman graph with two insertions of a $d=5$ operator is of order~$\smash{\ord{M^{-2}}}$ and therefore requires a counterterm of mass dimension~$d=6$. A diagram with two insertions of this counterterm would then require a $d=8$ counterterm and so on. Thus an infinite set of operators would be required to render the theory finite. However, the EFT comes with an associated expansion in powers of~$E/M$: if all terms with more than $k$~powers of this parameter are neglected, only a finite set of parameter remains and the theory can be renormalized in the usual sense. This means that all infinities up to terms of order~$\smash{\brackets{E/M}^k}$ can be canceled by a finite set of couplings, and that the corresponding renormalization group~(RG) equations can be derived.

The procedure of integrating out heavy particles as shown in Eq.~\eqref{eq:EFT_integrate-out} can be performed repeatedly. Suppose we have a theory with particles at several well separated mass scales~$\Lambda_1 \gg \Lambda_2 \gg \Lambda_3 \gg \ldots$. We can first integrate out the heavy particles at the scale~$\Lambda_1$ then compute the RG equations of the resulting EFT to run the theory down from~$\Lambda_1$ to~$\Lambda_2$. Next, we can integrate out the particles at the mass scale~$\Lambda_2$ obtaining a second EFT only containing practices with masses~$\lesssim \Lambda_3$. Again we can compute the RG equations of the new EFT to run down to the scale~$\Lambda_3$ and so on, until we reach the desired mass scale. The advantage of this multi-step procedure is the systematic resummation of large logarithms, that would appear in the matching steps if we would only do a single matching at the desired scale and integrate out all heavy particles at once.

The scenario as described above can be viewed as \textit{top-down} approach to EFTs: we start with a known theory at the high scale and integrate out the heavy particles. This is the adequate procedure when we strive to make precise predictions from a known theory with known UV behavior.
However, EFTs can also be a useful tool if the full theory at the high scale is unknown, but only some of its features. This was for instance the case for the strong interactions before the discovery of the $\mathrm{SU}(3)_c$~gauge theory which was helped by the work on current algebra and chiral perturbation theory. This scenario is often referred to as \textit{bottom-up approach}. This is also the case for  the present situation, where the Standard Model is known and one would like to understand the underlying theory. This is the approach of the SM effective theory. In this case the operators~$\smash{Q_i^{(d)}}$ in Eq.~\eqref{eq:EFT_Lagrangian} do not emerge in the matching procedure, but have to be constructed using symmetry arguments. Suppose we want to find the EFT operators for the~SM. In this case we have $\L_{d \leq 4}=\L_\mathrm{SM}$ and for the EFT operators~$\smash{Q_i^{(d)}}$ at a given mass dimension~$d$ we simply have to construct all structures that are invariant under the local and global symmetries of the theory of interest, i.e., the~SM. In this bottom-up setup we usually replace the explicit mass~$M$ in Eq.~\eqref{eq:EFT_Lagrangian} by a generic UV scale~$\Lambda$, that can be identified with a heavy BSM mass scale once the EFT is matched to some UV theory. In the remainder of this review we will focus on these EFT extensions of the SM, with particular emphasis on the so-called~SMEFT.


\subsection{The Standard Model as an effective theory}
\label{smeft}

As mentioned above, the Standard Model can be interpreted as the leading-order dimension-four piece of a larger effective theory. This EFT must have the same gauge  symmetries as the~SM. The gain of embedding the unknown physics into an effective theory is that it applies to all particle-physics processes and thus allows us to use a common framework to relate results of different experiments. There are actually two candidate EFTs that are distinguished only by their assumptions on the realization of the electroweak symmetry group. The Standard Model effective field theory~(SMEFT) assumes that the electroweak symmetry is realized linearly, whereas the Higgs effective field theory~(HEFT) allows us to consider the more general case of a
non-linear realization.\footnote{The HEFT is sometimes also called the electroweak chiral Lagrangian~(EWChL).} Within the SM both versions are equivalent as they are related by a field redefinition. However, they lead to different EFT descriptions  as in the EFT framework it is not always possible to find a field redefinition to go from a non-linear to a linear realization of the electroweak symmetry (we review this issue in more detail in Sec.~\ref{sec:HEFT}). 
The HEFT is thus a more general theory containing the~SMEFT as special case.
In particular, the HEFT scenario applies also to BSM theories where the Higgs is part of a strongly-interacting and not fully decoupled sector.

In this review we will focus mainly on the SMEFT, on the one hand because of its ``simplicity'', on the other hand because  present data on SM precision tests and Higgs couplings seem to favor a linearly realized electroweak symmetry, i.e., a fundamental (or quasi-fundamental) Higgs field transforming as as doublet of~$\mathrm{SU}(2)_L$.
For an extensive discussion about differences between  HEFT and SMEFT we refer to~\cite{Brivio:2017vri}.

Applying the general concepts of EFT discussed in the previous section, we can decompose the SMEFT Lagrangian as
\begin{samepage}
\begin{align}
\L_\mathrm{SMEFT}(\psi,H,A) &= \L_\mathrm{SM}(\psi,H,A) 
\label{eq:SMEFT_Lagrangian_general}
\\
&\quad+ \sum_{d=5}^{\infty} \sum_{i=1}^{n_d} \frac{C_i^{(d)}}{\Lambda^{(d-4)}} Q_i^{(d)}(\psi,H,A)\, .
\nonumber
\end{align}%
\end{samepage}%
Here, $\psi$, $H$, and~$A$ collectively denoted the SM fermion, Higgs, and gauge fields, respectively, as listed in Tab.~\ref{tab:SM_field-content}. 
The key assumption of this construction is indeed the hypothesis that physics beyond the SM is characterized by one or more heavy scales. 
As in most of the literature, we adopt the convention where 
the Wilson coefficients~$C_i^{(d)}$ are dimensionless quantities, this is why we
pull out explicitly the factor $\Lambda^{(4-d)}$ in the effective couplings. 
In principle, the sum on $d$ runs over all possible values; however, the majority of our discussion will be focused on operators up to dimension~six, and therefore we often drop the superscript~$(d)$ denoting the operator dimension.

After fixing the mass dimension up to which we expand the EFT, which is equivalent to determining the desired accuracy of our result, $\L_\mathrm{SMEFT}$ is capable of describing the low-energy signatures of generic UV completions of the~SM. 
One of the less trivial aspect of this approach is the construction 
of a suitable basis of operators 
at a given dimension. Not surprisingly, a long time passed from the initial formulation of a complete basis 
for the SMEFT at dimension~six by \textcite{Buchmuller:1985jz}, till the identification of a complete and non-redundant basis by \textcite{Grzadkowski:2010es}.
We will review in detail how this is done in general, 
and specifically for the SMEFT up to dimension~six, in Sec.~\ref{subsec:SMEFT_Operator-basis}.

In many realistic UV completions, 
the physics above the electroweak scale is characterized by several mass scales. 
What matters to determine the convergence of the EFT expansion is the lowest of such scales, that we can identify with~$\Lambda$. 
However, the presence of additional energy scales can play a role 
in determining the size of the~$\smash{C_i^{(d)}}$, given the conventional choice of assuming a 
unique normalization scale $\Lambda$ in (\ref{eq:SMEFT_Lagrangian_general}). We will come back to this point in more detail at the end of Sec.~\ref{sect:SMEFT} and in Sec.~\ref{sec:GlobalSymmetries}.

The two key assumptions of this construction in describing generic extensions of the SM is that no unknown light particles exist and the electroweak symmetry is linearly realized. 
Under these hypotheses, any experimental result on the search for new physics can be given in the framework of the SMEFT, i.e.,~in terms of bounds on the Wilson coefficients, if the energies probed in the experiment are well below the scale of new physics. At the same time, different models of new physics can be matched onto the SMEFT Lagrangian by integrating out the heavy particles in each theory. More interestingly, if a deviation from the SM emerges, the SMEFT can be used to test its consistency pointing out correlated observables and discriminating among large varieties of UV completions. Illustrating all this with concrete examples is the subject of Sec.~\ref{sect:practical}.

The absence of light new particles is definitely a strong hypothesis. Several examples of light new states, such as axion-like particles or the dilaton, are well motivated and can originate by physics at energies far beyond the weak scale. However, such new states are necessarily very weakly coupled to the SM fields (otherwise they would have already been discovered). This implies we can neglect their effect in a large class of observables, for which the description in terms of the SMEFT remains a very efficient tool. Of course, to describe in full generality these frameworks requires to add the corresponding light fields in the EFT. This can be done, case by case, according the nature of the new degrees of freedom, but is beyond the scope of this review. 

\section{Standard Model effective field theory}
\label{sect:SMEFT}
In this section, we provide a comprehensive introduction to the SMEFT.
We start presenting general arguments on how to find an operator basis and then focus on the construction of the commonly used \textit{Warsaw basis} \cite{Grzadkowski:2010es}. 
In Sec.~\ref{subsec:SMEFT_OperatorSize}, we analyze how the size of the different operator coefficients can be estimated using general theoretical considerations. We conclude
in Sec.~\ref{subsec:SMEFT_Constraints} analyzing some constraints on the Wilson coefficients and discussing the validity of the EFT approach to describe BSM physics.

\subsection{Operator bases}
\label{subsec:SMEFT_Operator-basis}
On general grounds, we consider the SMEFT in a
bottom-up EFT perspective: we know the low-energy limit of the theory, which is~the Standard Model, while we do not know its UV~completion. 
The goal is to find a general description, in terms of higher-dimensional operators, of the effects generated by integrating out heavy degrees of freedom that are a priori unknown. In the absence of a clear UV~theory to start with,  we constrain the set of operators using only symmetry arguments. The symmetries we assume are  Lorentz invariance, the SM gauge symmetry, $\mathcal{G}_\mathrm{SM}$,
and possible additional global symmetries, such as baryon and lepton number. With the known symmetries, it becomes a pure group theory exercise --although a non-trivial one-- to construct all the allowed operators. 

Concerning the global symmetries, it is not obvious if properties of the SM, such as baryon and lepton number, are fundamental symmetries of the underlying theory or approximate symmetries arising accidentally at low energies. We postpone a detailed discussion of this point to Sec.~\ref{sec:GlobalSymmetries}. On the other hand, there is no doubt that the SM local symmetry provides a useful and unambiguous tool to classify the higher-dimensional operators, since the UV theory must have a local symmetry group that includes~$\mathcal{G}_\mathrm{SM}$ as a subgroup. 

For the construction of an operator basis, we will restrict ourselves for now to work only up to mass-dimension six. To this end, we express the SMEFT Lagrangian as
\begin{align}
\L_\mathrm{SMEFT} &= \L_\mathrm{SM} + \frac{1}{\Lambda} \, \L_5 + \frac{1}{\Lambda^2} \, \L_6 + \ord{\Lambda^{-3}} \, , 
\end{align}
where $\L_{5(6)}$ contains all dimension-five~(-six) operators. 

As an illustration, we construct, following~\cite{Buchmuller:1985jz}, the dimension-five piece~$\L_5$, which consists of a single term: the so-called Weinberg operator \cite{Weinberg:1979sa}, and its hermitian conjugate. For dimensional reasons it is impossible to form a dimension-five operator only out of fermions or only out of field-strength tensors. It can also not be built only out of Higgs doublets~$H$ due to gauge invariance. For the same reason, or due to Lorentz invariance, it is also impossible to combine three scalars with a field-strength tensor. In principle the combination of a field-strength tensor and a fermion bilinear is of the right dimension, but for it to be Lorentz invariant the fermion bilinear would have to be a tensor current, which necessarily transforms as an $\SU{2}_L$~doublet, therefore violating gauge invariance. Thus, the only remaining option is to combine two scalars and two fermions. If we choose $H$ and $H^\ast$ as the scalars, the net hypercharge of the fermion product must vanish, which is only possible by choosing a fermion and its charge conjugate, but this combination does not yield a Lorentz scalar. Therefore, both scalars must be $H$ and combine into an $\SU{2}_L$~triplet, as the singlet combination vanishes. Then both fermions also have to be $\SU{2}_L$ doublets that combine into a triplet and carry no color to form a gauge invariant operator. The resulting operator can be written as 
\begin{align}
Q_\mathrm{Weinberg} &= \varepsilon^{ik} \varepsilon^{jl} H_k H_l \bar{\ell}^c_{i} \ell_j \, ,
\label{eq:Weinberg-Operator}
\end{align}
where we have explicitly shown the $\mathrm{SU}(2)_L$ indices~$(i,j,k,l)$ and 
suppressed the flavor ones.\footnote{The fully anti-symmetric rank-two tensor~$\varepsilon^{ij}$ is defined by $\smash{\varepsilon^{ij}=-\varepsilon^{ji}}$ and~$\varepsilon^{12}=\varepsilon_{12}=+1$, and 
the superscript~${}^c$ denotes 
the charge conjugate of a fermion given by $\psi^c=C{\overline{\psi}}^\intercal$ with the charge conjugation matrix~$C = i \gamma^2 \gamma^0$.} 
After electroweak symmetry breaking, the Weinberg operator introduces a Majorana mass for the left-handed neutrinos~$\nu_L$: 
$\langle\smash{Q_\mathrm{Weinberg}}\rangle = \smash{({v^2}/{2})\,\bar{\nu}_L{}^{\!c} \,  \nu_L}$, where $v/\sqrt{2}$~is the vacuum expectation value of~$H$. 
The  operator $Q_\mathrm{Weinberg}$ violates one of the global symmetries of the SM Lagrangian: 
it violates total lepton number by two units. As we shall discuss in more detail in Sec.~\ref{sec:GlobalSymmetries}, this fact could naturally justify its smallness and, correspondingly, the smallness of neutrino masses. 
Postponing a discussion about global symmetry violations to 
Sec.~\ref{sec:GlobalSymmetries}, in the rest to this section we 
focus on lepton and baryon number conserving operators, which start at dimension six.

Beside the continuous global symmetries mentioned above,
one can constrain the SMEFT structure also via the discrete global 
charge-parity~(CP) symmetry, that experimentally 
is violated only in specific 
flavor-changing processes, as predicted in the SM.
 Contrary to continuous symmetries, imposing 
 CP invariance does 
 not limit the operator structures, but rather the form of the allowed couplings: non-Hermitian operators are not allowed to 
appear in the Lagrangian with imaginary couplings. 
However, requiring only real Wilson coefficients 
does not offer a sufficient protection from CP violation, since CP-even operators can still interfere with the CP-violating phase of the SM. 
This form of indirect CP violation, also called opportunistic CP violation, allows us to derive additional constraints on CP-even operators from measurements of CP-violating observables.
For more details about CP violation in the SMEFT see \citep{Bonnefoy:2021tbt,Bonnefoy:2023bzx}.

The operators of~$\L_6$ can be obtained by considerations analogous to those presented 
to derive Eq.~(\ref{eq:Weinberg-Operator}).
We will list a minimal and independent set of them in Sec.~\ref{subsubsec:SMEFT_Warsaw-Basis}. The first complete SMEFT operator set up to dimension six was constructed in the original analysis by \textcite{Buchmuller:1985jz}.\footnote{In fact, one operator was missing in the printed version of this paper, but mentioned in \cite{Buchmuller:1987ur}.} 
Some extensive lists of previously known operators have already been given in~\cite{Leung:1984ni} and the references therein. However, these lists contain many redundant operators which have been eliminated in~\cite{Buchmuller:1985jz}, which however still did not provide a minimal basis. This goal was achieved later on in \cite{Grzadkowski:2010es}. In the following, we discuss general arguments on how different effective operators can be related and how an independent set can be obtained.

\subsubsection{Toward a non-redundant basis}\label{sec:redundant-operators}
A~set of effective operators constructed with the procedure illustrated in the example above
usually contains many redundancies.\footnote{The effective operators form a complex vector space and the redundancy in the operator choice is equivalent to the redundancy in defining a basis for this vector space \cite{Einhorn:2013kja}. We also call a minimal set of operators an operator basis.} Two or more operators or a larger set of operators are redundant if they yield the same contribution to all physical observables, hence some of them can be dropped with no physical consequences if the coefficients of the remaining operators are modified accordingly. Redundant operators can be eliminated using various techniques.
The most relevant ones are: {\em a})~Integration by parts; {\em b})~Field redefinitions (and equations of motion);
{\em c})~Fierz identities;
{\em d})~Dirac structure reduction. 
We now proceed discussing each of them in more detail.
Notice however that it might be necessary to perform further simplifications, for example by applying the Jacobi/Bianchi or the Chisholm identities, to obtain a minimal operator basis for the EFT. Furthermore, it might be required to exploit the internal symmetries of the EFT operators, such as (anti-)symmetric indices. Thus, the discussion below is not meant as a complete description for reducing a given operator set to a basis, but only highlights the most common methods used in this procedure.

\paragraph{Integration by parts.} Within QFT we commonly assume that total derivatives vanish, i.e., all fields vanish at infinity. Thus the action~$S$ of the theory,  $S=\int \dd^4 x \, \L$, is invariant under integration by parts~(IBP) identities. As a consequence, we can use IBP to relate different operators. In the SM this can, for example, be used to write the kinetic term for the Higgs in the two equivalent forms ${(D_\mu H)^\ast (D^\mu H)}$ and ${-H^\ast D^2 H}$. The same technique can be applied also to rewrite higher-dimensional effective operators in the~SMEFT.

\paragraph{Field redefinitions.} 
The probably most relevant form of equivalence among different effective operators is due to field redefinitions. According to the LSZ reduction formula \cite{Lehmann:1954rq} we are free to choose any form for the interpolating quantum fields of our theory without affecting physical observables, as long as the fields we use can create all the relevant states from the vacuum. This freedom allows us to perform field redefinitions for our effective Lagrangian modifying the operators and, in practice, reducing the operator basis, but leaving the physical observables invariant \cite{Politzer:1980me,Georgi:1991ch,Arzt:1993gz}. The field redefinitions of interest for the SMEFT are perturbative transformations of the type 
\begin{align}
\phi \rightarrow \tilde\phi (\phi) = \phi + \epsilon F(\phi) \, ,
\label{eq:SMEFT_field-redefinition}
\end{align}
where the new field~$\tilde\phi$ is given by the original field~$\phi$ plus some small~($\epsilon \ll 1$) perturbation~$F(\phi)$ that can depend not only on the field $\phi$ itself, but also on all the other fields of the~SM and their covariant derivatives. We furthermore assume that $F$ is an analytic function of the SM fields, their derivatives, and of~$\epsilon$. Usually for the SMEFT the expansion parameter~$\epsilon$ is related to some power~$n$ of the EFT expansion parameter~$(E/\Lambda)^n$, where $E$ is the typical energy scale for the process of interest. 

Following the work of \textcite{Criado:2018sdb}, we will now show that field redefinitions leave the $S$-matrix, and by that all observables, invariant. Let the generating functional of the SM be
\begin{align}
Z_\mathrm{SM}[J] &= \int \mathcal{D}\phi \, \exp\brackets{i S_\mathrm{SM}[\phi] + J\phi}
\label{eq:SMEFT_generating-functional}
\end{align}
with $\phi$ representing all SM fields collectively and $J$~being the corresponding 
source terms. Using that the field redefinition in Eq.~\eqref{eq:SMEFT_field-redefinition} is always invertible in a perturbative sense, we can perform a coordinate transformation for the path integral in Eq.~\eqref{eq:SMEFT_generating-functional}
\begin{align}
Z_\mathrm{SM}[J] &= \!\! \int \!\! \mathcal{D}\phi \, \det\!\brackets{\frac{\delta \tilde\phi (\phi)}{\delta \phi}} \exp\brackets{i S_\mathrm{SM}[\tilde\phi(\phi)] + J\tilde\phi(\phi)} .
\end{align}
Thus a field redefinition in the action $\tilde S_\mathrm{SM}[\phi] = S_\mathrm{SM}[\tilde\phi(\phi)]$ leaves the resulting generating functional invariant if it is accompanied by the Jacobian of the transformation and an appropriate transformation of the source terms. 

Using ghost fields~$\eta$ and~$\bar\eta$ we can write the Jacobian as
\begin{align}
\det\brackets{\frac{\delta \tilde\phi (\phi)}{\delta \phi}} &= \int \mathcal{D}\bar\eta \, \mathcal{D}\eta \, \exp\brackets{-i\bar\eta \frac{\delta\tilde\phi(\phi)}{\delta\phi}\eta} \, .
\end{align}
We can then simply add the ghost part to the action~$S_\mathrm{SM}$. Using Eq.~\eqref{eq:SMEFT_field-redefinition} we find that the ghost propagator is proportional to the identity and ghost loops can only depend on $\delta F(\phi)/\delta\phi$, which is a polynomial in the internal momenta since $\tilde\phi$ is analytic in the fields and their derivatives. In dimensional regularization, which we assume throughout this work,  
these scaleless loops thus vanish. Therefore, the Jacobian of the coordinate transformation is the identity and we can simply neglect the ghosts.

The modification of the source terms affects off-shell quantities, however, due to the LSZ formula~\cite{Lehmann:1954rq} the source terms do not alter the $S$-matrix and by that the physical observables. This means that the generating functional with the action obtained after the field transformation
\begin{align}
\tilde Z_\mathrm{SM}[J] &= \int \mathcal{D}\phi \, \exp\brackets{i \tilde S_\mathrm{SM}[\phi] + J\phi}
\end{align}
yields the same $S$-matrix as the original generating functional~$Z_\mathrm{SM}[J]$ and, therefore, they are physically equivalent. For a more detailed analysis and further information on the treatment of fields with non-zero vacuum expectation values and a discussion of the inclusion of renormalization see \cite{Criado:2018sdb}.

Next, we give a concrete example how field redefinition can be used to eliminate redundant operators from the~SMEFT. Consider the SM amended by the two effective operators
\begin{align}
[Q_{Dl}]_{pr} &= \big(\bar\ell_p (\overleftarrow{\slashed{D}} + \overrightarrow{\slashed{D}}) \ell_r\big)(H^\dagger H)\,, \\
[Q_{eH}]_{pr} &= (\bar\ell_p^i e_r) H_i (H^\dagger H)\,, 
\end{align}
with the corresponding Wilson coefficients $[C_{Dl}]_{pr}$ and~$[C_{eH}]_{pr}$. 
Here, $p$~and~$r$ are flavor indices and $i$ is a fundamental $\SU{2}_L$ index only shown when the contraction is nontrivial.
Our goal is to show that both of these operators are equivalent. We first notice that both operators are of mass-dimension six and are allowed by the SM symmetries. Moreover, $Q_{Dl}$ is hermitian contrary to~$Q_{eH}$. We can now write the part of the SMEFT Lagrangian relevant for this example:
\begin{align}
\begin{split}
\L_\mathrm{SMEFT} &\supset i (\bar\ell_p \slashed{D}\ell_p) - \brackets{[Y_e]_{pr} (\bar\ell_p^i e_r) H_i + \mathrm{h.c.}} 
\\
&+ \left(\frac{[C_{eH}]_{pr}}{\Lambda^2} [Q_{eH}]_{pr} + \mathrm{h.c.} \right) 
\\
&+ \frac{[C_{Dl}]_{pr}}{\Lambda^2} [Q_{Dl}]_{pr} + \mathcal{O}(\Lambda^{-4}) \, .
\end{split}
\label{eq:SMEFT_operator_reduction_example}
\end{align}
We now apply the perturbative field redefinitions
\begin{align}
\ell_{i p} &\rightarrow \ell_{i p} + \frac{1}{\Lambda^2} F_{i p}(\ell,H) \, , & \bar\ell_p^i &\rightarrow \bar\ell_p^i + \frac{1}{\Lambda^2} \overline{F_p^i(\ell,H)} \, ,
\end{align}
where $F$ is some analytic function of the fields~$\ell,\bar\ell,H$, and~$H^\dagger$ and their derivatives. Since $\ell$ is a complex field we also have to shift its charge conjugate, or equivalently~$\bar\ell$. Using that the field redefinition is perturbative in our EFT expansion, i.e., keeping a consistent truncation at mass-dimension six, we find
\begin{align}
\L_\text{SMEFT} &\to i (\bar\ell_p \slashed{D}\ell_p) + \frac{i}{\Lambda^2} \brackets{\overline{F}_p \overrightarrow{\slashed{D}} \ell_p - \bar{\ell}_p \overleftarrow{\slashed{D}} F_p} \\
&- \!\brackets{\![Y_e]_{pr} (\bar\ell_p^i e_r) H_i + \frac{1}{\Lambda^2} [Y_e]_{pr} \!\brackets{\overline{F}_p^i e_r}\! H_i + \mathrm{h.c.}\!} \nonumber\\
&+ \frac{[C_{Dl}]_{pr}}{\Lambda^2} \brackets{\bar\ell_p (\overleftarrow{\slashed{D}} + \overrightarrow{\slashed{D}}) \ell_r} (H^\dagger H) \nonumber\\
&+ \brackets{\frac{[C_{eH}]_{pr}}{\Lambda^2}(\bar\ell^i_p e_r)H_i (H^\dagger H) + \mathrm{h.c.}} + \ord{\Lambda^{-4}}  \nonumber,
\end{align}
where we used IBP for the last term of the first line to move the derivative away from the function~$F$.
We observe that by choosing $F_{i p} = -i [C_{Dl}]_{pr} \,\ell_{i r} (H^\dagger H)$ the two terms originating from shifting the kinetic term of the fermions cancel exactly the operator~$Q_{Dl}$. The final result we thus obtain reads
\begin{align}
\begin{split}
\L_\mathrm{SMEFT} &\supset i (\bar\ell_p \slashed{D}\ell_p) - \left( [Y_e]_{pr} (\bar\ell_p^i e_r) H_i + \mathrm{h.c.} \right) 
\\
&+ \!\left(\!\frac{[C_{eH}^\prime]_{pr}}{\Lambda^2} [Q_{eH}]_{pr} + \mathrm{h.c.} \!\right)\! + \ord{\Lambda^{-4}} .
\end{split}
\label{eq:SMEFT_non-redundant_Lagrangian}
\end{align}
We have found that the operator $Q_{Dl}$ is redundant and it is sufficient to only include~$Q_{eH}$ in the Lagrangian. The effect of removing the redundant operator~$Q_{Dl}$ in our example is a shift of the Wilson coefficient of the remaining operator~$Q_{eH}$ given by $[C_{eH}^\prime]_{pr} = [C_{eH}]_{pr} - i [C_{Dl}]_{ps} [Y_e]_{sr}$. Equally well we could also have removed $Q_{eH}$ in favor of $Q_{Dl}$ with the field redefinition $\ell_p \to \ell_p + [A]_{pr} \ell_r (H^\dagger H)\big/\Lambda^2$ where $A$ is the matrix defined by $[A]_{ps} [Y_e]_{sr} = [C_{eH}]_{pr}$. However, it is often more convenient to remove the operators with more derivatives in favor of operators with fewer derivatives, which is also the strategy we will pursue in the following. The procedure presented above can be used to eliminate any operator that is redundant due to field redefinitions. In the case where we remove an operator with derivatives it is always the shift of the kinetic term that cancels the redundant effective operator. 

In many cases, including the SMEFT, when only keeping effective operators of mass-dimension six, there is a simpler way of removing redundant operators than using field redefinitions. It can be shown that at leading power in the EFT expansion the use of equations of motion is equivalent to applying field redefinitions, which we will prove below.

Consider a Lagrangian~$\L$ depending on the fields~$\phi$, e.g., the SM Lagrangian depending on all the SM fields. We then perform a perturbative field redefinition of the form $\phi \to \smash{\tilde\phi} = \phi + \epsilon \delta\phi$ on the Lagrangian, where $\epsilon$ is again a small ($\epsilon \ll 1$) expansion parameter related to some power~$n$ of the EFT expansion~$(E/\Lambda)^n$. Expanding the shifted  action  around the original field configuration~$\phi$ we find
\begin{align}
S[\phi] \rightarrow S[\tilde\phi] 
&= \left. S[\tilde\phi]\right|_{\tilde\phi=\phi} + \epsilon \left. \frac{\delta S[\tilde\phi]}{\delta\tilde\phi} \right|_{\tilde\phi=\phi} \delta\phi + \ord{\epsilon^2} 
\nonumber\\[0.1cm]
&= S[\phi] + \epsilon \int \dd^4 x \, E[\phi] \delta\phi + \ord{\epsilon^2}
\label{eq:SMEFT_field-redefinitions_EOM}
\end{align}%
at leading order in~$\epsilon$, where $E[\phi] = \big({\delta \L[\tilde\phi]}\big/{\delta\tilde\phi}\big) \big|_{\tilde\phi=\phi}$ symbolized the equations of motion of the field~$\phi$. Therefore, instead of performing a field redefinition we can also add a term proportional to the equations of motion of a field to the Lagrangian, which at leading power has the same effect. Since we work up to order~$\ord{\epsilon}$ it is also sufficient to only use the leading piece of the equations of motion. That means for the SM we can just use the pure SM equations of motion dropping all contributions of higher-dimensional operators.  

We can now come back to our example from Eq.~\eqref{eq:SMEFT_operator_reduction_example} and find that the operator we removed before is indeed proportional to the equations of motion for the fields $\ell$ and~$\bar\ell$. We already know that in this case $\epsilon=\Lambda^{-2}$. Thus we can use the leading SM equations of motion
\begin{subequations}
\begin{align}
E[\bar\ell\,]_{i p} &= i\overrightarrow{\slashed{D}} \ell_{i p} - [Y_e]_{pr} e_r H_i + \ord{\Lambda^{-2}} \, ,
\\
E[\ell]_p^i &= -i\bar{\ell}_p^i \overleftarrow{\slashed{D}} - [Y_e^\ast]_{pr} \bar{e}_r {H^\ast}^i + \ord{\Lambda^{-2}} \, .
\end{align}
\end{subequations}
Then adding the term $\epsilon \overline{\delta\phi}_p^i E[\bar\ell\,]_{i p}  + \epsilon E[\ell]_p^i \delta\phi_{i p}$ with $\delta\phi_p = F_p(\ell,H)$ to the Lagrangian in Eq.~\eqref{eq:SMEFT_operator_reduction_example} yields the same result than using field redefinitions. Notice that since~$\ell$ is a complex field we need to use the equations of motions for both the field and its charge conjugate.
In practice it is easier to directly plug in the equations of motion in the effective operators we want to remove. In our example we could simply replace $\overrightarrow{\slashed{D}} \ell_{i p}$ and $\bar{\ell}_p^i \overleftarrow{\slashed{D}}$ in the operator~$Q_{Dl}$ by $- i [Y_e]_{pr} e_r H_i$ and $ i[Y_e^\ast]_{pr} \bar{e}_r {H^\ast}^i$, respectively, directly obtaining the result in Eq.~\eqref{eq:SMEFT_non-redundant_Lagrangian}.

In the literature it is often stated that some operators are removed by means of the equations of motion. This statement is not strictly correct in general, since the equations of motion can only be used at leading order in the EFT expansion. If we work at subleading power~$\epsilon^2$, e.g., include dimension-eight operators in our previous example, we should add the term 
\begin{align}
	\left. \frac{1}{2} \epsilon^2 \frac{\delta^2 S[\tilde\phi]}{\delta\tilde\phi^2} \right|_{\tilde\phi=\phi} \delta\phi^2
    \label{eq:non-linear-shift}
\end{align}
to Eq.~\eqref{eq:SMEFT_field-redefinitions_EOM} to obtain a consistent truncation of the EFT expansions up to order~$\ord{\epsilon^2}$. It is immediately clear that in this case the use of equations of motion is no longer equivalent to applying field redefinitions as the former do not capture the subleading shift of the fields in Eq.~\eqref{eq:non-linear-shift}. Therefore, when considering a Lagrangian with effective operators of different powers we must not use the equations of motion to remove redundancies but we have to apply the field redefinitions to obtain the correct result. For more details on the failure of equations of motion see \cite{Criado:2018sdb,Jenkins:2017dyc}.

The common approach is thus to first use IBP, if necessary, to bring an operator into the form of the equations of motion, and then use these to eliminate the operator in favor of other effective operators containing fewer derivatives. If we work at subleading power in the EFT, the equations of motion cannot be used and we have to apply field redefinitions instead. In this situation, we have to remove the redundant operators order by order starting with the lowest order operators, since a shift to eliminate an operator produces operators of the same or of higher mass dimension when shifting massless fields.\footnote{In the SMEFT, only the Higgs~$H$ has a mass term, thus shifting it to remove a redundant operator can introduce lower-dimensional operators.}

\paragraph{Fierz identities.} 
These identities follow from completeness relations on certain matrix spaces, and provide additional relations among operators. 
We start by discussing Fierz identities of the Lorentz group~\cite{fierz:1937}. These identities can be applied to four-fermion operators, allowing us to rearrange the ordering of the different spinors. 
For their derivation we follow the discussion in \cite{Nishi:2004st}.
When working with a chiral theory such as the SMEFT, it is usually most convenient to derive the Fierz identities in the chiral basis $\{\Gamma^n\}$ for the Dirac algebra in four spacetime dimensions which we define as
\begin{subequations}
\begin{align}
    \{\Gamma^n\} &= \left\{P_L,P_R,\gamma^\mu P_L,\gamma^\mu P_R, \sigma^{\mu\nu}\right\} \,,
    \label{eq:Dirac-basis-4D}
    \\
    \{\widetilde{\Gamma}_n\} &= \left\{P_L,P_R,\gamma_\mu P_R,\gamma_\mu P_L, \sigma_{\mu\nu}/2\right\} \,,
\end{align}
\end{subequations}
where $P_{R/L}=\frac{1}{2}(\mathds{1}\pm\gamma_5)$ are the chirality projectors and $\smash{\sigma^{\mu\nu}=\frac{i}{2}[\gamma^\mu,\gamma^\nu]}$ with $\mu<\nu$.
Moreover, we have also defined the dual basis~$\smash{\{\widetilde{\Gamma}_n\}}$. With this definition the orthogonality condition $\smash{\mathrm{tr}\{\Gamma^n \widetilde{\Gamma}_m\}}=2\delta^n_m$ is satisfied. Since $\{\Gamma^n\}$ forms a basis of all $4\times 4$~matrices we can write any such matrix~$X$ as $X=X_n \Gamma^n$ with $X_n=\smash{\frac{1}{2}\mathrm{tr}\{X\widetilde{\Gamma}_n\}}$, and thus $X=\smash{\frac{1}{2}\mathrm{tr}\{X\widetilde{\Gamma}_n\}\Gamma^n}$. Writing the latter equation in its components and inserting appropriate delta functions we obtain
\begin{align}
    \delta_{ij} \delta_{kl} &= \frac{1}{2} (\widetilde{\Gamma}_n)_{kj} (\Gamma^n)_{il}
    &
    &\text{or}
    &
    (~) \otimes [~] &= \frac{1}{2} (\widetilde{\Gamma}_n] \otimes [\Gamma^n)\,,
\end{align}
where in the last equation we schematically identified the indices with parenthesis as follows: $i\sim($, $j\sim)$, $k\sim[$, and $l\sim]$. Multiplying this equation by generic matrices~$X$ and~$Y$ we find
\begin{align}
    (X) \otimes [Y] &= \frac{1}{4} \, \mathrm{tr}\big\{X \widetilde{\Gamma}_n Y \widetilde{\Gamma}_m \big\} \, (\Gamma^m] \otimes [\Gamma^n)
    \label{eq:Fierz-projection}
\end{align}
which allows us to project any tensor product of two matrices onto a product of matrices from the chosen Dirac basis. In particular, by choosing $X,Y\in\{\Gamma^n\}$ we can derive the Fierz identities
\begin{subequations}
\begin{align}
    \begin{split}
    (P_{A}) \otimes [P_{A}] &= \frac{1}{2} (P_{A}] \otimes [P_{A}) \\
    &\quad +\frac{1}{8} (\sigma^{\mu\nu}P_{A}] \otimes [\sigma_{\mu\nu}P_{A}) \,,
    \label{eq:fierz-scalar-XX}
    \end{split}
    \\
    (P_{A}) \otimes [P_{B}] &= \frac{1}{2} (\gamma^\mu P_{B}] \otimes [\gamma_\mu P_{A}) \,,
    \\
    (\gamma^\mu P_{A}) \otimes [\gamma_\mu P_{A}] &= - (\gamma^\mu P_{A}] \otimes [\gamma_\mu P_{A}) \,,
    \\
    (\gamma^\mu P_{A}) \otimes [\gamma_\mu P_{B}] &= 2\, (P_{B}] \otimes [P_{A}) \,,
    \label{eq:Fierz-vector-AB}
    \\
    \begin{split}
    (\sigma^{\mu\nu} P_{A}) \otimes [\sigma^{\mu\nu} P_{A}] &= 6\, (P_{A}] \otimes [P_{A}) \\
    &\quad -\frac{1}{2} (\sigma^{\mu\nu}P_{A}] \otimes [\sigma_{\mu\nu}P_{A}) \,,
    \end{split}
    \\
    (\sigma^{\mu\nu} P_{A}) \otimes [\sigma^{\mu\nu} P_{B}] &= 0 \,,
\end{align}%
\label{eq:Fierz-ids}%
\end{subequations}%
where $A,B\in\{L,R\}$ but $A \neq B$. The above equations correspond only to relations among Dirac structures; however, when applying them to four-fermion operators we also anti-commute two spinors thus acquiring an additional minus sign with respect to Eq.~\eqref{eq:Fierz-ids}.
For example, Eq.~\eqref{eq:Fierz-vector-AB} allows us to rewrite the operator $(\bar\ell^i \gamma^\mu q_i)(\bar d \gamma_\mu e) = -2 (\bar\ell^i e)(\bar d q_i)$ which has the quarks and leptons in separate currents. 
Notice that we assumed the Dirac algebra in four spacetime dimensions to evaluate the traces in Eq.~\eqref{eq:Fierz-projection} and obtain the relations~\eqref{eq:Fierz-ids}. However, when working at the loop level, we encounter divergent integrals that we regulate using dimensional regularization in $D=4-2\epsilon$~dimensions, which is incompatible with the results obtained before. At the loop level, using the relations~\eqref{eq:Fierz-ids} while working in $D$~dimensions introduces so-called evanescent operators, i.e., operators that vanish in $D=4$. We will discuss these evanescent contributions in Sec.~\ref{sec:evanescent}.

Furthermore, we have the Fierz identity for the generators~$T^a$ of the fundamental representation of~$\SU{N}$ groups
\begin{align}
	(T^a)_{ij} (T^a)_{kl} = \frac{1}{2} \brackets{\delta_{il}\delta_{kj} - \frac{1}{N}\delta_{ij}\delta_{kl}} \, ,
    \label{eq:SUN-Fierz}
\end{align}
or in our notation 
\begin{align}
    (T^a) \otimes [T^a] = \frac{1}{2} (~] \otimes [~) - \frac{1}{2N} (~) \otimes [~] \,,
\end{align}
where the parenthesis now correspond to indices of the fundamental representation of~$\mathrm{SU}(N)$. For example, for~$\mathrm{SU}(2)_L$ this allows us to rewrite the Higgs operator
$(H^\dagger \tau^I H)(H^\dagger \tau^I H) = (H^\dagger H)^2$.

\paragraph{Dirac structure reduction.}
Equation~\eqref{eq:Dirac-basis-4D} constitutes a Dirac basis in $D=4$~dimensions and is therefore enough to construct an EFT operator basis in the physical four-dimensional limit. Nevertheless, we can write down operators with Dirac structures different than in~\eqref{eq:Dirac-basis-4D}, which we then have to project onto our chosen basis~$\{\Gamma^n\}$ using gamma-tensor reduction \cite{Buras:1989xd,Herrlich:1994kh,Tracas:1982gp}.
Following \cite{Fuentes-Martin:2022vvu} we write this projection as
\begin{align}
    X \otimes Y &= \sum_{n} b_n (X,Y) \, \Gamma^n \otimes \widetilde{\Gamma}_n + E(X,Y) .
    \label{eq:Dirac-projection}
\end{align}
Notice that in $D$~dimensions the Dirac algebra is infinite dimensional and thus it is not possible to project a generic structure onto the finite four-dimensional basis~$\{\Gamma^n\}$. As in the case of the Fierz identities, performing such a projection then introduces an evanescent operator~$E(X,Y)$, which is implicitly defined by Eq.~\eqref{eq:Dirac-projection}. Working at the tree level, which we assume for the moment, we can take the four-dimensional limit and therefore~$E(X,Y)$ vanishes. However, at the loop level this is not the case and the evanescent contributions can be treated similar to the discussion in Sec.~\ref{sec:evanescent} and \cite{Fuentes-Martin:2022vvu}.
The coefficients~$b_n(X,Y)$ can be determined by contracting Eq.~\eqref{eq:Dirac-projection} with the basis elements~$\Gamma^k$
\begin{align}
    \mathrm {tr} \left\{ \Gamma^k X \widetilde{\Gamma}_k Y \right\} &= \sum_n b_n(X,Y) \, \mathrm{tr} \left\{ \Gamma^k \Gamma^n \widetilde{\Gamma}_k \widetilde{\Gamma}_n \right\} + \mathcal{O}(\epsilon^2)
\end{align}
which for $k=1,\ldots,10$ yields a system of equations that we can solve to find the coefficients~$b_n(X,Y)$. To compute the traces above we use na\"ive dimensional regularization~(NDR) (see Appendix~\ref{app:gamma5}) defining our evanescent operator scheme. We find
\begin{widetext}
\begin{subequations}
\vspace*{-0.5cm}
\begin{align}
	\gamma^\mu \gamma^\nu P_A \otimes \gamma_\nu \gamma_\mu P_A &= ( 4- 2 \epsilon )\, P_A \otimes P_A + \sigma^{\mu\nu} P_A \otimes \sigma_{\mu\nu} P_A \, , 
    \\
	\gamma^\mu \gamma^\nu P_A \otimes \gamma_\nu \gamma_\mu P_B &= 4( 1-2\epsilon )\, P_A \otimes P_B + E_{AB}^{[2]} \, , 
    \\
	\gamma^\mu \gamma^\nu \gamma^\lambda P_A \otimes \gamma_\lambda \gamma_\nu \gamma_\mu P_A &= 4(1-2\epsilon )\, \gamma^\mu P_A \otimes \gamma_\mu P_A + E_{AA}^{[3]} \, , 
    \\
	\gamma^\mu \gamma^\nu \gamma^\lambda P_A \otimes \gamma_\lambda \gamma_\nu \gamma_\mu P_B &= 16( 1-\epsilon )\,\gamma^\mu P_A \otimes \gamma_\mu P_B + E_{AB}^{[3]} \, , 
    \\
	\gamma^\mu \gamma^\nu \sigma^{\lambda\rho} P_A \otimes \sigma_{\lambda\rho} \gamma_\nu \gamma_\mu P_A &= 16(3-5\epsilon )\,P_A \otimes P_A + 2(6-7\epsilon )\,\sigma^{\mu\nu} P_A \otimes \sigma_{\mu\nu} P_A + E_{AA}^{[4]} \, ,
\end{align}
\end{subequations}
\end{widetext}
implicitly defining the evanescent structures~$\smash{E_{AB}^{[2]}}$, $\smash{E_{AA}^{[3]}}$, $\smash{E_{AB}^{[3]}}$, and $\smash{E_{AA}^{[4]}}$, where $A,B\in\{L,R\}$ with~${A \neq B}$. 
Other schemes, and hence alternative definitions of the evanescent operators differing form our choice by $\mathcal{O}(\epsilon)$~terms, are also possible, see e.g. \cite{Herrlich:1994kh,Dekens:2019ept}.

\subsubsection{The Warsaw basis}
\label{subsubsec:SMEFT_Warsaw-Basis}
We can now apply the methods illustrated so far in this section to the set of all effective operators that are compatible with the symmetries of the~SM. By~that, we can construct a basis, i.e., a minimal set of effective operators of the SMEFT.\footnote{Notice that the term ``basis'' is not always used appropriately in the EFT literature. One should keep in mind that sometimes it is incorrectly used also for over-complete or even incomplete operator sets. Sometimes we will also refer to complete operator sets without redundancies as \textit{minimal bases}.}

As mentioned, a complete list of operators up to mass-dimension six was first given by \textcite{Buchmuller:1985jz}. Besides proving that at dimension five there is a single operator, namely $\smash{Q_\mathrm{Weinberg}}$ in Eq.~\eqref{eq:Weinberg-Operator}, 
they identified 80~independent operators at dimension six (up to the flavor structure) that conserve baryon and lepton number. However, some redundancies still remained in this set of operators as pointed out in \cite{Grzadkowski:2003tf,Fox:2007in,Aguilar-Saavedra:2008nuh,Aguilar-Saavedra:2009ygx}. Only in 2010 the first minimal basis for dimension-six operators in the SMEFT was derived by Grzadkowski, Iskrzy\'nski, Misiak, and Rosiek \cite{Grzadkowski:2010es}. It contains only 59~dimension-six operators that conserve baryon and lepton number. Considering the flavor structure of the operators this amounts to 2499~couplings out of which 1350 are CP-even and 1149 are CP-odd \cite{Alonso:2013hga}. 
The basis is known as the \textit{Warsaw basis} and is the most commonly used basis for the $d=6$~SMEFT. Table~\ref{tab:Warsaw-basis} list all baryon and lepton number conserving $d=6$ operators of the Warsaw basis. For the non-hermitian operators the hermitian conjugate is understood to be included. The operators are divided into classes according to their field content and chirality as in \cite{Grzadkowski:2010es,Alonso:2013hga}, which we follow in our classification of the operators below. The underlying algorithm used to construct the Warsaw basis can be summarized as:
\begin{enumerate}
	\item use IBP and equations of motion to remove operators with more derivatives in favor of operators with fewer derivatives,
	\item use the Fierz identities~\eqref{eq:Fierz-ids} and~\eqref{eq:SUN-Fierz} such that:
	\begin{enumerate}
		\item leptons and quarks do not appear in the same fermion currents,
		\item the gauge indices of the largest gauge group are contracted within each bilinear,
		\item each current is a Hypercharge singlet.
	\end{enumerate} 	 
\end{enumerate} 

\begin{table*}[t]
	\centering
	\newcommand{\OpScale}{.85} 
	\renewcommand{\arraystretch}{1.5}
	\scalebox{\OpScale}{
	\centering
	\begin{tabular}{| lc || lc || lc | lc |}
		\multicolumn{8}{c}{1--4: Bosonic Operators} \\[.1cm] \hline
		\multicolumn{2}{|c||}{1: $X^3$ \small[LG]} & 
		\multicolumn{2}{c||}{2: $H^6$ \small[PTG]} &
		\multicolumn{4}{c|}{4: $X^2 H^2$ \small[LG]} \\ \hline
		$Q_{G}$ & $f^{ABC} G_\mu^{A\nu} G_\nu^{B\rho} G_\rho^{C\mu}$ &	
		$Q_{H}$ & $(H^\dagger H)^3$ &
		$Q_{HG}$ & $(H^\dagger H) G_{\mu\nu}^A G^{A\mu\nu}$ &
		$Q_{H B}$ & $(H^\dagger H) B_{\mu\nu} B^{\mu\nu}$ \\ \cline{3-4}
		$Q_{\widetilde G}$ & $f^{ABC} \widetilde G_\mu^{A\nu} G_\nu^{B\rho} G_\rho^{C\mu}$ &
		\multicolumn{2}{c||}{3: $H^4 D^2$ \small[PTG]} &
		$Q_{H \widetilde G}$ & $(H^\dagger H) \widetilde G_{\mu\nu}^A G^{A\mu\nu}$ &
		$Q_{H \widetilde B}$ & $(H^\dagger H) \widetilde B_{\mu\nu} B^{\mu\nu}$ \\ \cline{3-4}
		$Q_{W}$ & $\varepsilon^{IJK} W_\mu^{I\nu} W_\nu^{J\rho} W_\rho^{K\mu}$ &	
		$Q_{H\Box}$ & $(H^\dagger H) \Box (H^\dagger H)$ & 
		$Q_{HW}$ & $(H^\dagger H) W_{\mu\nu}^I W^{I\mu\nu}$ &
		$Q_{H W B}$ & $(H^\dagger \tau^I H) W_{\mu\nu}^I B^{\mu\nu}$ \\ 
		$Q_{\widetilde W}$ & $\varepsilon^{IJK} \widetilde W_\mu^{I\nu} W_\nu^{J\rho} W_\rho^{K\mu}$ & 
		$Q_{HD}$ & $(H^\dagger D_\mu H)^\ast (H^\dagger D^\mu H)$ & 
		$Q_{H \widetilde{W}}$ & $(H^\dagger H) \widetilde W_{\mu\nu}^I W^{I\mu\nu}$ & 
		$Q_{H \widetilde W B}$ & $(H^\dagger \tau^I H) \widetilde W_{\mu\nu}^I B^{\mu\nu}$ \\
		\hline
	\end{tabular}
	}
	\vspace{0.3cm}
	\\ \centering
	\scalebox{\OpScale}{
	\centering
	\begin{tabular}{| lc || lc | lc | lc |}
		\multicolumn{8}{c}{5--7: Fermion Bilinears $(\psi^2)$} \\[.1cm] \hline
		\multicolumn{8}{|c|}{non-hermitian $(\bar L R)$} \\ \hline
		\multicolumn{2}{|c||}{5: $\psi^2 H^3$ + h.c. \small [PTG]} & \multicolumn{6}{c|}{6: $\psi^2 X H$ + h.c. \small [LG]}  \\ \hline
		$Q_{eH}$ & $(H^\dagger H)(\bar\ell_p e_r H)$ &	
		$Q_{eW}$ & $(\bar\ell_p \sigma^{\mu\nu}e_r)\tau^I H W_{\mu\nu}^I$ &
		$Q_{uG}$ & $(\bar q_p \sigma^{\mu\nu}T^A u_r)\widetilde{H}G_{\mu\nu}^A$ &
		$Q_{dG}$ & $(\bar q_p \sigma^{\mu\nu}T^A d_r)H G_{\mu\nu}^A$ \\ 
		$Q_{uH}$ & $(H^\dagger H)(\bar q_p u_r \widetilde{H})$ &
		$Q_{eB}$ & $(\bar\ell_p \sigma^{\mu\nu}e_r) H B_{\mu\nu}$ &
		$Q_{uW}$ & $(\bar q_p \sigma^{\mu\nu}u_r)\tau^I \widetilde{H}W_{\mu\nu}^I$ &
		$Q_{dW}$ & $(\bar q_p \sigma^{\mu\nu}d_r)\tau^I H W_{\mu\nu}^I$ \\ 
		$Q_{dH}$ & $(H^\dagger H)(\bar q_p d_r H)$ &		
		& & 
		$Q_{uB}$ & $(\bar q_p \sigma^{\mu\nu}u_r)\widetilde{H}B_{\mu\nu}$ &
		$Q_{dB}$ & $(\bar q_p \sigma^{\mu\nu}d_r) H B_{\mu\nu}$ \\ \hline
		\end{tabular}
	}
	\vspace{0.1cm} \newline \centering
	\scalebox{\OpScale}{
	\begin{tabular}{| lc | lc | lc |}
		\hline 
		\multicolumn{6}{|c|}{7: $\psi^2 H^2 D$ ~~~--~~~ hermitian + $Q_{Hud}$  \small [PTG]} \\ \hline
		\multicolumn{2}{|c|}{$(\bar L L)$} & 
		\multicolumn{2}{c|}{$(\bar R R)$} & 
		\multicolumn{2}{c|}{$(\bar R R^\prime)$ + h.c.} \\ \hline
		$Q_{H\ell}^{(1)}$ & $(H^\dagger i \overleftrightarrow{D}_\mu H)(\bar\ell_p \gamma^\mu \ell_r)$ & 
		$Q_{H e}$ & $(H^\dagger i \overleftrightarrow{D}_\mu H)(\bar e_p \gamma^\mu e_r)$ & 
		$Q_{Hud}$ & $i(\widetilde{H}^\dagger D_\mu H)(\bar u_p \gamma^\mu d_r)$ \\
		$Q_{H\ell}^{(3)}$ & $(H^\dagger i \overleftrightarrow{D}_\mu^I H)(\bar\ell_p \tau^I\gamma^\mu \ell_r)$ & 
		$Q_{H u}$ & $(H^\dagger i \overleftrightarrow{D}_\mu H)(\bar u_p \gamma^\mu u_r)$ & 
		& \\
		$Q_{Hq}^{(1)}$ & $(H^\dagger i \overleftrightarrow{D}_\mu H)(\bar q_p \gamma^\mu q_r)$ & 
		$Q_{H d}$ & $(H^\dagger i \overleftrightarrow{D}_\mu H)(\bar d_p \gamma^\mu d_r)$ & 
		& \\
		$Q_{Hq}^{(3)}$ & $(H^\dagger i \overleftrightarrow{D}_\mu^I H)(\bar q_p \tau^I\gamma^\mu q_r)$ & 
		& & 
		& \\ \hline
	\end{tabular}
	}
	\vspace{0.3cm}
	\\ \centering
	\scalebox{\OpScale}{
	\setlength{\tabcolsep}{1.8mm}
	\begin{tabular}{| lc | lc | lc || lc |}
		\multicolumn{8}{c}{8: Fermion Quadrilinears $(\psi^4)$ \small [PTG]} \\[.1cm] \hline
		\multicolumn{6}{|c||}{hermitian} & \multicolumn{2}{c|}{non-hermitian} \\ \hline
		\multicolumn{2}{|c|}{$(\bar L L)(\bar L L)$} & 
		\multicolumn{2}{c|}{$(\bar R R)(\bar R R)$} & 
		\multicolumn{2}{c||}{$(\bar L L)(\bar R R)$} &
		\multicolumn{2}{c|}{$(\bar L R)(\bar L R)$ + h.c.}
		\\ \hline
		$Q_{\ell\ell}$ & $(\bar\ell_p \gamma_\mu \ell_r)(\bar\ell_s \gamma^\mu \ell_t)$ & 
		$Q_{ee}$ & $(\bar e_p \gamma_\mu e_r)(\bar e_s \gamma^\mu e_t)$ & 
		$Q_{\ell e}$ & $(\bar\ell_p \gamma_\mu \ell_r)(\bar e_s \gamma^\mu e_t)$ & 
		$Q_{quqd}^{(1)}$ & $(\bar q_p^i u_r)\varepsilon_{ij}(\bar q_s^j d_t)$ \\
		$Q_{qq}^{(1)}$ & $(\bar q_p \gamma_\mu q_r)(\bar q_s \gamma^\mu q_t)$ & 
		$Q_{uu}$ & $(\bar u_p \gamma_\mu u_r)(\bar u_s \gamma^\mu u_t)$ & 
		$Q_{\ell u}$ & $(\bar\ell_p \gamma_\mu \ell_r)(\bar u_s \gamma^\mu u_t)$ &
		$Q_{quqd}^{(8)}$ & $(\bar q_p^i T^A u_r)\varepsilon_{ij}(\bar q_s^j T^A d_t)$ \\
		$Q_{qq}^{(3)}$ & $(\bar q_p \gamma_\mu \tau^I q_r)(\bar q_s \gamma^\mu \tau^I q_t)$ & 
		$Q_{dd}$ & $(\bar d_p \gamma_\mu d_r)(\bar d_s \gamma^\mu d_t)$ & 
		$Q_{\ell d}$ & $(\bar\ell_p \gamma_\mu \ell_r)(\bar d_s \gamma^\mu d_t)$ &
		$Q_{\ell equ}^{(1)}$ & $(\bar \ell_p^i e_r)\varepsilon_{ij}(\bar q_s^j u_t)$ \\
		$Q_{\ell q}^{(1)}$ & $(\bar\ell_p \gamma_\mu \ell_r)(\bar q_s \gamma^\mu q_t)$ & 
		$Q_{eu}$ & $(\bar e_p \gamma_\mu e_r)(\bar u_s \gamma^\mu u_t)$ & 
		$Q_{q e}$ & $(\bar q_p \gamma_\mu q_r)(\bar e_s \gamma^\mu e_t)$ &
		$Q_{\ell equ}^{(3)}$ & $(\bar \ell_p^i \sigma_{\mu\nu} e_r)\varepsilon_{ij}(\bar q_s^j \sigma^{\mu\nu} u_t)$ \\
		$Q_{\ell q}^{(3)}$ & $(\bar\ell_p \gamma_\mu \tau^I \ell_r)(\bar q_s \gamma^\mu \tau^I q_t)$ & 
		$Q_{ed}$ & $(\bar e_p \gamma_\mu e_r)(\bar d_s \gamma^\mu d_t)$ & 
		$Q_{qu}^{(1)}$ & $(\bar q_p \gamma_\mu q_r)(\bar u_s \gamma^\mu u_t)$ & 
		& \\ 
		& & 
		$Q_{ud}^{(1)}$ & $(\bar u_p \gamma_\mu u_r)(\bar d_s \gamma^\mu d_t)$ & 
		$Q_{qu}^{(8)}$ & $(\bar q_p \gamma_\mu T^A q_r)(\bar u_s \gamma^\mu T^A u_t)$ & & \\ \cline{7-8}
		& & 
		$Q_{ud}^{(8)}$ & $(\bar u_p \gamma_\mu T^A u_r)(\bar d_s \gamma^\mu T^A d_t)$ & 
		$Q_{qd}^{(1)}$ & $(\bar q_p \gamma_\mu q_r)(\bar d_s \gamma^\mu d_t)$ &
		\multicolumn{2}{c|}{$(\bar L R)(\bar R L)$ + h.c.} \\ \cline{7-8}
		& & 
		& & 
		$Q_{qd}^{(8)}$ & $(\bar q_p \gamma_\mu T^A q_r)(\bar d_s \gamma^\mu T^A d_t)$ &
		$Q_{\ell e d q }$ & $(\bar \ell _p^i e_r)(\bar d_s q_{ti})$ \\ \hline
	\end{tabular}
	}
	\caption{List of all baryon and lepton number conserving SMEFT operators at mass-dimension six in the Warsaw basis \cite{Grzadkowski:2010es}. The division into {classes~1--8} is adopted from \cite{Alonso:2013hga} and further refined according to the chirality of the fields. It is also marked which classes are \textit{potentially tree-generated}~[PTG] and which are \textit{loop-generated}~[LG] according to \cite{Einhorn:2013kja,Arzt:1994gp}.
	\label{tab:Warsaw-basis}
	}
\end{table*}

The purely bosonic operators are built out of combinations of field-strength tensors $X_{\mu\nu} \in \{G_{\mu\nu}, W_{\mu\nu}, B_{\mu\nu}\}$, the Higgs doublet~$H$ and covariant derivatives~$D_\mu$. Due to $\SU{2}_L$ and Lorentz invariance the Higgs fields and the covariant derivatives must both occur in even numbers in the operators. After constructing all allowed operators and removing the redundant ones, four classes of bosonic operators remain: 
\begin{itemize}
	\item 4~pure gauge operators containing three field strength tensors (class~1:~$X^3$),
	\item 1~pure scalar operator with six Higgs doublets (class~2:~$H^6$),
	\item 2~operators with four Higgs fields and two covariant derivatives (class~3:~$H^4 D^2$),
	\item 8~mixed operators with two Higgs fields and two field strength tensors (class~4:~$X^2 H^2$).
\end{itemize}

For operators with two fermion fields we have three types of fermion currents: scalar $\smash{(\overline{\psi}_{L/R} \psi_{R/L})}$, vector $\smash{(\overline{\psi}_{L/R} \gamma^\mu \psi_{L/R})}$, and tensor~$\smash{(\overline{\psi}_{L/R} \sigma^{\mu\nu} \psi_{R/L})}$. After removing the redundant operators we obtain one class of operators for each type of current:
\begin{itemize}
	\item 3~non-hermitian Yukawa like operators with a scalar fermion current, and three Higgs (class~5:~$\psi^2 H^3$),
	\item 8~non-hermitian dipole operators with a tensor current, one Higgs, and one field-strength tensor (class~6:~$\psi^2 X H$),
	\item 8~operators (all hermitian except for~$Q_{Hud}$) with a vector current, two Higgs fields, and a covariant derivative (class~7:~$\psi^2 H^2 D$).
\end{itemize}

Last, we have 25 four-fermion operators in class~8 subdivided according to their chiral structures ${(\bar L L)(\bar L L)}$, ${(\bar R R)(\bar R R)}$, ${(\bar L L)(\bar R R)}$, ${(\bar L R)(\bar L R)}$, and~${(\bar L R)(\bar R L)}$. To see explicitly how other types of operator classes can be removed see the discussion in \cite{Grzadkowski:2010es}.\footnote{The Feynman rules for the SMEFT in the Warsaw basis in the $R_\xi$-gauges are given in \cite{Dedes:2017zog}.}

\subsubsection{Other bases}
The Warsaw basis is of course only one viable choice of basis and other options are possible. Although the Warsaw basis is most commonly used, other bases can be of advantage when considering specific sets of observables.
A~commonly adopted set of dimension-six operators in phenomenological analyses is the so-called strongly-interacting light Higgs (SILH) ``basis'' \cite{Giudice:2007fh}. 
However, although called basis, it does not represent a complete set at dimension six \cite{Brivio:2017bnu}. 
The same is also true for the HISZ basis \cite{Hagiwara:1993ck}. 
A~full and minimal basis, containing the operators of the original SILH set, was constructed in \cite{Elias-Miro:2013eta}, 
see also \cite{Contino:2013kra}. An extensive discussion about the basis choice in the SMEFT can be found in \cite{Passarino:2016saj}.

The \textit{Green's basis} \cite{Gherardi:2020det} is another common set of SMEFT operators. Although constituting a complete set of operators, it is not a ``minimal'' basis as it contains redundancies. The Green's basis is an extension of the Warsaw basis by all the operators that are removed from the latter by the equations of motion. Therefore, the operators in the Green's basis are only independent under IBP but not under field redefinitions. This basis is often convenient for SMEFT matching computations.
In functional matching the effective Lagrangian obtained by integrating out some heavy particles is usually in the Green's basis (up to IBP). Also the diagrammatic off-shell matching procedure involves the operators of this basis. (See Sec.~\ref{sec:matching} for more details.)
The results from the matching computations in the Green's basis can then be converted to the minimal Warsaw basis using the basis reduction relations given in the appendix of \cite{Gherardi:2020det}.
See also \cite{Ren:2022tvi} for a derivation of a Green's basis of the SMEFT at dimension eight.

\subsubsection{Higher-dimensional operators}
\label{sec:d8-operators}
As already discussed, at mass-dimension five there is only a single operator, i.e., the Weinberg operator~\cite{Weinberg:1979sa}, and it is violating lepton number. Higher-dimensional operators that also do not conserve baryon and lepton number were derived in~\cite{Weinberg:1980bf}. The first full set of dimension-seven operators was given in \cite{Lehman:2014jma} finding a total of 20 independent operators. However, in \cite{Liao:2016hru} it is shown that two of these operators are redundant, thus obtaining a basis of 18 operators. All of these contain either two or four fermions and do not conserve lepton number. Seven of these operators furthermore violate baryon number as well.
An important point to note is that all odd mass-dimension operators in the SMEFT violate either baryon or lepton numbers or both \cite{Kobach:2016ami,Helset:2019eyc}. Due to the stringent experimental bounds on processes that do not conserve these symmetries, the scale generating such violating process must be very high (see Sec.~\ref{sec:GlobalSymmetries}). 
Given these are exact global symmetries of the SM Lagrangian, it is common to assume that they are also exact or almost-exact symmetries of the SMEFT, and operators that violate baryon or lepton number 
are often neglected but for specific analyses devoted to the corresponding symmetry-violating processes. 

More recently, the first complete bases of dimension-eight operators \cite{Murphy:2020rsh,Li:2020gnx} have been derived, finding 1029~independent structures up to different flavor contractions \cite{Murphy:2020rsh}.\footnote{For earlier attempts at deriving dimension-eight operator see \cite{Lehman:2015coa}.} Although these operators are suppressed by four powers of the new physics scale~$\Lambda^{-4}$ they can still be relevant for phenomenological studies. This is in particular the case for UV theories that do not generate dimension-six operators contributing to a given set of observables and  
the leading contribution starts at dimension eight. More generally, dimension-eight terms can be relevant for observables where the dimension-six operators do not interfere (or have a suppressed interference) with the SM amplitude (see Sec.~\ref{sect:dim8}).
Furthermore, also a basis for the SMEFT at dimension nine is known \cite{Liao:2020jmn,Li:2020xlh}. 
An all-order approach to constructing bases of EFTs, has been presented in \cite{Henning:2017fpj}. 
The same authors also counted the number of independent effective operators for the SMEFT present at different higher dimensions using the Hilbert series in \cite{Henning:2015alf}.
See also \cite{Fonseca:2017lem,Fonseca:2019yya,Li:2022tec} for computer tools helping with the construction of higher-dimensional operator bases in generic EFTs.

\subsubsection{Evanescent operators}
\label{sec:evanescent}
Nowadays, nearly all loop computations in the SMEFT are performed using dimensional regularization working in $D=4-2\epsilon$ dimensions. This leads to another subtlety when reducing redundant operators to a specific basis. 
As already mentioned, in non-integer dimensions the Lorentz algebra is infinite-dimensional, whereas in $D=4$ dimensions it is finite. Now, consider a $D$-dimensional BSM Lagrangian obtained, e.g., through a one-loop matching computation (see Sec.~\ref{sec:matching}). When we want to reduce it to a physical four-dimensional basis, such as the Warsaw basis, we necessarily introduce additional operators called evanescent, due to the mismatch of the dimensionality of the bases. Schematically we can write
\begin{align}
    R \xrightarrow{~\mathcal{P}~} Q + E \,,
\end{align}
where $R$ denotes a redundant operator, $Q$~an operator part of the physical four-dimensional basis, and $E$ an evanescent operator.
The projection~$\mathcal{P}$ is performed using, e.g., Fierz identities or Dirac algebra reduction identities, as discussed before, which are intrinsically four-dimensional. The evanescent operator can then be implicitly defined as ${E} \equiv {R}-{Q}$. It is formally of rank~$\epsilon$ and thus vanishes in the four-dimensional limit. However, when inserting an evanescent operator in a UV divergent one-loop diagram, the operator can combine with a $1/\epsilon$~pole, resulting in a finite contribution to a one-loop matrix element. Therefore, despite vanishing in four dimensions, evanescent operators still yield physical contributions. However, these contributions are local, since the UV poles of any one-loop diagram are so. Thus, the one-loop effect of evanescent operators can be interpreted as finite shifts of the Wilson coefficients of the physical basis. Therefore, their physical effects can be absorbed by introducing finite counterterms. The resulting renormalization scheme is free of evanescent operators, but notably does not agree with the $\smash{\overline{\mathrm{MS}}}$~scheme.

Evanescent contributions were first studied in the context of next-to-leading-order~(NLO) computations of the anomalous dimension of the \textit{weak effective Hamiltonian} \cite{Dugan:1990df,Buras:1989xd,Herrlich:1994kh}, and recently extended to the low-energy effective field theory~(LEFT) \cite{Aebischer:2022aze,Aebischer:2022tvz,Aebischer:2022rxf},\footnote{See also \cite{Aebischer:2021raf,Aebischer:2020dsw} for previous works on evanescent operators in $\Delta F=1,2$ transitions.} and the SMEFT \cite{Fuentes-Martin:2022vvu}.

The latter reference introduced an alternative but equivalent projection prescription to handle evanescent operators: let $\smash{S_\mathrm{R}}$ be the action of the EFT containing redundant operators. Now, reducing the operators in $\smash{S_\mathrm{R}}$ to the Warsaw basis (or any other physical basis) using four-dimensional identities (such as Fierzing or Dirac structure reduction) we obtain the action~$\smash{S_\mathrm{W}^\prime}$. As discussed before, $\smash{S_\mathrm{R}}$ and $\smash{S_\mathrm{W}^\prime}$ do not reproduce the same physics and the difference is given by evanescent operators. However, we have seen that their effects can be absorbed by finite one-loop shifts of the Wilson coefficients in~$\smash{S_\mathrm{W}^\prime}$.
Thus, take the action~$\smash{S_\mathrm{W}}$ which contains the same operators as~$\smash{S_\mathrm{W}^\prime}$, and we fix the Wilson coefficients of $\smash{S_\mathrm{W}}$ by requiring that it describes the same physics as~$\smash{S_\mathrm{R}}$. We can achieve this by requiring the corresponding quantum effective actions to agree~$\smash{\Gamma_\mathrm{W}=\Gamma_\mathrm{R}}$. 
We can express the effective action as
\begin{align}
    \Gamma_\mathrm{X} &= S_\mathrm{X}^{(0)} + S_\mathrm{X}^{(1)} + \overline{\Gamma}_\mathrm{X}^{(1)}  + \ldots \,,
\end{align}
where $\smash{S_\mathrm{X}^{(0,1)}}$ contains only local operators and their corresponding tree-level or one-loop Wilson coefficients, respectively. Furthermore, $\smash{S_\mathrm{X}^{(1)}}$ contains the counterterms, and the ellipses denote higher-loop contributions. The term $\smash{\overline{\Gamma}_\mathrm{X}^{(1)}}$ represents the contributions by all one-loop diagrams built with insertions of operators from~$\smash{S_\mathrm{X}^{(0)}}$.
We then find that the physical evanescent-free action describing the same physics as~$\smash{S_\mathrm{R}}$ is given by
\begin{align}
    S_\mathrm{W}^{(0)} &= \mathcal{P} S_\mathrm{R}^{(0)} \,,
    \label{eq:evanescent-projection-tree-level}
    \\
    S_\mathrm{W}^{(1)} &= \mathcal{P} S_\mathrm{R}^{(1)} + \underbrace{\mathcal{P} \left[ \overline{\Gamma}_\mathrm{R}^{(1)} - \overline{\Gamma}_\mathrm{W}^{(1)} \right]}_{\equiv \Delta S^{(1)}} \,,
\end{align}
where $\smash{\overline{\Gamma}_\mathrm{R}^{(1)} - \overline{\Gamma}_\mathrm{W}^{(1)}}$ is the sum of all one-loop diagrams containing an evanescent operator. 
Since this term is already of one-loop order, we can simply apply the four-dimensional identities to project~$(\mathcal{P})$ it back to the Warsaw basis.\footnote{Different definition of the projection operator~$\mathcal{P}$ are possible, differing by $\mathcal{O}(\epsilon)$~terms. These define different prescription for the evanescent operators, and we have to follow one prescription consistently. For more details see \cite{Fuentes-Martin:2022vvu}.}
Any effect of evanescent operators in this projections would yield a two-loop effect and can be neglected at the desired order.\footnote{Notice that physical operators can flow into evanescent operators at two-loop order. Thus, leading to a non-vanishing coefficient for the latter even if we started with zero coupling for the evanescent operators, which could then possibly flow back into the physical coefficients. However, as observed by \textcite{Dugan:1990df,Herrlich:1994kh}, the running of the physical coefficients can be made independent of the evanescent ones by an appropriate finite compensation of the evanescent couplings.}
The action~$S_\mathrm{W}$ thus obtained is free of evanescent operators and reproduces the same physics as the original action with redundant operators~$S_\mathrm{R}$.

\begin{figure}[t]
    \centering
    \resizebox{0.95\linewidth}{!}{%
    \includegraphics{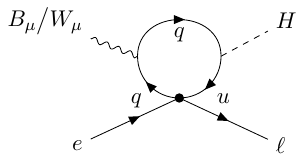}
    \includegraphics{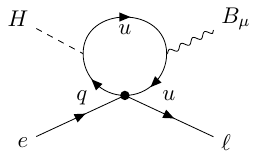}
    }
    \caption{One-loop SMEFT diagrams allowing for the insertion of the evanescent operator~$E_{u^c e l q^c}$ and contributing to the leptonic dipole operators.}
    \label{fig:evanescent-dipole}
\end{figure}

For example, consider the redundant operator
\begin{align}
    [R_{u^c e l q^c}]_{prst} &= \big(\overline{u}^c_p e_r\big) \varepsilon_{ij} \big(\overline{\ell}{}_{s}^i {q_t^c}^j\big)
\end{align}
which, as we will see in Sec.~\ref{sec:matching}, is generated at tree level by integrating out an $S_1$~leptoquark.
It can be projected onto the Warsaw basis by applying the four-dimensional Fierz identity~\eqref{eq:fierz-scalar-XX}
\begin{align}
    [R_{u^c e l q^c}]_{prst} &\overset{(d=4)}{=} -\frac{1}{2} [Q_{lequ}^{(1)}]_{srtp} +\frac{1}{8} [Q_{lequ}^{(3)}]_{srtp} \,.
    \label{eq:4d-fierz}
\end{align}
The evanescent operator introduced by this can be written as $\smash{E_{u^c e l q^c} \equiv R_{u^c e l q^c} - \big( -\frac{1}{2} Q_{lequ}^{(1)} + \frac{1}{8} Q_{lequ}^{(3)} \big)}$ schematically. 
The tree-level action can be directly obtained from Eq.~\eqref{eq:4d-fierz}. 
However, this introduces the finite shift~$\Delta S^{(1)}$ in the one-loop action of the evanescent-free scheme. 
To determine it, we would have to compute all one-loop diagrams with the insertion of the operators $R_{u^c e l q^c}$ or~$\smash{Q_{lequ}^{(1,3)}}$. 
For simplicity, we only consider the leptonic dipole contributions here, which are due to the diagrams shown in Fig.~\ref{fig:evanescent-dipole}. 
Computing the corresponding amplitudes we find
\begin{align}
    \Delta S^{(1)} = 
    &-\frac{1}{16\pi^2} \frac{5}{8} g_1 [Y_u^\ast]_{pr} (1-\xi_\mathrm{rp}) [C_{u^c e l q^c}^{(R)}]_{rtsp} [Q_{eB}]_{st}
    \nonumber\\
    &+\frac{1}{16\pi^2} \frac{3}{8} g_2 [Y_u^\ast]_{pr} (1-\xi_\mathrm{rp}) [C_{u^c e l q^c}^{(R)}]_{rtsp} [Q_{eW}]_{st}
    \nonumber\\
    &+\ldots
    \label{eq:evanescent-dipole-shift}
\end{align}
where $C_{u^c e l q^c}^{(R)}$ is the Wilson coefficient of $R_{u^c e l q^c}$, and the ellipses denote other operators than the leptonic dipoles. 
The diagrams involving the $\smash{Q_{lequ}^{(3)}}$~operator are particularly complicated since they involve closed fermion loops giving a Dirac trace of the form
\begin{align}
    \mathrm{tr} \left[ \gamma^\mu \gamma^\nu \gamma^\rho \gamma_\mu \gamma^\sigma \gamma^\delta \gamma^5 \right]
    \label{eq:6gamma-trace}
\end{align}
which is not well defined in dimensional regularization.
This is attributed to the commonly known problem of extending~$\gamma^5$, which is an intrinsically four-dimensional object, to $D$~dimensions.
Here, we choose to work in the na\"ive dimensional regularization~(NDR), where the cyclicity of Dirac traces of the type given in Eq.~\eqref{eq:6gamma-trace} is lost.
Therefore, these traces exhibit a so-called \textit{reading point ambiguity}: the results of these Dirac traces depend on where we start reading the closed fermion loops, i.e., which vertex or propagator comes first in the trace. 
This reading point ambiguity is parametrized by~$\xi_\mathrm{rp}$ in Eq.~\eqref{eq:evanescent-dipole-shift}, which takes on different values depending on where we start the trace.
In our case, we have $\xi_\mathrm{rp}=0$ when the Dirac trace is read starting from the Higgs interaction vertex (or the propagator coming after it). 
For all other reading points we find $\xi_\mathrm{rp}=1$, therefore leading to a vanishing of this particular evanescent contribution. 
Nevertheless, removing $R_{u^c e l q^c}$ in favor of~$\smash{Q_{lequ}^{(1,3)}}$ will still yield non-vanishing evanescent contributions to other operators than the dipoles, but we do not consider these here.
We can use any prescription for choosing the reading point of this Dirac trace to compute the evanescent contribution in this basis change, given we apply this prescription consistently in all subsequent computations within the EFT, i.e., for calculating all one-loop matrix elements involving~$\smash{Q_{lequ}^{(3)}}$.
More details are provided in \cite{Fuentes-Martin:2022vvu} and in Appendix~\ref{app:gamma5}.

\subsection{How large are the Wilson coefficients?}
\label{subsec:SMEFT_OperatorSize}
The value of the Wilson coefficients in an EFT is determined by the matching condition to the corresponding UV~theory. However, in the bottom-up approach of SMEFT, the underlying BSM model is  unknown. In this case the operator coefficients can only be determined by experiment. Nevertheless, it is still possible to derive some information about the size of the Wilson coefficients from general theoretical arguments. 

One way of estimating the coefficients is to use more elaborate versions of dimensional analysis. 
A~second option is understanding if an operator can be generated at the tree level, or only through loops by the full BSM theory. A~third possibility is using global (approximate) symmetries of the underlying theory. We discuss the first two options below, while the case of 
global symmetries will be discussed in Sec.~\ref{sec:GlobalSymmetries}.

\subsubsection{Power counting and dimensional analysis}
\label{sec:powercounting}
Up to now we only estimated the size of the coefficient of an effective operator using its mass/energy dimension.  As it is well known, in $D=4$~spacetime dimensions each Lagrangian term must be of mass-dimension four. Thus a mass-dimension~$d$ operator must be suppressed by a factor of $\Lambda^{4-d}$ yielding its approximate size. There is, however, an alternative option for estimating the size of coefficients called na\"ive dimensional analysis~(NDA) first developed in the context of chiral perturbation theory in~\cite{Manohar:1983md}. It combines the EFT expansion in the new-physics scale~$\Lambda$ with an expansion in factors of~$4\pi$, or equivalently in~$\hbar$ coming from the loop-expansion factor~$\hbar/(4\pi)^2$. It was later applied to general EFTs and the NDA master formula for a term in the SMEFT Lagrangian is \cite{Gavela:2016bzc}

\begin{widetext}
\begin{align}
\frac{\Lambda^4}{(4 \pi)^2} 
\squarebrackets{\frac{\partial}{\Lambda}}^{N_p}
\squarebrackets{\frac{4\pi H}{\Lambda}}^{N_H}
\squarebrackets{\frac{4\pi A}{\Lambda}}^{N_A}
\squarebrackets{\frac{4\pi \psi}{\Lambda^{3/2}}}^{N_\psi}
\squarebrackets{\frac{g}{4\pi}}^{N_g}
\squarebrackets{\frac{y}{4\pi}}^{N_y}
\squarebrackets{\frac{\lambda}{(4\pi)^2}}^{N_\lambda}
\, ,
\label{eq:SMEFT_NDA_master-formula}
\end{align}
\end{widetext}
where $\partial$ is a derivative, $H$~the Higgs doublet, $A\in\{G,W,B\}$ a vector field, $\psi$~one of the SM fermion fields, $g\in\{g_3,g_2,g_1\}$, $y$~a~Yukawa coupling, and $\lambda$ the quartic Higgs coupling. The numbers~$N_i$ give the power for each factor that is included in the Lagrangian term. 
The NDA scaling of all operator classes in the Warsaw basis is shown in Tab.~\ref{Tab:NDA}. We can now compare the SMEFT Lagrangian with the conventional normalization
\begin{align}
\begin{split}
\L \supset &\brackets{D_\mu H}^\dagger \brackets{D^\mu H} + m^2 H^\dagger H - \frac{\lambda}{2} \brackets{H^\dagger H}^2 \\
&+ \frac{C_H}{\Lambda^2} \brackets{H^\dagger H}^3 + \ldots
\end{split}
\end{align}
to the Lagrangian rewritten using~NDA
\begin{align}
\hat{\L} \supset &\brackets{D_\mu H}^\dagger \brackets{D^\mu H} + \hat{m}^2 \Lambda^2 H^\dagger H - \frac{\hat{\lambda}}{2} (4\pi)^2 \brackets{H^\dagger H}^2 \nonumber\\
&+ \frac{(4\pi)^4 \hat{C}_H}{\Lambda^2} \brackets{H^\dagger H}^3 + \ldots
\end{align}
and we do not write out all other terms explicitly for simplicity. Since NDA does not modify the Lagrangian, i.e., we have $\hat\L = \L$, we can identify the coefficients as follows
\begin{align}
\hat{m} &= \frac{m}{\Lambda} \, ,
&
\hat{\lambda} &= \frac{\lambda}{(4\pi)^2} \, ,
&
\hat{C}_H &= \frac{1}{(4\pi)^4} C_H \, .
\end{align}
Following the discussion in~\cite{Gavela:2016bzc}, we can now consider the one-loop contribution to $C_H$
\begin{align}
	\Delta C_H \sim 
    \vcenter{\hbox{\includegraphics[scale=0.9]{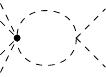}}}
  \sim \frac{\lambda}{(4\pi)^2} C_H \, ,
\end{align}
where we assume that the loop comes with a suppression factor of~$1/16\pi^2$. Using NDA instead we find
\begin{align}
	\Delta \hat{C}_H \sim 
    \vcenter{\hbox{\includegraphics[scale=0.9]{figures/Delta-CH_eq_2-38_and_eq_2-39.pdf}}}
  \sim \hat{\lambda}\hat{C}_H
\end{align}
without any factors of~$4\pi$. The form of the equation above is universal and holds in general, independently of the loop order, for NDA~\cite{Gavela:2016bzc}
\begin{align}
	\Delta \hat{C}_i \sim \prod_k \hat{C}_{i_k} \, .
\end{align}
It also holds for both strongly and weakly coupled theories. For strongly coupled theories we have $\Delta\hat{C}\lesssim 1$ \cite{Manohar:1983md}, whereas for weakly coupled theories we can have~$\Delta\hat{C}\ll 1$. Only $\Delta\hat{C} \gg 1$ is not allowed as in this case the higher-order correction~$\Delta\hat{C}$ would be larger than~$\hat{C}$ itself. Thus interactions become strongly coupled if~$\hat{C} \sim 1$. Therefore, the Wilson coefficients~$\hat{C}$ in the NDA formalism directly indicate how close a theory is to the strong coupling regime without any factors of~$4\pi$. In the usual normalization, not using NDA, the strong coupling regime is reached for $C_H \sim (4\pi)^4$ in the example above or for the SM gauge couplings at $g \sim 4\pi$ as can be seen from Eq.~\eqref{eq:SMEFT_NDA_master-formula}.

Note that the NDA master formula~\eqref{eq:SMEFT_NDA_master-formula} only dictates the maximally allowed size of an operator. Smaller or even vanishing coefficients are always possible. For example, this happens in the case where certain operators are forbidden or suppressed by some (global) symmetry, as we will discuss in Sec.~\ref{sec:GlobalSymmetries}.

\begin{table*}
\centering
\begin{tabular}{c|c|c|c|c|c|c|c}
1: $X^3$ & 2: $H^6$ & 3: $H^4 D^2$ & 4: $X^2 H^2$ & 5: $\psi^2 H^3$ & 6: $\psi^2 X H$ & 7: $\psi^2 H^2 D$ & 8: $\psi^4$
\\\hline
&&&&&&&\\[-0.3cm]
$\dfrac{4\pi}{\Lambda^2} X^3$ & $\dfrac{(4\pi)^4}{\Lambda^2} H^6$ & $\dfrac{(4\pi)^2}{\Lambda^2} H^4 D^2$ & $\dfrac{(4\pi)^2}{\Lambda^2} X^2 H^2$ & $\dfrac{(4\pi)^3}{\Lambda^2} \psi^2 H^3$ & $\dfrac{(4\pi)^2}{\Lambda^2} \psi^2 X H$ & $\dfrac{(4\pi)^2}{\Lambda^2} \psi^2 H^2 D$ & $\dfrac{(4\pi)^2}{\Lambda^2} \psi^4$
\end{tabular}
\caption{NDA scaling of the operator classes in the Warsaw basis.
\label{Tab:NDA}
}
\end{table*}

\subsubsection{Loop- versus tree-level generated operators}
\label{sec:tree-loop-generated-operators}

\begin{figure}
\centering
\includegraphics{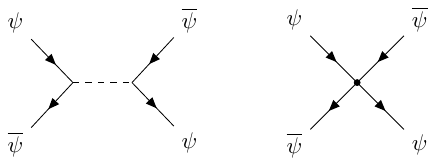}
\caption{The Feynman diagram on the left-hand side shows a process in the UV~theory generating the effective four-fermion operator shown on the right-hand side. The solid lines represent SM fermions, whereas the dashed line denotes a heavy bosons (either vector or scalar).
\label{fig:loop-tree_generation}}
\end{figure}

In principle, BSM theories, when matched to the SMEFT, can generate effective operators at different orders in their loop expansion. If the UV theory contains a tree-level process that produces a specific effective operator after integrating out the heavy states this operator is called \textit{tree-generated}. Contrary if there is no tree-level contribution, but a contribution at the loop level, then we call the operator \textit{loop-generated}. Different UV theories can generate certain operators at different orders in the loop expansion. As it turns out, even though the SMEFT is constructed to allow for a description of generic UV~completions of the~SM, it is impossible to generate certain effective operators at tree level, simply because no possible UV~extension exists producing these operators at leading order. The only assumption for the proof of this statement in \cite{Arzt:1994gp} is that the underlying UV~extension of the SM is a weakly coupled gauge theory built out of a finite (small) number of scalars, vectors, and fermions. For example, all four-fermion operators can, in principle, be generated by the exchange of either a heavy scalar or a heavy vector boson coupling to both fermion currents in the~UV, as shown in Fig.~\ref{fig:loop-tree_generation}. Therefore, we call them \textit{potentially tree-generated}~[PTG] as it is still possible to find specific models in which they are produced at the loop level and not by tree graphs.

A~counter-example are the operators of the type~$X^3$ with three field-strength tensors. It is simply impossible to generate them in any gauge theory at the tree level. These operators are therefore called \textit{loop-generated}~[LG] and their coefficients come with an additional suppression factor of~$(16\pi^2)^{-n}$, where $n$ is the loop order, if they are produced by a weakly coupled UV~theory. The classification of the SMEFT operators according to tree and loop generation was worked out in \cite{Arzt:1994gp} and later adapted to the Warsaw basis in \cite{Einhorn:2013kja}. In the latter reference the authors also argue that, when constructing a basis of effective operators for an EFT and having a set of equivalent operators, where some are~PTG and others are~LG, it is always preferable to remove the LG~operators since the PTG~operators potentially come with larger coefficients and are therefore phenomenologically more relevant. If, on the contrary, one would remove a PTG~ operator in favor of a LG~operator, the coefficient of the latter could potentially gain a tree-level contribution through the corresponding field redefinition \cite{Arzt:1994gp} depending on the specific UV~model. This condition of removing LG~operators in favor of PTG~operators whenever possible is also satisfied by the Warsaw basis \cite{Einhorn:2013kja}. 
As an example, the application of the tree/loop classification to the dimension-six operators of the SMEFT contributing to the renormalization of $h \to \gamma\gamma$ and $h \to \gamma Z$ is presented in \cite{Elias-Miro:2013gya}. 
The role of the strong coupling assumption is illustrated by a model discussed in \cite{Manohar:2013rga}: in the limit of infinitely many heavy particles, equivalent to the strong coupling limit, the leading terms in fact are loop-generated. 
For a further discussion of the tree/loop classification see \cite{Jenkins:2013fya,Boggia:2016asg}.

All UV completions of the SM containing general heavy scalar, spinor and vector fields with arbitrary interactions, that contribute to the dimension-six SMEFT Wilson coefficients at the tree level, have been classified in \cite{deBlas:2017xtg} and references therein.
This work also reports all tree-level matching conditions for these models.
Therefore, it presents a complete tree-level UV/IR~dictionary for the $d=6$ SMEFT, allowing to figure out which SMEFT operator is generated by which UV model at the tree level and consequently to determine all other operators induced in this UV scenario at leading order.
This greatly simplifies phenomenological analyses when a deviation in the experimental data is observed, and the possibly contributing SMEFT operators have been identified.
Generalizations of this dictionary to higher dimensions and to the one-loop level can be found in \cite{Craig:2019wmo,Li:2022abx} and \cite{Guedes:2023azv}, respectively.

\subsection{Constraints and validity}
\label{subsec:SMEFT_Constraints}

We have seen in the previous section that the scaling of Wilson coefficients can be constrained by purely theoretical arguments. Further constraints on the whole structure of the theory and its validity can be derived from additional general theoretical considerations. 
A~powerful constraint follows from unitarity: operators with arbitrary coefficients can lead to an uncontrolled growth of scattering amplitudes with energy, violating unitarity, and hinting at possible inconsistencies in the UV or a breakdown of the EFT expansion. 
More general constraints on the EFT coefficients follow from the combined requirement of 
analyticity and unitarity of the $S$-matrix. In this section, we briefly review these arguments, together with some general considerations about the convergence of the operator expansion and the validity of the SMEFT.

\subsubsection{Convergence of the \texorpdfstring{$1/\Lambda$}{1/Lambda} expansion
and validity range}
\label{sect:dim8}

The EFT expansion can be done on two different levels: at the amplitude (or Lagrangian level), and  at the level of the observables, which are proportional to the square of a given transition amplitude.
To obtain results that have a consistent expansion in powers of the UV cutoff, 
it is necessary to truncate consistently the expansion of the observables. 
For example, if we want to work up to dimension eight, we can write the Lagrangian as
\begin{align}
	\L &= \L_\mathrm{SM} + \frac{1}{\Lambda^2} C_6 Q_6 + \frac{1}{\Lambda^4} C_8 Q_8 + \ord{\Lambda^{-6}} \, ,
\end{align}
where $Q_{6(8)}$ represents a generic dimension-six (-eight) operator with corresponding Wilson coefficient~$C_{6(8)}$. Using this Lagrangian to compute some observable $O$ we find schematically
\begin{equation}
    O 
	\sim \mathrm{SM}^2 
	+ \frac{1}{\Lambda^2}C_6 \times \mathrm{SM}
	+ \frac{1}{\Lambda^4}C_6^2 
	+ \frac{1}{\Lambda^4}C_8 \times \mathrm{SM}
	+ \ord{\Lambda^{-6}} ,
\label{eq:observable-EFT-series}
\end{equation}
where $\mathrm{SM}$ denotes the Standard Model contribution. 
The first term is the pure SM contribution to the observable of interest. 
The second term is the interference of dimension-six terms with the SM and the only term of order~$\ord{\Lambda^{-2}}$. 
Thus if we would have chosen to work up to dimension six instead, these first two terms would be the only ones contributing. 
However, one should note that working to $\cO(\Lambda^{-2})$ at observable level can, in principle, lead to negative cross sections if the interference term is sizable and negative. 
To ensure a positive cross section, one should include the third term in Eq.~\eqref{eq:observable-EFT-series} which is a ``new-physics--squared'' contribution of a combination of two dimension-six operators, thus being of order~$\ord{\Lambda^{-4}}$. 
In principle, the last term, which is the interference of a dimension-eight operator with the~SM, is also of order~$\cO(\Lambda^{-4})$. 
In many phenomenological applications these contributions are neglected, which is consistent with the truncation of the EFT series on amplitude level.
Only such truncation ensures positive cross sections.

Besides the pure scaling with inverse powers of~$\Lambda$, 
care must be taken about the size of the interference terms with the SM amplitude, 
which can easily be suppressed with respect to the formally leading terms.
As pointed out in \cite{azatov:2016sqh}, helicity selection rules imply that in a large
fraction of $2\to 2$ scattering processes at high energy, the $1/\Lambda^2$ terms in~\eqref{eq:observable-EFT-series} vanish, and the contribution from dimension-eight 
operators can be quite relevant. More generally, dimension-eight operators as well
as dimension-six--squared terms, can become relevant for searches at high-$p_T$,
due to the energy growth of the corresponding 
contribution to the cross section (see Sec.~\ref{sect:unit}).

Studies about the impact of dimension-eight operators in the SMEFT can be found in \cite{Corbett:2021eux} analyzing the effect on electroweak precision data, and \cite{Hays:2018zze} investigating the impact on Higgs measurements. A~comparison of the effect of dimension-six and dimension-eight operators in more general terms can be found in \cite{Hays:2020scx}. 

The issue of the convergence of the $1/\Lambda$~expansion, and the growth with energy of the cross section is intimately related to the applicability range of the EFT approach. 
On the one hand, it is clear that the momentum expansion cannot be trusted if~$E/\Lambda =\ord{1}$, 
such that all terms in the operator-product expansion become of the same order. 
On the other hand, in a bottom-up approach, it is not obvious how to determine the precise validity range of the EFT, given the intrinsic ambiguity in  determining the value of~$\Lambda$. 
More precisely, the new-physics scale~$\Lambda$ is not an independent parameter in the EFT. Only ratios of Wilson coefficients over the new-physics scale can be determined, i.e., $C^{(d)}/\Lambda^{d-4}$ for a dimension~$d$ operator. Therefore, Wilson coefficients are also often defined as dimensionful quantities in the literature $\mathsf{C}{}^{(d)}=C^{(d)}/\Lambda^{d-4}$, such that $[\mathsf{C}{}^{(d)}]=M^{4-d}$. However, throughout this review, we use dimensionless coefficients for the benefit of having an explicit EFT power counting. 
Consistency conditions for specific classes of reactions, ensuring data is analyzed in a kinematical range where the SMEFT approach is valid, 
have been discussed in \cite{Contino:2016jqw,Baglio:2020oqu,Boughezal:2021tih,Lang:2021hnd}. 
See also \cite{Brivio:2022pyi}.

\subsubsection{Unitarity violation and positivity constraints}
\label{sect:unit}

The high-energy behavior of scattering amplitudes in the SM is governed by a subtle set of cancellations among different contributions. These protect the theory from unitarity violations due to the unbounded growth of amplitudes with energy. 
When working with the low-energy degrees of freedom, i.e., the massive physical states after electroweak symmetry breaking, the gauge symmetries 
responsible for these cancellations are obscured, although still guaranteeing the same protection at high energies. A~well known example in the SM is the scattering of longitudinally polarized $W$-bosons, $W_L W_L \to W_L W_L$ \cite{LlewellynSmith:1973yud,Lee:1977eg,Lee:1977yc}. If one does not include the quartic self-interaction of the gauge bosons, required by the non-Abelian nature of the gauge symmetry, the corresponding amplitude grows with the energy~$E$ as~$E^4$. Including the quartic contact interaction dampens the energy growth to~$E^2$ but still leads to unitarity violation. Only after also considering the contribution from the Higgs and Goldstone bosons, and by that restoring the relations imposed by a linear realization of the 
$\mathrm{SU}(2)_L$ symmetry breaking via the vev of the Higgs field, we find the correct energy behavior of the amplitude, no longer growing with energy.

The additional effective operators in the SMEFT can modify the SM interactions or generate new Lorentz structures after electroweak symmetry breaking. Both can alter the energy growth of scattering amplitudes and potentially lead to unitarity violating effects, despite the SMEFT still respecting the same gauge symmetry as the~SM \cite{Distler:2006if,Maltoni:2019aot,Maltoni:2019pau,Corbett:2014ora,Corbett:2017qgl}.
To this purpose, we note that all SMEFT operators, except for the four-fermion operators, contain more than one interaction vertex. An operator containing a certain number of Higgs doublets can have different multiplicities of vev insertions, and operators with field-strength tensors can lead to interactions with different numbers of gauge fields. Therefore, as in the~SM, very different scattering processes can be related by the underlying gauge symmetry.

A~contact interaction~$Q_d$ of mass-dimension~$d$ must have a coupling of dimension~$4-d$ in four spacetime dimensions $\L \supset Q_d / \Lambda^{d-4}$. The scattering amplitude for a $2 \to N$~process has the mass-dimension~$2-N$. 
The contact interaction~$Q_d$ thus leads to a contribution to the $2 \to N$~amplitude with the maximum energy scaling 
\begin{align}
\delta \mathcal{A} &= \frac{1}{\Lambda^{d-4}}E^{d-N-2} \, .
\end{align}
This implies that at~$d=6$ the maximal energy growth is~$E^2$, as expected by general dimensional considerations, and it occurs 
in $2 \to 2$~scattering. The amplitude with the maximal energy growth induced by a specific operator originates from the highest-point contact interaction that the operator includes. Lower-point interaction, e.g., obtained by vev insertions or picking the Abelian part of a field-strength tensor instead of the non-Abelian piece, usually come with lower energy scaling.\footnote{Although a longitudinally polarized gauge boson can compensate a vev insertion and bring an additional scaling with the energy~$E$.}
The energy scaling of amplitudes contributing to various scattering processes measurable 
at the~LHC, including SMEFT contributions at~$d=6$, and corresponding constraints
imposed by avoiding (perturbative) violations of unitarity have been discussed in \cite{Maltoni:2019aot,Corbett:2014ora,Corbett:2017qgl}. 


A~more general class of constraints on the SMEFT coefficients is following from the 
general requirement of analyticity and unitarity of the $S$-matrix \cite{Adams:2006sv}.
The corresponding bounds, which appear in the form of constraints on the sign of certain combinations of Wilson coefficients, are commonly known as positivity bounds.
The basic idea behind these constraints is the following:
cross sections, which are necessarily positive, can be related to the imaginary part of a forwards scattering amplitude using the optical theorem (i.e.~exploiting unitarity). 
The imaginary part of the scattering amplitude, in turn, is determined by the analytical structure of the amplitude containing isolated poles and branch cuts. 
We can then use Cauchy's integration formula to relate the amplitude to the Lagrangian parameters, allowing us to determine certain combinations of Wilson coefficients to be positive.

\begin{figure}[tb!]
    \centering
    \includegraphics[width=0.95\linewidth]{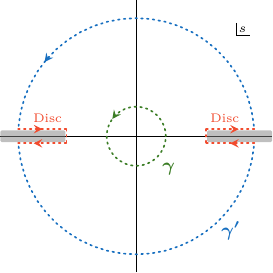}
    \caption{Analytical structure of a scattering amplitude containing a branch cut (gray shaded region) on the real $s$~axis. We also show certain integration contours that we use to determine positivity constraints on the coefficients entering the amplitude. See text for more details.}
    \label{fig:positivity-countour}
\end{figure}

For concreteness, consider a $2 \to 2$~scattering process, for which the optical theorem reads
\begin{align}
    \mathrm{Im} \, \mathcal{A}(s) &= s \, \sigma(s) \,,
    \label{eq:optical-theorem}
\end{align}
where $s=(p_1+p_2)^2$ is the Mandelstam variable, $\mathcal{A}$ is the corresponding forward scattering amplitude, and $\sigma$ is the $2 \to N$ cross section  ($\sigma \geq 0$).
After analytically continuing $s$ to the complex plane, the analytic structure of   $\mathcal{A}(s)$ is determined by isolated poles, due to intermediate single-particle on-shell production, and branch cuts, due to multi-particle on-shell production.
This allows for a power expansion of the amplitude: $\mathcal{A}(s)=\sum_k \lambda_k s^k$.
To isolate individual expansion coefficients, we can then apply Cauchy's integral formula
\begin{align}
    \lambda_n &= \frac{1}{2 \pi i} \oint_\gamma \frac{\dd s}{s^{n+1}} \mathcal{A}(s) \,,
\end{align}
where we integrate along a suitable contour~$\gamma$ as indicated by the green/inner dashed line in Fig.~\ref{fig:positivity-countour}.
For simplicity, we consider only a branch cut on the real axis for ${|s|>s_0}$ (gray shaded regions).
Enlarging the radius of our contour~$\gamma$ (green/inner dashed line), we can deform it to a new contour containing a circle $\gamma^\prime$ (blue/outer dashed line) and an integration around the branch cuts (orange dashed lines marked by ``Disc'').
We find
\begin{align}
\begin{split}
    \lambda_n &= \frac{1}{2 \pi i} \oint_{\gamma^\prime} \frac{\dd s}{s^{n+1}} \mathcal{A}(s) \\
    &\quad + \frac{1}{2 \pi i} \left( \int_{-\infty}^{-s_0} + \int_{s_0}^{\infty} \right) \frac{\dd s}{s^{n+1}} \, \mathrm{Disc} \, \mathcal{A}(s) \,,
    \label{eq:positivity-disc}
\end{split}
\end{align}
where we defined the discontinuity on the real axis by $\mathrm{Disc} \, \mathcal{A}(s) = \lim_{\epsilon \to 0} \left[ \mathcal{A}(s+i\epsilon) - \mathcal{A}(s-i\epsilon) \right] = 2 i \, \mathrm{Im} \, \mathcal{A}(s)$, using the Schwarz reflection principle, $\mathcal{A}(s^\ast)=\mathcal{A}(s)^\ast$.
Assuming that $\mathcal{A}(s)$ falls off sufficiently rapidly at infinity, such that the integral along~$\gamma^\prime$ vanishes when the radius of the circle is taken to infinity, we obtain
\begin{align}
\begin{split}
    \lambda_n &= \frac{1}{\pi} \big[ 1 + (-1)^n \big] \int_{s_0}^\infty \frac{\dd s}{s^{n+1}} \, \mathrm{Im} \, \mathcal{A}(s)
    \\
    &= \frac{1}{\pi} \big[ 1 + (-1)^n \big] \int_{s_0}^\infty \frac{\dd s}{s^{n}} \, \sigma (s) \,,
\end{split}
\end{align}
where we changed variables $s \to -s$ in the first integral of the second line in Eq.~\eqref{eq:positivity-disc} and used $\mathcal{A}(-s)=\mathcal{A}(s)$, which holds due to the crossing symmetry of the forward scattering amplitude, from which we can deduce $\mathrm{Disc}\,\mathcal{A}(-s)=-\mathrm{Disc}\,\mathcal{A}(s)$.
Furthermore, we directly applied the optical theorem~\eqref{eq:optical-theorem} in the second equality.
We thus find $\lambda_n=0$ for odd~$n$, whereas 
\begin{align}
    \lambda_n &= \frac{2}{\pi} \int_{s_0}^\infty \frac{\dd s}{s^n} \, \sigma(s) \geq 0 
    \label{eq:positivity-master-formula}
\end{align}
holds for even~$n$. 
The latter condition provides a positivity constraint on a combination of Wilson coefficients.
For more details on this topic see \cite{Adams:2006sv}.

As a concrete but simple example, consider an EFT of a real massless scalar~$\phi$, whose Lagrangian contains a single $d=8$ interaction term:
\begin{align}
    \mathcal{L} &= \frac{1}{2} (\partial_\mu \phi) (\partial^\mu \phi) + \frac{C_8}{2 \Lambda^4} \left[ (\partial_\mu \phi) (\partial^\mu \phi) \right]^2 \,.
\end{align}
The $2 \to 2$ scattering amplitude~$\mathcal{M}$ for this theory reads $\mathcal{M}(s,t)=2\frac{C_8}{\Lambda^4} (s^2 + t^2 + st)$, where we used $s+t+u=0$. Thus, we obtain the forward amplitude 
\begin{align}
    \mathcal{A}(s)= \lim_{t \to 0} \mathcal{M}(s,t)=2\frac{C_8}{\Lambda^4} s^2 \,. 
\end{align}
Realizing that $\lambda_2 = 2 C_8 / \Lambda^4$ and using our previous result in Eq.~\eqref{eq:positivity-master-formula} we find $C_8 > 0$. 
Therefore, we find that the Wilson coefficient~$C_8$ must be positive based only on unitarity and analyticity.
More details on this specific example can be found in \cite{Remmen:2019cyz}.

This type of bounds can also be exploited for more complicated theories, such as the SMEFT.
Of course, in general the $\lambda_n$ depend on several different coefficients and we can thus only determine certain combinations of Wilson coefficients that must be positive.
In fact, there has been a lot of progress on this front recently \cite{Remmen:2019cyz,Remmen:2020uze,Yamashita:2020gtt,Zhang:2018shp,Chala:2021wpj,Bellazzini:2020cot,Dvali:2012zc}, narrowing down the possible range of SMEFT coefficients, at both dimension six and eight.


\subsubsection{Gauge anomalies and reparametrization invariance}

To conclude this section, we mention two additional aspects of the SMEFT coefficients 
related to the symmetry properties of the underlying theory. 

The first aspect deals with gauge anomalies. As is well known, in (classically) renormalizable theories the 
criterion for the absence of gauge anomalies relies entirely on the charges of the fermion fields
under the local symmetry \cite{Georgi:1972bb}. When moving from the renormalizable case to the non-renormalizable one, 
this property is less obvious. In particular, doubts have been raised if the request of anomaly cancellations does
impose any additional constraint on the SMEFT Wilson coefficients. This issue has been clarified 
recently in \cite{Feruglio:2020kfq}, where it has been shown that the dependence of the anomaly 
on the non-renormalizable part of the Lagrangian 
can be removed by adding a local counterterm to the theory. 
As a result, the condition for gauge anomaly cancellation is controlled only 
by the charge assignment of the fermion sector, exactly as in the renormalizable theory.
In other words, no additional constraints can be derived on the SMEFT by requesting anomaly cancellations.

The second aspect is the so-called  reparametrization invariance of the 
dimension-six coefficients appearing in $\bar\psi \psi \to \bar\psi \psi$ 
scattering amplitudes \cite{Brivio:2017bnu}. In the Warsaw basis, 
the operators contributing $\bar\psi \psi \to \bar\psi \psi$ scattering 
give rise to a flat direction. For some time, this created confusion in 
global SMEFT fits, given the central role played by $\bar\psi \psi \to \bar\psi \psi$ data
in constraining the parameter space.
As pointed out in \cite{Brivio:2017bnu}, this fact is a consequence of 
the combined action of a field redefinition (for the vector fields) 
together with a shift of the vector--fermion couplings. This transformation 
leaves all the (physical) $\bar\psi \psi \to \bar\psi \psi$ amplitudes unchanged.
However, this is not a complete degeneracy of the theory, and indeed it is lifted when 
considering other amplitudes, such as $\bar\psi \psi \to \bar\psi \psi \bar\psi \psi$.
This property illustrates well the importance of considering complete sets of data, 
and a complete operator basis, when performing bottom-up analyses of the SMEFT parameter space.

\section{Global symmetries}
\label{sec:GlobalSymmetries}

\subsection{The role of accidental symmetries}
\label{sect:SymmA}
A key concept in any EFT is that of {\em accidental symmetries},
i.e., symmetries that arise in the lowest-dimensional operators as
indirect consequences of the field content and the symmetries explicitly imposed on the theory.
Within the SMEFT, two well-known examples are baryon number~($B$) and lepton number~($L$).
These are exact accidental global symmetries of the $d=4$ part of the Lagrangian, or the SM:
they do not need to be imposed in the SM because gauge invariance forbids to write any 
$d=4$~operator violating $B$~or~$L$. 

If the accidental symmetries are not respected by the underlying UV~completion,
we expect them to be violated by the higher-dimensional operators. 
The strong bounds on $B$-violating terms from proton stability, and the tiny coefficient of the $L$-violating 
Weinberg operator in Eq.~\eqref{eq:Weinberg-Operator} from neutrino masses, indicate that such symmetries 
remain almost unbroken in the SMEFT. This observation can be interpreted in a natural way assuming that the fundamental 
interactions responsible for $B$ and $L$ violation appear at very high energy scales, therefore, assuming 
a very high cutoff scale for these operators. 
This is not in contradiction with the possibility of having a lower cutoff scale 
for the $d=6$~SMEFT operators preserving $B$~and~$L$, since the symmetry-preserving sector cannot 
induce violations of the global symmetries. In other words, accidental global symmetries allow us to 
define a stable partition of the tower of effective operators into different sectors characterized 
by different cutoff scales, reflecting a possible multi-scale structure of the underlying theory.
The key point is that this partition is stable with respect to quantum corrections. 

Besides $B$~and~$L$, the SM Lagrangian (or better the SMEFT at~$d=4$) has two additional exact accidental
global symmetries related to the individual lepton flavor, that we can conventionally choose as 
$L_{e-\mu}$ and~$L_{\mu-\tau}$ (combined with~$L$, these correspond to the conservation of each 
individual lepton flavor). However, a much larger number of {\em approximate accidental symmetries} 
appears in the limit where we neglect the tiny Yukawa couplings of the light families and the small
off-diagonal entries of the Cabibbo-Kobayashi-Maskawa matrix. These approximate flavor symmetries 
are responsible for the smallness of flavor-changing neutral-current (FCNC) processes, such as $B$--$\overline{B}$
and $K$--$\overline{K}$ mixing, which are severely constrained by data. 
Despite the precision and the energy scales involved are very different, the situation is 
similar to that of $B$~and~$L$: the experimental bounds on FCNC processes 
imply high cutoff scales for the $d=6$~operators violating the approximate SM flavor symmetries. 
Similarly to the case of exact accidental symmetries, also the approximate accidental symmetries allow us to conceive an underlying multi-scale structure,
separating the symmetry-preserving and symmetry-breaking sectors of the theory
(the maximal scale separation being limited by the size of the explicit symmetry-breaking terms). 
This implies that  the scale of the symmetry-preserving sector of the SMEFT can be as low as few TeV, if at that  
scale not only $B$~and~$L$, but also the tightly constrained accidental flavor symmetries, remain valid,  or are broken only by small symmetry-breaking terms. 

The technical implementation of the concept of small symmetry-breaking terms, in presence of approximate 
(or exact) symmetries in the low-energy sector of the EFT, is obtained via the {\em spurion} technique, 
discussed in Sec.~\ref{sect:Flavor}. Generalizing the case of exact accidental symmetries, this technique 
can be viewed as a consistent partitioning of the tower of effective operators, reflecting a possible 
underlying multi-scale structure. This classification is particularly important in the SMEFT, given the 
large number of flavor-violating operators at~$d=6$, and the very different bounds on the 
symmetry-preserving and symmetry-breaking terms. If we do not conceive an underlying multi-scale 
structure, we are unavoidably led to the conclusion that the cutoff scale of the SMEFT is extremely high,
preventing the observation of any deviation from the SM except in rare $B$- or $L$-violating processes.

\subsection{Baryon and lepton number}
As already discussed in Sec.~\ref{subsec:SMEFT_Operator-basis},  
the unique $d=5$~operator of the SMEFT is the $L$-violating term in Eq.~\eqref{eq:Weinberg-Operator}.
This operator provides a clear illustration of the general concept of accidental symmetries discussed above:
it describes well all phenomena related to neutrino masses,
hence it provides an (indirect) evidence\footnote{Alternative descriptions of neutrino masses
not involving the Weinberg operator and preserving~$L$ are possible, but require the enlargement of the field 
content or the inclusion of operators of even higher dimension, see e.g. \cite{Gonzalez-Garcia:2007dlo}.} that~$L$ is violated beyond~$d=4$.
On the other hand, its coupling inferred from neutrino masses points to a very high effective scale:
$10^{14}\,{\rm TeV} < \Lambda < 10^{15}\,{\rm TeV}$ for $\mathcal{O}(1)$~coefficients
(following from $0.03\,{\rm eV} < \sum m_\nu  < 0.3\,{\rm eV}$).

Possible baryon number violating terms appear first at~$d=6$. The complete list of the $B$~(and~$L$) violating $d=6$~operators is shown in Tab.~\ref{Tab:Warsaw-basis_BV}. These operators satisfy the SM gauge symmetries because of the $\SU{3}_c$~property $\boldsymbol{3} \otimes \boldsymbol{3} \otimes \boldsymbol{3} \sim \boldsymbol{1}$. This is also the reason why there are no baryon and lepton number violating operators with three leptons and one quark at dimension six, and why $B-L$ is conserved at this order. The strong bound from proton decay implies severe bounds on some of these operators: $\Lambda > 10^{16}\,\mathrm{GeV}$, for $\mathcal{O}(1)$~coefficients, for terms involving only first-generation fermions.  
The constraints are significantly weaker for operators involving heavy fermions which cannot contribute to proton decay at the tree 
level \cite{Nikolidakis:2007fc}.

\begin{table}[t]
\centering
\renewcommand{\arraystretch}{1.5}
\begin{tabular}{| lc |}
	\hline
	\multicolumn{2}{| c |}{Baryon number violating $\psi^4$ operators}
	\\ \hline
	$Q_{duql}$ & $\varepsilon^{abc}\varepsilon^{ij} \squarebrackets{{(d_{a p})}^\intercal C u_{b r}} \squarebrackets{{(q_{c i s})}^\intercal C \ell_{j t}}$ 
	\\
	$Q_{qque}$ & $\varepsilon^{abc}\varepsilon^{ij} \squarebrackets{{(q_{a i p})}^\intercal C q_{b j r}} \squarebrackets{{(u_{c s})}^\intercal C e_t}$
	\\
	$Q_{qqql}$ & $\varepsilon^{abc}\varepsilon^{il}\varepsilon^{jk} \squarebrackets{{(q_{a i p})}^\intercal C q_{b j r}} \squarebrackets{{(q_{c k s})}^\intercal C \ell_{l t}}$
	\\
	$Q_{duue}$ & $\varepsilon^{abc} \squarebrackets{{(d_{a p})}^\intercal C u_{b r}} \squarebrackets{{(u_{c s})}^\intercal C e_t}$ 
	\\[0.1cm]
	\hline
\end{tabular}
\caption{Baryon number violating dimension-six operators in the Warsaw basis \cite{Grzadkowski:2010es}, with the operator labels adopted from \cite{Alonso:2014zka}. The color indices are labeled~$\{a,b,c\}$, the indices of $\SU{2}_L$ are~$\{i,j,k,l\}$, the flavor indices read~$\{p,r,s,t\}$, the charge conjugation matrix is~$C=i\gamma^2\gamma^0$, and $\varepsilon$ denotes the totally antisymmetric rank two or three tensor respectively.
\label{Tab:Warsaw-basis_BV}}
\end{table}

\subsection{Flavor symmetries}
\label{sect:Flavor}

After imposing exact $B$~and~$L$ conservation, the number of independent electroweak structures at~$d=6$ amounts 
to the $59$~terms listed in Tab.~\ref{tab:Warsaw-basis}.
The huge proliferation in the number of independent coefficients in the SMEFT at~$d=6$ occurs when all the possible flavor structures for these 
terms are taken into account: in absence of any flavor symmetry, they amount to 1350 CP-even and 1149 CP-odd independent coefficients for the dimension-six operators \cite{Alonso:2013hga}.  

Among these couplings, those contributing at tree level to flavor-violating observables, in particular 
meson--antimeson mixing and lepton-flavor violating processes are strongly constrained: 
these set bounds of $\mathcal{O}(10^{5}\,\text{TeV})$ on~$\Lambda$ for $\mathcal{O}(1)$~coefficients \cite{Isidori:2010kg}. 
If this high scale were the overall cutoff scale of the SMEFT, it would imply that all the other $d=6$~operators play an irrelevant 
role in current experiments, making the whole construction not very interesting from the phenomenological point of view.  
On the other hand, from the known structure of the SM Yukawa couplings, we know that flavor is highly non-generic, 
at least in the $d=4$~sector of the SMEFT. As anticipated, it is conceivable to assume this being the result of an underlying 
multi-scale structure, leading to approximate flavor symmetries in the whole SMEFT also beyond~$d=4$.
This assumption allows us to reduce, in a consistent way, the number of relevant parameters, making the whole 
construction more consistent and  more interesting from the phenomenological point of view, with competing constraints from  
flavor-conserving and flavor-violating processes on a given effective operator.

The price to pay to achieve this goal is the choice of the flavor symmetry and symmetry-breaking sector, which necessarily introduces some model dependence, given there is no exact flavor symmetry to start with (contrary to the case of $B$~and~$L$). 
If we are interested in symmetries and symmetry-breaking patterns able to successfully reproduce the SM Yukawa couplings and, at the same time, suppress non-standard contributions to flavor-violating observables, the choice is limited.  
Here, we analyze in some detail two cases which are particularly motivated from this point of view: the flavor symmetries $\mathrm{U}(3)^5$ and~$\mathrm{U}(2)^5$, with possible minor variations. In both cases the starting point is the flavor symmetry allowed by the SM gauge group. 

The $\mathrm{U}(3)^5$~symmetry is the maximal flavor symmetry allowed by the SM gauge group, while $\mathrm{U}(2)^5$ is the corresponding  subgroup acting only on the first two (light) generations. The $\mathrm{U}(3)^5$~symmetry allows us to implement the minimal flavor violation~(MFV) hypothesis \cite{Chivukula:1987py,DAmbrosio:2002vsn}, which is the most restrictive consistent hypothesis we can utilize in the SMEFT  to suppress non-standard contributions to flavor-violating observables \cite{DAmbrosio:2002vsn}. The $\mathrm{U}(2)^5$ symmetry with minimal breaking \cite{Barbieri:2011ci,Barbieri:2012uh,Blankenburg:2012nx} is quite interesting since it retains most of the MFV virtues, but it allows us to have a much richer structure as far as third-generation dynamics is concerned.

\subsubsection{\texorpdfstring{$\mathrm{U}(3)^5$}{U(3)5} and minimal flavor violation}
\label{sec:U3}
The largest group of global symmetry transformations of the SM fermions, compatible with 
the gauge symmetries of the SM Lagrangian, is \cite{Gerard:1982mm,Chivukula:1987py}
\bea
	\cG_f &=&  \mathrm{U}(3)_\ell \times \mathrm{U}(3)_q \times \mathrm{U}(3)_e \times \mathrm{U}(3)_u \times \mathrm{U}(3)_d \no\\
&\equiv& \mathrm{U}(3)^5 = \mathrm{SU}(3)^5 \times \mathrm{U}(1)^5 \,.
\eea
Within the SM, the Yukawa couplings ($Y_{e,u,d}$) are the only 
source of breaking of~$\cG_f$. They break this global symmetry as follows 
\bea
\cG_f = \left\{
\ba{l}
\mathrm{SU}(3)^5 \\[3pt]   \mathrm{U}(1)^5 
\ea \right. \!\!
\stackrel{ Y_{e,u,d}\neq0 }{\longrightarrow} 
\ba{l}
 \mathrm{U}(1)_{e - \mu} \times \mathrm{U}(1)_{\tau - \mu}  \\
 \mathrm{U}(1)_B \times \mathrm{U}(1)_L \times \mathrm{U}(1)_Y  \\
\ea \,, \quad
\eea
where we separated explicitly the flavor--universal and flavor--non-universal subgroups.
The three unbroken flavor--universal $\mathrm{U}(1)$ groups are baryon number, lepton number, and 
hypercharge.

Most of the $d=6$~SMEFT operators can be viewed as independent $\cG_f$--breaking terms,
hence they can be classified according to their transformation properties under~$\cG_f$.
To start with, let us consider the limit of unbroken~$\cG_f$: retaining only the $\cG_f$~invariant operators at~$d=6$ 
is not a fully consistent hypothesis, since $\cG_f$
is broken in the $d=4$~sector. However, it is a useful starting point for the classification of the operators,
and it is a coherent hypothesis to be implemented in the SMEFT in the limit where we neglect $\cG_f$--breaking terms also in the SM sector,
i.e., in the limit where we neglect the SM Yukawa couplings.

The number of independent $d=6$~terms respecting~$\cG_f$ is reported in Tab.~\ref{tab:U3new}
under the ``Exact''~$\mathrm{U}(3)^5$ column: the left (right) value in each entry indicates 
the number of CP-even (CP-odd) coefficients. For comparison, the counting of independent
coefficients if no symmetry is imposed, or if a single generation of fermions is considered, is also shown. 
As can be seen, the number of independent coefficients respecting the $\cG_f$~symmetry is smaller 
than in the single-generation case: this is because $\cG_f$ forbids bilinear couplings of fermions with 
different gauge quantum numbers, such as those appearing in the Yukawa couplings. 

\begin{table*}
\centering
\begin{center}
	\renewcommand{\arraystretch}{1.2} 
	\begin{tabular}{c   l   ||   rr   |   rr   ||    rr   |   rr  |    rr  ||   rr  |   rr  |   rr   }
	& &   \multicolumn{4}{c||}{	No symmetry }  &  \multicolumn{6}{c||}{ $U(3)^5$} & \multicolumn{6}{c}{ $U(2)^5$}   \\
   Class & Operators & \multicolumn{2}{c|}{  3 Gen.}  & \multicolumn{2}{c||}{  1 Gen.}  &  \multicolumn{2}{c|}{  Exact } &  \multicolumn{2}{c|}{ $\cO(Y_{e,d,u}^1)$ }   &  \multicolumn{2}{c||}{ $\cO(Y_{e}^1, Y_d^1 Y^2_u)$} & \multicolumn{2}{c|}{Exact} & \multicolumn{2}{c|}{$\mathcal{O}(V^1)$} & \multicolumn{2}{c}{$\mathcal{O}(V^2,\Delta^1)$}   \\  \hline
		1--4 	& $X^3$, 	$H^6$, $H^4 D^2$, $X^2 H^2$		& 9 		& 6 		& 9	& 6 	& 9	& 6  	& 9	& 6   &  9  &  6  	& 9 	& 6	& 9	& 6	 & 9		& 6	\\	\hline
		5 					& $\psi^2 H^3$ 			& 27 		& 27 		& 3	& 3	& -- 	& -- 	& 3	& 3	&  4	& 4	& 3 	& 3	& 6	& 6	 & 9		& 9	\\	
		6 					& $\psi^2 X H$ 			& 72		& 72		& 8	& 8	& --	& --	& 8	& 8	&  11	& 11	& 8	& 8	& 16	& 16	 & 24		& 24	\\
		7 					& $\psi^2 H^2 D$		& 51		& 30		& 8	& 1	& 7	& --	& 7	& --   & 11	& 1  	& 15	& 1	& 19	& 5	 & 23		& 5		\\ 	\hline
							& $(\bar{L}L)(\bar{L}L)$	& 171	& 126	& 5	& --	& 8	& --    & 8	& --	& 14 &-- 	& 23	& --	& 40	& 17	 & 67		& 24	  	\\
							& $(\bar{R}R)(\bar{R}R)$	& 255	& 195	& 7	& --	& 9	& --	& 9	& --	& 14 &-- 	& 29	& --	& 29	& --	 & 29		& -- 		\\
		8					& $(\bar{L}L)(\bar{R}R)$	& 360	& 288	& 8	& --	& 8	& --	& 8	& --	& 18 &-- 	& 32	& --	& 48	& 16  & 69		& 21			\\
							& $(\bar{L}R)(\bar{R}L)$ 	& 81		& 81		& 1	& 1	& --	& --	& --	& --	& --	& --	& 1	& 1	& 3	& 3	 & 6		& 6	\\	
							& $(\bar{L}R)(\bar{L}R)$ 	& 324	& 324	& 4	& 4	& --	& --	& --	& --	& 4	& 4	& 4	& 4	& 12	& 12	 & 28		& 28		\\ 	\hline
		\multicolumn{2}{c||}{\bf total:}					&1350	&1149	& 53	& 23 & 41	& 6	& 52 & 17	& 85 & 26 & 124& 23	&182	& 81	 & 264	& 123
			\end{tabular}
	\caption{Number of independent $d=6$~SMEFT operators without any symmetry for three and one generation(s), and when imposing a
	$\mathrm{U}(3)^5$ or $\mathrm{U}(2)^5$ flavor symmetry with different powers of symmetry-breaking terms \cite{Faroughy:2020ina}.
	In each column the left (right) number corresponds to the number of CP-even (CP-odd) coefficients.  
	$\cO(X^n)$ stands for including terms up to $\cO(X^n)$. 
	\label{tab:U3new}}
	\end{center}
\end{table*}

The MFV~hypothesis is the assumption that the SM Yukawa couplings are the only sources of $\mathrm{U}(3)^5$~breaking \cite{Chivukula:1987py,DAmbrosio:2002vsn}. 
The exact $\mathrm{U}(3)^5$~limit can be viewed as employing the MFV~hypothesis and working to zeroth order in the symmetry-breaking terms.
To go beyond the leading order, we promote the SM Yukawa couplings to become $\mathrm{U}(3)^5$~spurions, i.e., non-dynamical fields with 
well-defined transformation properties under~$\mathrm{U}(3)^5$. The latter are deduced by the 
structure of the SM Lagrangian \cite{DAmbrosio:2002vsn}:
\begin{align}
    Y_u&=\left( \boldsymbol{1},\boldsymbol{3},\boldsymbol{1}, \bar{\boldsymbol{3}},\boldsymbol{1} \right)\,,
    &
    Y_d&=\left( \boldsymbol{1},\boldsymbol{3},\boldsymbol{1},\boldsymbol{1},\bar{\boldsymbol{3}} \right)\,, 
    \nonumber\\
	Y_e&=\left( \boldsymbol{3},\boldsymbol{1},\bar{\boldsymbol{3}},\boldsymbol{1},\boldsymbol{1} \right)\,.
\end{align}
With these transformation properties, the $d=4$~sector of the theory is formally invariant under~$\mathrm{U}(3)^5$.
The MFV~hypothesis consist in constructing the higher-dimensional operators using SM fields and 
spurions, such that the EFT remains formally invariant under~$\mathrm{U}(3)^5$ to all orders, 
and the breaking occurs only via the appropriate insertions of the spurions~$Y_{u,d,e}$.

In principle, the spurions can appear with arbitrary powers both in the renormalizable ($d=4$)~part of the Lagrangian 
and in the dimension-six effective operators. 
However, via a suitable redefinition of both fermion fields and spurions, we can always put the $d=4$~Lagrangian 
to its standard expression, identifying the spurions with the SM Yukawa couplings. 
This implies we can always choose a flavor basis where the spurions are 
completely determined in terms of fermion masses and the Cabibbo-Kobayashi-Maskawa~(CKM) matrix,~$V_{\rm CKM}$. A~representative 
example is the down-quark mass-eigenstate basis, where
\bea
&& Y_e =  \textrm{diag}(y_e,y_\mu,y_\tau)\,, \qquad
Y_d  = \textrm{diag}(y_d,y_s,y_b)\,,  \no \\
&& Y_u =  V_{\rm CKM}^\dagger \times \textrm{diag}(y_u,y_c,y_t)\,.
\label{eq:d-basis}
\eea
The key point is that there are no free (observable) parameters in the structure of the MFV spurions. 
A~related important point is the fact that, knowing the structure of the spurions, 
we know that they are all small but for the top Yukawa~$y_t$. We can thus limit the spurion expansion to a few terms. 

The overall number of independent terms allowed by the MFV hypothesis with at most one ``small'' Yukawa coupling, namely 
$Y_d$ or~$Y_e$, and up to two powers of~$Y_u$ is shown in the last column of Tab.~\ref{tab:U3new} \cite{Faroughy:2020ina}.
As can be seen, this number is almost two orders of magnitudes smaller than what is obtained 
in absence of any symmetry (for three generations) and quite close to the single generation case.
With the corresponding set of operators we can describe the SM spectrum and possible
deviations from the SM in a series of rare flavor-violating processes \cite{DAmbrosio:2002vsn}.
A~representative set of these operators is shown in Tab.~\ref{tab:MFV}.

\begin{table}[t]
\begin{center}
\resizebox{\linewidth}{!}{%
\begin{tabular}{c | c | c}
EW type & possible MFV form & ~Bound on $\Lambda$ \\
\hline
$Q_{dB}$ 			& $ [\bar q_r (Y_u Y_u^\dagger Y_d)_{rp}   \sigma^{\mu\nu} d_p] H (g_1 B_{\mu\nu})  \phantom{\frac{P^X}{2}}$ 	& $6.1$~TeV  \\[2pt]   
$Q_{dG}$ 		& $ [\bar q_r (Y_u Y_u^\dagger Y_d)_{rp}  \sigma^{\mu\nu} T^A   d_p)  H (g_3 G_{\mu\nu}^A)$ 					& $3.4$~TeV  \\[2pt] 
$Q_{Hq}^{(1)}$   	& $(H^\dagger i \overleftrightarrow{D}_\mu H)  [{\bar q}_r (Y_u Y_u^\dagger)_{rp}  \gamma_\mu q_p]$ 			& $2.3$~TeV  \\[2pt]   
$Q_{qq}^{(1)}$ 		& $[\bar{q}_r  (Y_u Y_u^\dagger)_{rp} \gamma_{\mu} q_p][\bar{q}_r  (Y_u Y_u^\dagger)_{rp} \gamma_{\mu} q_p]$ 	& $6.0$~TeV  \\[2pt]     
$Q_{q e}$   		& $[{\bar q}_r (Y_u Y_u^\dagger)_{rp} \gamma_\mu q_p] [ {\bar e}_s \gamma_\mu e_s]$  					& $2.7$~TeV  \\[2pt]   
$Q_{\ell q}^{(1)}$	& $[{\bar q}_r (Y_u Y_u^\dagger)_{rp}  \gamma_\mu q_p]  [\bar{\ell}_s \gamma_\mu \ell_s]$ 					& $1.7$~TeV  \\[2pt]   
\end{tabular}
}
\end{center}
\caption{\label{tab:MFV} Representative set of SMEFT operators with their flavor structure determined according to the MFV~hypothesis. 
Each electroweak structure (first column) can admit different MFV implementations: in the second column we indicate the one more constrained 
by flavor-violating processes in the quark sector. The corresponding bounds on the effective scale set by $B$- and $K$-meson physics measurements is reported in the 
third column (95\%\,C.L.~bound, assuming an effective coupling $\sim\pm 1/\Lambda^2$, considering each operator separately).}
\end{table}

The number of insertions of the (large) $Y_u$~spurions has been limited to two since,  
in the reference basis~\eqref{eq:d-basis}, one gets 
\be
\left[  Y_u (Y_u)^\dagger \right]^n_{r\not = p} ~\approx~
y_t^{2n} V^*_{tr} V_{tp} \propto  [Y_u (Y_u)^\dagger]_{r\not = p} \,.
\label{eq:basicspurion}
\ee
This result implies that within the MFV hypothesis rare FCNC processes that, within the SM, 
are not helicity suppressed and are
dominated by virtual top-quark contributions 
(such as $B^0$--$\overline{B}{}^0$ and $K^0$--$\overline{K}{}^0$ mixing, $b\to s \gamma$, $b\to s \ell^+\ell^-$, \ldots),
receive exactly the same CKM suppression as in the SM:
\begin{align}
\mathcal{A}(d^i \to d^j)_{\rm MFV} &=  (V^*_{ti} V_{tj}) \mathcal{A}^{(\Delta F=1)}_{\rm SM}
\!\!\left[ 1 + a_1 \frac{ 16 \pi^2 M^2_W }{ \Lambda^2 } \right], \!\!\!\! \no \\  
\mathcal{A}(M_{ij}\!-\!{\overline{M}}_{ij})_{\rm MFV}  &= \! (V^*_{ti} V_{tj})^2
\mathcal{A}^{(\Delta F=2)}_{\rm SM} \!\!\left[ 1 + a_2 \frac{ 16 \pi^2 M^2_W }{ \Lambda^2 } \!\right]\!,
\label{eq:FC}
\end{align}
where $\mathcal{A}^{(i)}_{\rm SM}$ denote the SM loop amplitudes and $a_i$ are $\mathcal{O}(1)$~parameters.
The~$a_i$ depend on different combinations of SMEFT coefficients but are flavor independent.  
Actually, Eq.~(\ref{eq:FC}) can be used 
to defined in an operative way the MFV hypothesis for a large class of flavor-changing processes, as proposed in \cite{Buras:2003jf}.


\subsubsection{The \texorpdfstring{$\mathrm{U}(2)^5$}{U(2)5} symmetry}
\label{sect:U2}

The $\mathrm{U}(2)^5$~flavor symmetry is the subgroup of the $\mathrm{U}(3)^5$ global symmetry   
that, by construction, distinguishes the first two generations of fermions 
from the third one \cite{Barbieri:2011ci,Barbieri:2012uh,Blankenburg:2012nx}.
For each set of SM fermions with the same gauge quantum numbers, the first two generations form a doublet of a given $\mathrm{U}(2)$ subgroup, whereas the third one transforms as a singlet. 
Denoting the five independent flavor doublets as $L,Q,E,U,D$, the flavor symmetry decomposes as 
\be
	\mathrm{U}(2)^5 = \mathrm{U}(2)_L \times \mathrm{U}(2)_Q \times \mathrm{U}(2)_E \times \mathrm{U}(2)_U \times \mathrm{U}(2)_D \,.
 \label{eqU25}
\ee
In the limit of unbroken~$\mathrm{U}(2)^5$, only third-generation 
fermions can have non-vanishing Yukawa couplings,
which is an excellent first-order approximation for the SM Lagrangian. 
This is why, contrary to the MFV case, the $\mathrm{U}(2)^5$~symmetry allows us to build an EFT where all the symmetry-breaking terms are small.

A~$\mathrm{U}(2)^3$~symmetry in the quark sector can be viewed as the result of a generalized MFV framework,
taking into account arbitrary insertions of the third-generation Yukawa couplings
without suppression (the so-called non-linear representation of MFV \cite{Feldmann:2008ja} or general MFV \cite{Kagan:2009bn}
hypothesis). However, this interpretation is less motivated in the lepton sector and it also 
implies a rather strict structure for the symmetry-breaking terms. On the other hand, the symmetry group in~\eqref{eqU25} with 
the symmetry-breaking terms discussed below, can be viewed as an effective way to describe in general terms 
the large class of SM extensions where the third-generation of fermions plays a special role.

\paragraph{Yukawa couplings and spurion structures.}
A~set of symmetry-breaking terms sufficient to reproduce 
the complete structure of the SM Yukawa couplings is \cite{Barbieri:2011ci}
\begin{align}
V_\ell &\sim\left(\boldsymbol{2},\boldsymbol{1},\boldsymbol{1},\boldsymbol{1},\boldsymbol{1}\right)\,,
&
V_q &\sim\left(\boldsymbol{1},\boldsymbol{2},\boldsymbol{1},\boldsymbol{1},\boldsymbol{1}\right)\,,  \no 
\\ 
\Delta_e &\sim\left(\boldsymbol{2},\boldsymbol{1},\bar{\boldsymbol{2}},\boldsymbol{1},\boldsymbol{1}\right)\,,
&
\Delta_{u(d)} &\sim\left(\boldsymbol{1},\boldsymbol{2},\boldsymbol{1},\bar{\boldsymbol{2}}(\boldsymbol{1}),\boldsymbol{1}(\bar{\boldsymbol{2}})\right)\,.
\label{eq:U2spur}
\end{align}
By construction, $V_{q,\ell}$ are complex two-vectors and $\Delta_{e,u,d}$ are complex $2\times 2$~matrices. In terms of these spurions, we can express the Yukawa couplings as
\bea
&&	Y_e = y_\tau\left(\begin{matrix}
		\Delta_e	 & x_\tau V_\ell \\
		0			 & 1
	\end{matrix}\right)\,, \quad
	 Y_u = y_t\left(\begin{matrix}
		\Delta_u	 & x_t V_q \\
		0			 & 1
	\end{matrix}\right)\,,  \no\\
&&	 Y_d = y_b\left(\begin{matrix}
		\Delta_d	 & x_b V_q \\
		0			 & 1
	\end{matrix}\right),
	\label{eq:YU2_5}
\eea
where $y_{\tau,t,b}$ and $x_{\tau,t,b}$ are free complex parameters expected to be of order~$\mathcal{O}(1)$. 

The spurion set in Eq.~\eqref{eq:YU2_5} is minimal in terms of 
independent $\mathrm{U}(2)^5$~structures (at least as far as the quark sector is concerned), 
and leads to spurions which are small and hierarchical in size. 
Contrary to the MFV~framework, in this case 
we cannot determine completely the spurions in terms of SM parameters. 
However, we can constrain their size requiring no tuning in the  $\cO(1)$~parameters. 
In particular, from the $2 \leftrightarrow 3$~mixing in the CKM matrix we deduce 
$|V_q| = \cO(|V_{cb}|)$, while light-quark and lepton masses imply $|\Delta_{u,d,e}|_{ij} \ll |V_{q}|$.

There are no unambiguous constraints about the size of~$V_\ell$. Actually, the SM lepton Yukawa coupling 
can be reproduced even setting~$V_\ell=0$. On the other hand, assuming 
a common structure for the three Yukawa couplings, as suggested by the similar hierarchies 
observed in the eigenvalues, it is natural to assume~$|V_\ell|\sim |V_q|$.
The assumption that $V_{q,\ell}$ are the leading 
$\mathrm{U}(2)^5$--breaking spurions ensures a suppression of 
flavor-violating terms in the quark sector, via higher-dimensional operators, as effective as the one implied by the MFV hypothesis.

It is convenient to define as reference (or interaction) basis the flavor basis 
in $\mathrm{U}(2)^5$~space where $ V_{q,\ell}=  |V_{q,\ell}| \times \vec n$,  with $\vec n ={(0,1)}^\intercal$, and $\Delta_{u,d,e}^\dagger \Delta_{u,d,e}$ 
are diagonal.
After the $\mathrm{U}(2)^5$~symmetry is broken, the residual flavor symmetry implies that the Yukawa matrices 
in the interaction basis can be written 
in the following form \cite{Fuentes-Martin:2019mun} 
\begin{subequations}%
\begin{align}
Y_u &= |y_t|
\begin{pmatrix}
U_q^\dagger O_u^\intercal\, \hat\Delta_u & |V_q|\,|x_t|\,e^{i\phi_q}\,\vec n\\
0 & 1
\end{pmatrix}
\,, \\
Y_d &=|y_b|
\begin{pmatrix}
\;\;\;U_q^\dagger \hat\Delta_d& |V_q|\,|x_b|\,e^{i\phi_q}\,\vec n\\
\;\;\;0 & 1
\end{pmatrix}
\,, \\
Y_e &=|y_\tau|
\begin{pmatrix}
\;\;O_e^\intercal\,\hat\Delta_e\;\;& |V_\ell|\,|x_\tau|\,\vec n\\
\;\;0 & 1
\end{pmatrix}
\,, 
\end{align}\label{eq:U2param}%
\end{subequations}%
where $\hat\Delta_{u,d,e}$ are $2\times 2$~diagonal positive matrices, $O_{u,e}$ are $2\times2$ orthogonal matrices, 
and $U_q$ is a complex unitary matrix. The unitary matrices that diagonalize the above Yukawa matrices 
 can be found in \cite{Fuentes-Martin:2019mun}.
After expressing the free parameters in terms of fermion masses and CKM elements, 
the residual terms which cannot be determined in terms of SM parameters are:
\begin{itemize}
\item{} {\em quark sector}:  $2\leftrightarrow 3$  mixing angle in the down sector, $s_b \approx |x_b|\,|V_q|$, and CP-violating phase $\phi_q$;
\item{} {\em lepton sector}: $2\leftrightarrow 3$ mixing angle $s_\tau \approx |x_\tau|\,|V_\ell|$ and $1\leftrightarrow 2$ mixing angle $s_e$ (which appears in $O_e$).
\end{itemize}

As pointed out in \cite{Greljo:2022cah}, the parametrization in~\eqref{eq:U2param} is redundant and all the 
(non-SM) parameters listed above can be eliminated via a suitable change of basis consistent within the $\mathrm{U}(2)^5$~framework. 
For instance, in the quark sector both~$s_b$ and~$\phi_q$ can be eliminated by a  
transformation mixing the $\mathrm{U}(2)_Q$~singlet field with the $\mathrm{U}(2)_Q$~doublet appropriately contracted with spurions. 
While this is certainly correct, this change of basis implies a shift in the tower of higher-dimensional operators. In a pure bottom-up approach, this shift has no practical consequences, hence the redundancy can safely be removed. On the other hand, keeping the redundant formulation in~\eqref{eq:U2param}
is particularly useful when matching to a specific UV theory: it highlights the fact that the third generation, with a special role in the UV, is not unambiguously determined by the SM Yukawa couplings.  

\paragraph{Higher-dimensional operators.}
The exact $\mathrm{U}(2)^5$~symmetry is the natural (and unavoidable) starting point to 
describe all processes where we can neglect light-fermion masses. 
This is why the SMEFT with unbroken~$\mathrm{U}(2)^5$ is employed to
describe  top-quark physics and related processes at colliders \cite{Aguilar-Saavedra:2018ksv}. 
The number of relevant operators is listed in Tab.~\ref{tab:U3new}.

\begin{figure}[t]
\begin{center}
\includegraphics[angle=-90,width=1.0\linewidth]{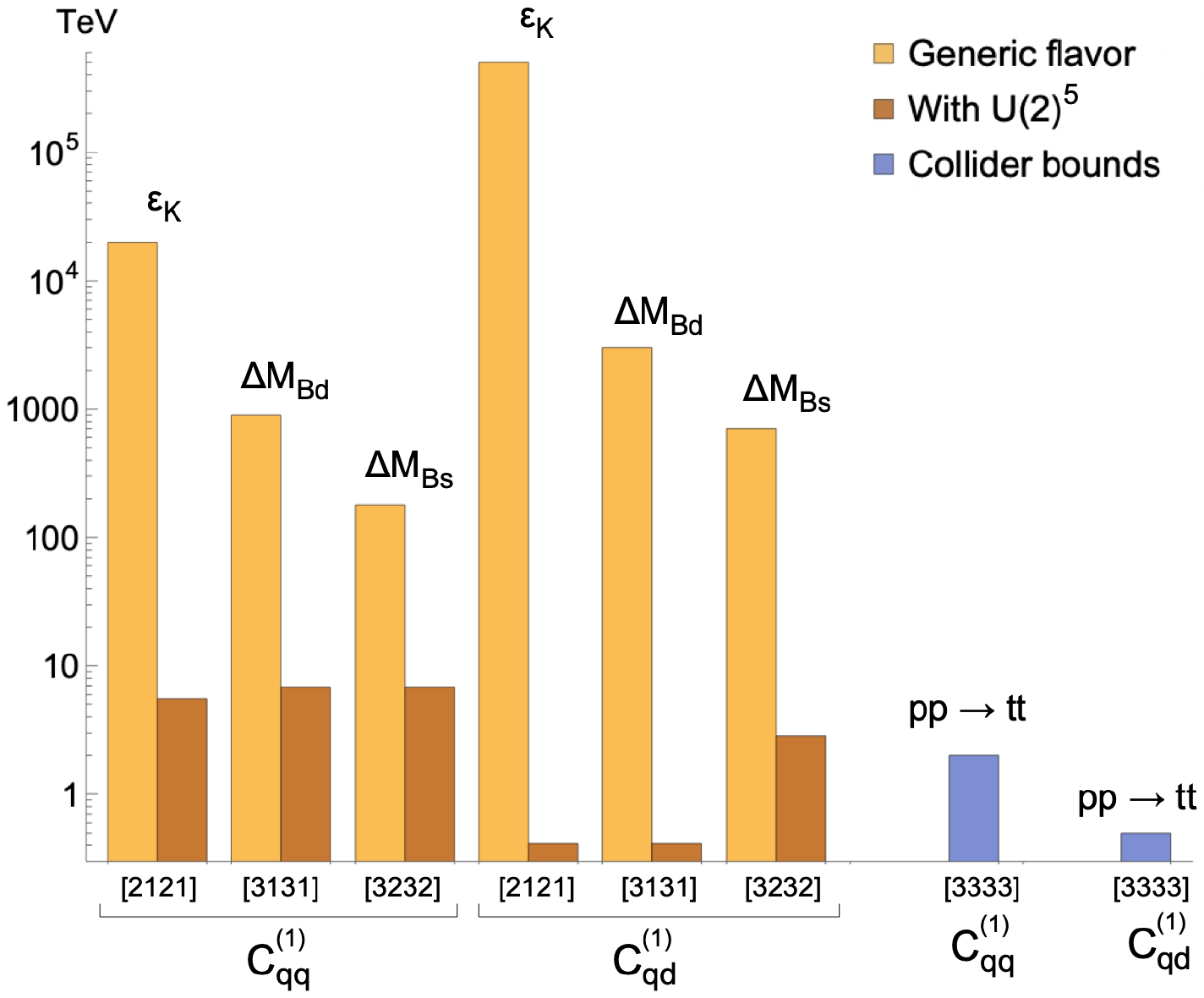}
\caption{Bounds on the effective 
scales of the SMEFT four-quark operators
$Q_{qq}^{(1)}$ and $Q_{qd}^{(1)}$, for different 
flavor indices, as reported
on the horizontal axis between square brackets (the bounds 
are 95\%\,C.L.~limits for effective scales defined as in Table~\ref{tab:MFV}). 
For left-handed fields, the flavor indices refer to the down-quark mass-eigenstate basis. 
The bounds ``with $\mathrm{U}(2)^5$'' (brown bars)
are obtained incorporating
in the operators one or more $\mathrm{U}(2)^5$ breaking terms, 
according to the rules discussed in Sect.~\ref{sect:U2}.
The observables used to derive the bounds are also indicated.
}
\label{fig:flavorbounds}
\end{center}
\end{figure}

In the same table also the terms obtained with one $V$~spurion, or two of them and one $\Delta$~spurion are shown. 
The higher-dimensional operators built in terms of a single $V_{q,\ell}$~spurion 
contribute to flavor-violating transitions which involve only left-handed fields and connect 
only the $2\leftrightarrow 3$~sectors in the interaction basis.
Considering terms with two $V_q$~spurions is the analog of considering two $Y_u$~insertions in~MFV.
Compared to the latter case, the $\mathrm{U}(2)^5$~hypothesis leads to more freedom (differentiating, for instance, 
effective operators contributing to flavor-violating process in $B$- and $K$-meson physics)
but also more terms. This latter statement can be understood by looking at the 
number of independent invariant $\bar{q}_r \gamma^{\mu} q_p$ bilinears 
in the two cases:\footnote{Here the flavor indices $\{r,p\}$ run from 1 to 3, whereas $\{i,j\}$ only between 1 and 2.}
\be
\left.
\begin{array}{l}
  	\bar{q}_r \gamma^{\mu} q_r  \\  
  	\bar{q}_r  (Y_u Y_u^\dagger)_{rp} \gamma^{\mu} q_p 
\end{array}
\right|_{\mathrm{U}(3)^5}
 	\to
\left.
\begin{array}{l}
	\bar{q}_3 \gamma^{\mu} q_3 \\  
	\bar{q}_i \gamma^{\mu} q_i  \\
	{\bar q}_i (V_q)_{i}  \gamma^\mu q_3 + \textrm{h.c.}\\  
	{\bar q}_i (V_q)_i \gamma^\mu (V_q^\dagger)_{j} q_j  
\end{array}	 
\right|_{\mathrm{U}(2)^5}
\nonumber
\ee

In Fig.~\ref{fig:flavorbounds} we illustrate more concretely some of these features 
showing bounds on two representative four-quark SMEFT operators,
$Q_{qq}^{(1)}$ and $Q_{qd}^{(1)}$ (see 
Tab.~\ref{tab:Warsaw-basis}), for different flavor indices. The strong bounds on the effective scales exceeding 100~TeV (light yellow bars) are those obtained without any symmetry hypothesis. They correspond to flavor combinations 
leading to un-suppressed tree-level contributions to specific meson-antimeson mixing amplitudes.
By contrast, once the suppression due the $\mathrm{U}(2)^5$ breaking spurions is taken into account, the same observables leads to bounds  on the effective scales 
below 10~TeV.
It is interesting to note that these bounds are 
comparable to those obtained by direct searches 
for flavor-conserving combinations involving only third-generation fermions, which are the most severely constrained by high-energy LHC data.

We stress that the hypothesis of a $\mathrm{U}(2)^5$~flavor symmetry broken by the minimal set of 
spurions in Eq.~\eqref{eq:YU2_5} naturally implies lepton flavor violation in charged leptons. This is controlled by 
the size of~$V_\ell$ and~$s_e$, which are left unconstrained by the SM Yukawa couplings.
This is one of the most evident differences between the genuine $\mathrm{U}(2)^5$~approach and the 
non-linear MFV hypothesis \cite{Feldmann:2008ja,Kagan:2009bn}.

\subsubsection{Other options and running}
The $\mathrm{U}(2)^5$~case discussed above is the prototype of a series of 
symmetry groups providing a suppression similar to MFV in the quark sector, but allowing 
more general breaking terms. The common ground is the presence of the (chiral) non-Abelian 
group~$\mathrm{U}(2)^3$ acting in the quark sector. The variations come from obtaining this 
group as a subgroup of possible larger symmetries, such as $\mathrm{U}(2)^2\times \mathrm{U}(3)_d$
or $\mathrm{U}(2)^3\times \mathrm{U}(1)_d$ \cite{Greljo:2022cah,Faroughy:2020ina}.
Given the smaller set of phenomenological constraints,
a larger set of variations have been proposed in the lepton sector \cite{Greljo:2022cah}.

A somehow different approach is that of using only $\mathrm{U}(1)$~groups, as originally proposed 
by \textcite{Froggatt:1978nt}. Recent analyses of this type can be found in \cite{Smolkovic:2019jow,Bordone:2019uzc}.

To conclude the discussion about flavor symmetries, it is worth mentioning that the approximate symmetries present in the SM are responsible for
a series of powerful (approximate) selection rules in the renormalization group evolution of the SMEFT~\cite{Feldmann:2015nia,Machado:2022ozb}. 
These are nothing but the manifestations of the statement made in 
Sec.~\ref{sect:SymmA}
that the 
partitioning of the EFT due to global symmetries is
stable with respect to quantum corrections. 
These selection rules become manifest when working in a basis of flavor invariants, where the apparently large  anomalous dimension matrix of dimension-six current-current operators is reduced to a block-diagonal structure with several blocks of small dimension~\cite{Machado:2022ozb}.

\subsection{Custodial symmetry}
\label{sect:custodial}
The large number of fermions in the SM implies that most of the exact or approximate  global symmetries of the theory are related to the fermion sector, as discussed so far. However, there is one important symmetry that involves mainly (but not only) the scalar sector.

Custodial symmetry is an exact symmetry of the pure Higgs sector of the~SM,
\be
\cL_H = \partial_\mu H^\dagger \partial^\mu H - V(H)\,,
\ee
with the scalar potential defined as in (\ref{eq:Hpotential}). 
The simplest way  to realize the global symmetry of~$\cL_H$ 
is to write the complex Higgs doublet in terms 
of four independent real scalar components~$\phi^i$ as in~\eqref{eq:Hdec},
\begin{align}
	H = \frac{1}{\sqrt{2}} 
	\begin{pmatrix}
		\phi^2 + i \phi^1 \\
		\phi^4 - i \phi^3
	\end{pmatrix}\,.
\label{eq:geoSMEFT_H_to_phi}
\end{align}
We find
\begin{align}
\begin{split}
    \L_{\phi}
	&= \frac{1}{2} \brackets{\partial_\mu \boldsymbol{\phi}} \cdot \brackets{\partial^\mu \boldsymbol{\phi}} 
	+ \frac{m^2}{2} \boldsymbol{\phi} \cdot \boldsymbol{\phi} - \frac{\lambda}{8} \left( \boldsymbol{\phi} \cdot \boldsymbol{\phi} \right)^2 \, ,
\end{split}	
\label{eq:geoSMEFT_SM_Lagrangian}
\end{align}
where we have defined
\begin{align}
	\boldsymbol{\phi}=\begin{pmatrix}
		\phi^1 \\ \phi^2 \\ \phi^3 \\ \phi^4
	\end{pmatrix} = \begin{pmatrix}
		\varphi^1 \\ \varphi^2 \\ \varphi^3 \\ v + h
	\end{pmatrix} 
\label{eq:geoSMEFT_Cartesian_coordinates}
\end{align}
with the Higgs vev~$v$, the physical Higgs bosons~$h$, and $\varphi^a$ being the Goldstone bosons of electroweak symmetry breaking.
It is easy to verify that $\cL_H$~or~$\cL_\phi$ depend only 
on $\boldsymbol{\phi}\cdot\boldsymbol{\phi} = 2H^\dagger H$ and are thus invariant under a global 
$\mathrm{O}(4)$~symmetry, with
the symmetry transformation $\boldsymbol{\phi} \to O \, \boldsymbol{\phi}$ for $O \in \mathrm{O}(4)$.
The minimum of the Higgs potential is the three-sphere~$S^3$ with radius~$v$, defined by $\langle\boldsymbol{\phi}\cdot\boldsymbol{\phi} \rangle = v^2$. Hence, the $\mathrm{O}(4)$~global symmetry
of~$\cL_H$
is spontaneously broken by the Higgs~vev to its subgroup~$\mathrm{O}(3)$. The corresponding Goldstone bosons $\boldsymbol{\varphi}=\smash{\left(\varphi^1,\varphi^2,\varphi^3\right)^\intercal}$ transform under this group as $\boldsymbol{\varphi} \to \tilde{O}\,\boldsymbol{\varphi}$, where $\tilde{O} \in \mathrm{O}(3)$.

This $\mathrm{O}(3)$~global symmetry of the Higgs sector after electroweak symmetry breaking 
is responsible, among other things, for the tree-level relation~$\rho=1$, where 
\be 
\rho  \equiv \frac{m^2_W}{m_Z^2} \frac{g_1^2 + g_2^2}{g_2^2}\,.
\ee
This relation is tested to the permil level finding good 
agreement with the SM prediction, after taking into account 
the small deviations generated beyond the tree level. 
On the other hand, adding to~$\cL_H$ 
generic dimension-six operators compatible only with the gauge symmetry of the SM,
one would expect $\rho-1 = \mathcal{O}(v^2/\Lambda^2)$. 

Custodial symmetry is explicitly broken in the SM,
both by the electroweak gauge symmetry, which acts differently on the different $\phi^i$~components,
and by the Yukawa interactions. These breaking terms are responsible for the deviation from~$\rho=1$ generated beyond the tree level.  In particular, the leading contribution induced by the top Yukawa coupling reads
\be 
(\rho - 1)^{y_t}_{\rm SM} = \frac{ 3 y_t^2}{32 \pi^2}
 \approx 1\%\,.
 \ee
Given the strong constraint on the SMEFT imposed by the $\rho$~parameter, 
it is interesting to conceive the case of new-physics models 
where the breaking of custodial symmetry is small as in the SM,
originating only from the gauge and the Yukawa sector. 
In other words, in close analogy to the flavor symmetries discussed above,
it is interesting to treat custodial symmetry as an approximate global symmetry 
of the SMEFT broken by a well-defined set of spurion terms. 

In order to describe the explicit breaking of custodial symmetry occurring in 
the~SM, it is more convenient to express the symmetry
in a different way, taking into account the (local) equivalence
of the $\mathrm{SO}(4)$~group with the product of two $\mathrm{SU}(2)$~groups:
\begin{align}
   \mathrm{O}(4) \simeq \SU{2}_L \times \SU{2}_R \, .
   \label{eq:custSU2}
\end{align}
To see how the 
$\mathrm{SU}(2)$~groups in~\eqref{eq:custSU2} act on the Higgs field, we can combine~$H$ and its conjugate, $\widetilde H = \varepsilon H^\ast$, where~$\varepsilon=i\tau^2$ is the totally anti-symmetric $\mathrm{SU}(2)$~tensor,
to form the $2\times 2$~matrix field~$\Sigma$, 
transforming as a $(\boldsymbol{2}_L,  \bar{\boldsymbol{2}}_R)$ 
under~\eqref{eq:custSU2}, namely 
\begin{align}
    \Sigma \equiv \left(\widetilde{H},H\right) \rightarrow V_L \, \Sigma \, V_R^\dagger
    \label{eq:Higgs-matrix-field}
\end{align}
with $V_{L(R)} \in \mathrm{SU}(2)_{L(R)}$.
We find $\smash{\mathrm{tr}\left[\Sigma^\dagger \Sigma\right]}=\smash{2\,H^\dagger \! H}=\smash{\boldsymbol{\phi}\cdot\boldsymbol{\phi}}$ allowing us to write the Higgs Lagrangian~$\mathcal{L}_H$ as 
\begin{align}
    \mathcal{L}_\Sigma
    &= 
    \frac{1}{2} \mathrm{tr} \left[ \left(\partial_\mu\Sigma\right)^\dagger \! \left(\partial^\mu\Sigma\right) \right] 
    +\frac{m^2}{2}\mathrm{tr}\left[\Sigma^\dagger \Sigma\right]
    -\frac{\lambda}{8} \left(\mathrm{tr}\left[\Sigma^\dagger \Sigma\right]\right)^2 \!\! .
    \label{eq:L-linear-sigma-model}
\end{align}
With this notation it is easy to verify that $\cL_H$~or~$\cL_\Sigma$ is invariant 
under $\SU{2}_L \times \SU{2}_R$~global transformations, and that the spontaneous symmetry breaking due to the Higgs vev corresponds to $\mathrm{O}(4) \simeq \mathrm{SU}(2)_L \times \mathrm{SU}(2)_R \to \mathrm{SU}(2)_{L+R} \simeq \mathrm{O}(3)$.
While the $\mathrm{SU}(2)_L$~group is fully gauged in the SM, only a part of the $\SU{2}_R$~group is gauged, 
leading to an explicit breaking of custodial symmetry. More precisely, gauging hypercharge in the Higgs sector 
is equivalent to gauging only the $\mathrm{U}(1)$~subgroup of~$\SU{2}_R$ corresponding to the diagonal generator~$T^3_R$.

Up to this level, i.e., when considering only the gauge sector, the identification of the explicit breaking of 
custodial symmetry from a general EFT point of view is unambiguous.
An ambiguity arises when considering also the fermion sector, given the action of~$T^3_R$ is not sufficient to 
describe fermion hypercharges. On general grounds, we can extend the symmetry to \cite{Elias-Miro:2013mua}
\be
\cG_{\rm cust} = \mathrm{SU}(2)_L \times \SU{2}_R \times \mathrm{U}(1)_X\,,
\ee 
such that hypercharge reads
\be 
Y = T^3_R +X\,.
\ee 
However, different embeddings of the SM fermions in~$\cG_{\rm cust}$ are possible. The simplest one corresponds to the choice
$X=(B-L)/2$. In such case, all right-handed fermions belong to doublets of~$\SU{2}_R$, with an incomplete 
doublet in the lepton sector due to the absence of right-handed neutrinos, while all left-handed fermions are assumed to be singlets of~$\SU{2}_R$. But other options are also possible \cite{Elias-Miro:2013mua}.

Once a representation of the SM fermions under~$\cG_{\rm cust}$ is chosen, we have all the ingredients to define 
a consistent EFT based on the hypothesis of minimal breaking of custodial symmetry, 
able to reproduce all SM properties.  
First, all representations of~$\cG_{\rm cust}$ including some SM fermions are promoted to be complete representations 
by introducing appropriate spurions (unphysical) fields which are set to zero in physical processes \cite{Elias-Miro:2013mua}.
Second, $\SU{2}_R$~breaking terms in the Yukawa couplings, such as the one responsible for the 
top-bottom splitting when $t_R$~and~$b_R$ are embedded in the same $\SU{2}_R$~multiplet,
are also promoted to be spurion fields \cite{Isidori:2009ww}. Finally, also spurion gauge bosons are introduced, 
so that the whole group~$\cG_{\rm cust}$ is formally gauged, and the SM is recovered as the limit obtained 
setting the spurion fields to zero \cite{Gonzalez-Alonso:2014eva}.

The consequences of these hypotheses for various subsets of
SMEFT operators (or physical amplitudes evaluated at~$d=6$ 
in the SMEFT) have been discussed in \cite{Contino:2013kra,Elias-Miro:2013mua,Gonzalez-Alonso:2014eva}.

\section{Non-linear realization of electroweak symmetry breaking}
\label{sec:HEFT}

Within the SM,  the spontaneous breaking of the electroweak symmetry occurs through the non-vanishing 
vacuum expectation value of the $\mathrm{SU}(2)_{L}$--doublet scalar $H$.
Expanding around the minimum of $H$
as in  Eq.~(\ref{eq:Hdec}), one identifies the 
massive field with $h$ and the three 
Goldstone bosons with $\varphi_{1,2,3}$.
From measurements of various electroweak observables
and high-energy processes, the existence of the three Goldstone bosons and the massive scalar~$h$ is well established. However, it is not yet evident that they are necessarily embedded into the four components of a single 
$\mathrm{SU}(2)_L$--doublet~$H$, as in Eq.~(\ref{eq:Hdec}).
We refer to this embedding as 
the linear realization of the 
electroweak symmetry breaking mechanism. 
As we shall discuss shortly,
in principle, other embeddings are still viable. 

The inclusion of the scalar state~$h$ in the theory, and the relations among the different couplings provided by embedding $h$ and the three~$\varphi_{1,2,3}$ into the doublet~$H$, is essential to ensure the unitarity of scattering amplitudes for longitudinally polarized electroweak gauge bosons at high energies.  
However, within an EFT approach the loss of unitarity is  not a problem as long as it happens above the cutoff scale of the theory  [see e.g.~\cite{Brivio:2017vri}]. As a consequence, in a general EFT 
approach to physics beyond the SM we are allowed to relax the strict
constraints following from the linear embedding of 
$h$ and the three~$\varphi_{1,2,3}$ into~$H$ and consider a more general structure. 

To this end, we employ the Callan-Coleman-Wess-Zumino~(CCWZ) formalism \cite{Coleman:1969sm,Callan:1969sn} and proceed similarly to the construction of chiral perturbation theory. In other words, we construct an EFT with a non-linear realization of the electroweak symmetry breaking mechanism. 
To make contact with the SM Lagrangian, it is convenient to decompose the matrix field~$\Sigma$ introduced in Eq.~\eqref{eq:Higgs-matrix-field} as
\begin{align}
    \Sigma(x) &= \frac{v+\hat{h}(x)}{\sqrt{2}} U(x) \quad\text{with}\quad U(x) = \exp \left( i\frac{\boldsymbol{\tau}\cdot\boldsymbol{\pi}(x)}{v} \right) ,
    \label{eq:HEFT-non-linear-field-redefinition}
\end{align}
where we introduced the unitary dimensionless field~$U(x)$, and $\boldsymbol{\tau}$ denotes the three-vector of Pauli matrices. The new fields~$\smash{\hat{h}}$ and $\boldsymbol{\pi}=\smash{\left(\pi_1,\pi_2,\pi_3\right)^\intercal}$, where the latter are the counterpart of the pions in two-flavor QCD, are related to the original fields $h$ and $\boldsymbol{\varphi}$ by a non-linear field redefinition.  We can still interpret~$\smash{\hat{h}}$ as the physical Higgs boson and $\boldsymbol{\pi}$ as the vector containing the three Goldstone bosons. The fields transform under the custodial symmetry group~\eqref{eq:custSU2} as
\begin{align}
    \hat{h}(x) &\rightarrow \hat{h}(x) \,,
    &
    &\text{and}
    &
    U(x) &\rightarrow V_L \, U(x) \, V_R^\dagger \,.
\end{align}
Following the discussion in \cite{Longhitano:1980tm,Longhitano:1980iz,Appelquist:1980vg,Appelquist:1980ae,Feruglio:1992wf,Grinstein:2007iv,Stoffer:EFT} and substituting Eq.~\eqref{eq:HEFT-non-linear-field-redefinition} into \eqref{eq:L-linear-sigma-model}, we can express the scalar part of the SM Lagrangian, including also the interactions with the weak gauge bosons,  as
\begin{align}
\begin{split}
    \mathcal{L}_{p^2}^\mathrm{scalar} &= \frac{1}{2} \left(\partial_\mu\hat{h}\right)\left(\partial^\mu\hat{h}\right) - \frac{1}{2} m_h^2 \hat{h}^2 
    \\
    &+ \frac{v^2}{4} \mathcal{F}\left(\frac{\hat{h}}{v}\right) \mathrm{tr}\left[ (D_\mu U)^\dagger (D^\mu U) \right] - V\left( \frac{\hat{h}}{v} \right)
\end{split}
\label{eq:non-linear-SM-Lagrangian}
\end{align}
with the mass~$m_h^2=2m^2=\lambda v^2$ of the physical Higgs boson~$\hat{h}$, and where we have defined
\begin{align}
    \mathcal{F}\left(\frac{\hat{h}}{v}\right) &= \left( 1 + \frac{\hat{h}}{v} \right)^{\!\!2} \,,
    \label{eq:F-heft}
    \\
    V\left( \frac{\hat{h}}{v} \right) &= v^4 \left[ \frac{m_h^2}{2v^2} \left( \frac{\hat{h}}{v} \right)^{\!\!3} + \frac{m_h^2}{8v^2} \left( \frac{\hat{h}}{v} \right)^{\!\!4} \right].
    \label{eq:V-heft}
\end{align}
We now have a non-linear formulation of the custodial symmetry breaking in close analogy to chiral perturbation theory, the only difference being the presence of the additional singlet state~$\hat{h}$. To write down Eq.~\eqref{eq:non-linear-SM-Lagrangian} we promoted the global custodial symmetry to a local one by introducing two ${2 \times 2}$~matrix spurion fields~$\hat{W}_\mu$ and~$\hat{B}_\mu$ as the gauge bosons of the chiral $\mathrm{SU}(2)_{L}$ and $\mathrm{SU}(2)_{R}$ groups, respectively. These fields must transform in the adjoint representation of the chiral groups\footnote{The transformation rules read $\smash{\hat{W}_\mu \to V_L \hat{W}_\mu V_L^\dagger + i V_L (\partial_\mu V_L)^\dagger}$, and $\smash{\hat{B}_\mu \to V_R \hat{B}_\mu V_R^\dagger + i V_R (\partial_\mu V_R)^\dagger}$, respectively.} to make~\eqref{eq:non-linear-SM-Lagrangian} formally invariant. The covariant derivative then reads
\begin{align}
    D_\mu U &= \partial_\mu U - i \hat{W}_\mu U + i U \hat{B}_\mu
\end{align}
and we obtain the SM case by fixing the spurions to 
\begin{align}
    \hat{W}_\mu &\to g_2 \frac{\tau^I}{2} W_\mu^I \,,
    &
    \hat{B}_\mu &\to g_1 \frac{\tau^3}{2} B_\mu \,,
\end{align}
where $W_\mu$ and $B_\mu$ are the weak gauge bosons of the~SM.
This breaks the chiral symmetry down to its gauged SM subgroup
\begin{align}
    \mathrm{SU}(2)_L \times \mathrm{SU}(2)_R \rightarrow \mathrm{SU}(2)_L \times \mathrm{U}(1)_Y \,.
\end{align}

\subsection{The Higgs effective field theory}
\label{sect:HEFT1}
After adding the fermion and gauge sectors to the chiral Lagrangian \cite{Longhitano:1980iz,Feruglio:1992wf} as given in Eqs.~\eqref{eq:non-linear-SM-Lagrangian}--\eqref{eq:V-heft} it is equivalent to the usual SM~Lagrangian written in terms of the Higgs doublet~$H$ shown in Eq.~\eqref{eq:SM_Lagrangian}. The two Lagrangians are related by the non-linear field redefinition~\eqref{eq:HEFT-non-linear-field-redefinition} which leaves the physical observables invariant.

However, going beyond the SM, i.e., considering the EFT extension of~\eqref{eq:non-linear-SM-Lagrangian}, the equivalence between the linear and non-linear realization can be broken as we will discuss shortly. Amending the Lagrangian~\eqref{eq:non-linear-SM-Lagrangian} to an EFT, we can express the functions~$\mathcal{F}$ and~$V$ as generic power series in their argument~$\hat{h}/v$
\begin{align}
    \mathcal{F}\left(\frac{\hat{h}}{v}\right) &= 1 + \sum_{n=1}^\infty a_n \left(\frac{\hat{h}}{v}\right)^{\!\!n} \,,
    \\
    V\left(\frac{\hat{h}}{v}\right) &= v^4 \sum_{n=3}^\infty b_n \left(\frac{\hat{h}}{v}\right)^{\!\!n} \,,
\end{align}
where we reproduce the SM by choosing $a_1=2$, $a_2=1$, $b_3=\lambda/2$, and $b_4=\lambda/8$ with all other coefficients vanishing. Given $\hat{h}$ is a singlet, in this generic EFT approach 
the coefficients~$a_n$ and~$b_n$ are free parameters, not fixed by the symmetries of the theory, and have to be determined experimentally. Since their determination requires the measurement of multi-Higgs processes, the current experimental constraints on these parameters are rather imprecise. 
Of course, allowing for generic coefficients $a_n$ and~$b_n$ is incompatible with the field redefinition~\eqref{eq:HEFT-non-linear-field-redefinition} that we used to relate the chiral SM Lagrangian~\eqref{eq:non-linear-SM-Lagrangian} to the linear one. In general, it is not possible to find a field redefinition that brings the generic non-linear EFT case
back to the linear one, while the reverse processes is always possible.

The above statement  implies that  the EFT constructed from the chiral Lagrangian~\eqref{eq:non-linear-SM-Lagrangian} is more general than the SMEFT which is built under the assumption of a linear realization of electroweak symmetry 
breaking mechanism. 
This effective theory is known as the Higgs effective field theory, or HEFT \cite{Feruglio:1992wf,Alonso:2012px,Buchalla:2013rka,Pich:2016lew}. The HEFT contains the SMEFT as the special case 
where a non-linear field redefinition can be found to map the scalar components ($\smash{\hat{h}}$~and~$\pi_{1,2,3}$) into a single $\mathrm{SU}(2)_{L}$~doublet~($H$). For more details on distinguishing SMEFT and HEFT see the discussion in Sec.~\ref{sec:geometric-formulation} and \cite{Falkowski:2019tft,Cohen:2020xca}.
When adding the Yukawa interactions to the EFT, these can of course also be multiplied by an arbitrary power of~$\hat{h}/v$, thus leading to an expansion of these similar to Eqs.~(\ref{eq:F-heft})--(\ref{eq:V-heft}). The NLO~Lagrangian can be constructed in close analogy to chiral perturbation theory, with the only difference being the presence of the additional singlet~$\hat{h}$ that allows every operator to be multiplied by a generic function of~$\hat{h}/v$.

The HEFT is therefore a combination of the fermionic and gauge sectors of the SMEFT, with their power counting in canonical mass dimension, and the scalar sector of chiral perturbation theory with a chiral power counting. The HEFT is based on the same gauge symmetry as the SMEFT and contains the same degrees of freedom apart from the Higgs doublet~$H$ which is replaced by the scalar singlet~$\hat{h}$ and the Goldstone boson matrix~$U$. This decorrelates the interactions of (multiple)~$\hat{h}$ and the Goldstone bosons, therefore allowing to cover more general BSM scenarios. 
A~complication in HEFT is that the matrix field~$U$ as well as $\hat{h}/v$ are adimensional: $[U]=[\hat{h}/v]=1$. Therefore, the EFT series cannot be truncated only according to the canonical mass dimension as in the SMEFT case, but we also need to consider a chiral power counting, i.e., an expansion in the number of derivatives. This is possible since $U^\dagger U=\mathds{1}$, and thus the field~$U$ must always be derivatively coupled, as generically expected for Goldstone bosons.  
Due to the mixture of the different countings, there is no unique way to define a consistent power counting for the HEFT, but a commonly used counting \cite{Gavela:2016bzc} is NDA (c.f.~Sec.~\ref{sec:powercounting}). 
This counting has been employed for the construction of the NLO HEFT basis in \cite{Brivio:2016fzo}, see also \cite{Sun:2022ssa,Sun:2022snw}. A discussion of the counting of the HEFT operators using the Hilbert series technique can be found in \cite{Sun:2022aag,Graf:2022rco}.

\subsection{Geometric interpretation for the scalar sector}
\label{sec:geometric-formulation}

An interesting approach to better appreciate the
difference between SMEFT and HEFT is the 
geometric interpretation of the scalar 
sectors of these EFTs, which we review in this section.
This technique,  initially developed 
in the context of non-linear sigma models,
has been extensively applied to analyze the 
scalar sectors of the SMEFT and HEFT \cite{Alonso:2015fsp,Alonso:2016btr,Alonso:2016oah,Helset:2018fgq,Corbett:2019cwl,Helset:2020yio}. 

\begin{figure}
\centering
\includegraphics[width=0.98\linewidth]{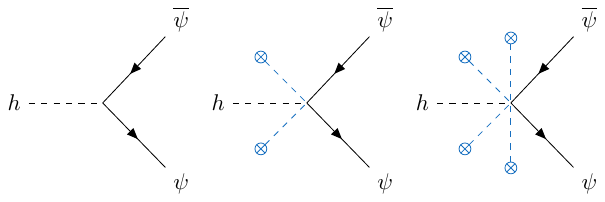}
\caption{Feynman diagrams contributing to the (effective) Yukawa interactions in the SMEFT after taking the \vev $v_T$ symbolized by the crossed dot {\color{\myBlue}$\otimes$} in the diagrams by replacing $\langle H^\dagger H \rangle = v^2_T / 2$.
\label{fig:geoSMEFT}
}
\end{figure}

The starting point of the geometric formulation is the observation that, after spontaneous symmetry breaking, a tower of higher-dimensional operators collapses into a single composite operator form \cite{Helset:2020yio}.
Consider for instance the SMEFT, with the Higgs vev defined by $v_T \equiv \sqrt{2 \langle H^\dagger H \rangle}$.\footnote{Note that, in general, we have $v \neq v_T$ due to the presence of higher-dimensional operators in the scalar potential of the SMEFT as we will discuss in Sec.~\ref{sect:LEFT}.} An example for the breakdown of a tower of higher-dimensional operators can be observed in the effective Yukawa interactions.
The operators of interest are of the form $\brackets{H^\dagger H}^n \brackets{\overline{\psi}_L H \psi_R}$ for some $n \in \mathbb{N}$. 
When the Higgs field acquires a vev, $\langle H^\dagger H\rangle \to v_T^2/2$, these higher-dimensional operators collapse into a number multiplying the (effective) SM~Yukawa operator, as illustrated in Fig.~\ref{fig:geoSMEFT}. A~similar breakdown of higher-dimensional operators is perceptible for other interactions as well. Thus, the interactions of all the particles, including the physical Higgs~$h$ itself, can be thought of as taking place in a Higgs-medium \cite{Helset:2020yio}. This medium can be described by a scalar field manifold~$\mathcal{M}$ with coordinates defined by the scalar fields. The $S$-matrix of the theory is invariant under scalar field redefinitions which in this case are equivalent to coordinate transformations on~$\mathcal{M}$. These coordinate redefinitions also leave invariant the geometry of the scalar manifold~$\mathcal{M}$. Therefore, the $S$-matrix, and thus all physical observables, only depend on the geometric properties\footnote{We are considering here the geometry of the scalar field space~$\mathcal{M}$, contrary to general relativity which considers the geometry of spacetime. However, many of the concepts are similar.} of~$\mathcal{M}$, but not on the choice of coordinates \cite{Alonso:2016oah}.

The geometric formulation leads to a factorization of the EFT power counting expansions. In the SMEFT there are two distinct expansions that are often not properly distinguished. The first expansion~$(i)$ is in the ratio of the electroweak scale~$v_T$ to the new-physics scale~$\Lambda$, whereas the second expansion~$(ii)$ is in the ratio of the kinematical scale~$p$ for the process of interest relative to the new-physics scale
\begin{align*}
	(i): & \quad \frac{v_T}{\Lambda} \, , &
	(ii): & \quad \frac{p^2}{\Lambda^2} \, ,
\end{align*}
where $p^2$ is some kinematic Lorentz invariant. In the geometric formulation, expansion $(i)$~is largely factorized out, as it can be linked to the curvature of the scalar manifold, whereas $(ii)$~is determined by the derivative expansion \cite{Alonso:2016oah}. This factorization of the power counting allows to define the SM Lagrangian parameters to all orders in the SMEFT power counting as shown in \cite{Helset:2020yio}. 

The geometric interpretation of the SM is particularly
simple. As we have already seen in Sec.~\ref{sect:custodial}, the scalar sector of the SM 
is invariant under a global $\mathrm{O}(4)$~symmetry,
and the minimum of the scalar potential~$V(\phi)$ 
defines a three-sphere~$S^3$ with 
radius~$v= \sqrt{\langle\boldsymbol{\phi}\cdot\boldsymbol{\phi}\rangle}$.
Conventionally, we align the vev of $\boldsymbol{\phi}$ to its fourth component, i.e., $\langle \boldsymbol{\phi} \rangle = \brackets{0,0,0,v}^\intercal$. This triggers the breaking of the custodial symmetry group~$\mathcal{G}=\mathrm{O}(4)$ down to the subgroup~$\mathcal{H}=\mathrm{O}(3)$, acting only on the first three components of~$\boldsymbol{\phi}$.
Expressing~$\phi$ in terms of the radial component~$h$ and the 
three Goldstone bosons as in~\eqref{eq:geoSMEFT_Cartesian_coordinates},
the scalar Lagrangian~\eqref{eq:geoSMEFT_SM_Lagrangian} assumes the form
\begin{align}
\begin{split}
	\L_\varphi 
	\!=\! \frac{1}{2} \!\brackets{D_\mu \boldsymbol{\varphi}}\! \cdot\! \brackets{D_\mu \boldsymbol{\varphi}} 
	+\! \frac{1}{2}\! \brackets{\partial_\mu h}^2
    -\! \frac{\lambda}{8}\! \brackets{h^2 + 2hv + \boldsymbol{\varphi}\!\cdot\!\boldsymbol{\varphi}}^2 \!.
\end{split}	
\label{eq:geoSMEFT_SM_Lagrangian_2}
\end{align}
The real scalar fields~$\varphi^a$ with $a \in \{1,2,3\}$ transform in the vector representation of~$\mathcal{H}$, whereas the physical Higgs~$h$ transforms as a singlet under~$\mathcal{H}$. Together these four real scalar fields constitute coordinates in the scalar field space of the~SM.

\subsubsection{Geometric formulation of the SMEFT}
\label{sec:geometric_formulation_SMEFT}
\begin{figure}
\centering
\includegraphics[width=0.75\linewidth]{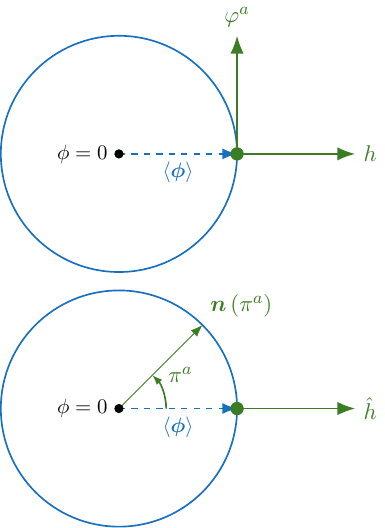}
\caption{Illustration of the SM flat scalar field manifold $\mathcal{M}_\mathrm{SM}=\mathbb{R}^4$ adapted from figure~1 in \cite{Alonso:2016oah}. The vacuum manifold~$S^3=\mathcal{G}/\mathcal{H}$ is represented by the blue circle with radius~$\langle\phi\rangle=v$, the green dot on this circle represents the physical vacuum, and the solid green axes symbolize the scalar field coordinate system.
In the top figure Cartesian coordinates $\{\varphi^1,\, \varphi^2,\, \varphi^3,\, h\}$ centered at the physical vacuum are chosen, whereas in the figure on the bottom polar coordinates $\{\pi^1,\, \pi^2,\, \pi^3,\, \hat{h}\}$ are used, where $\hat{h}$~is the radial coordinate and the three~$\pi^a$ form a vector~$\boldsymbol{n}(\pi^a) \in S^3$.
\label{fig:geoSMEFT_SM_manifold}}
\end{figure}

The generic kinetic term for a scalar field~$\Phi^i$ in a general scalar field space is
\begin{align}
	\L_\mathrm{kin} 
	= \frac{1}{2} g_{ij}(\Phi) \brackets{D_\mu \Phi}^i \brackets{D_\mu \Phi}^j \, ,
	\label{eq:geoSMEFT_SM_kinetic}
\end{align}
where $g_{ij}(\Phi)$ is the metric of the scalar field space. Comparing Eq.~\eqref{eq:geoSMEFT_SM_Lagrangian} with~\eqref{eq:geoSMEFT_SM_kinetic}, by choosing $\Phi^i=\phi^i$ and promoting partial to covariant derivatives, we find ${g_{ij}(\boldsymbol{\phi})=\delta_{ij}}$. Therefore, the scalar field manifold of the Standard Model is $\mathcal{M}_\mathrm{SM}=\mathbb{R}^4$, i.e., the scalar field space is flat four-dimensional Euclidean space and the fields~$\phi^i$ (with $i\in\{1,2,3,4\}$) or equivalently $\varphi^a$ and~$h$ (with $a\in\{1,2,3\}$) define a Cartesian coordinates system on~$\mathcal{M}_\mathrm{SM}$, as shown at the top of Fig.~\ref{fig:geoSMEFT_SM_manifold}. The black dot in the center represents $\boldsymbol{\phi}=\boldsymbol{0}$ or equivalently~$H=0$. The blue circle symbolizes the Goldstone boson vacuum manifold~$S^3$ given by the coset space~$\mathcal{G}/\mathcal{H}$. The dashed blue arrow points to the physical vacuum denoted by the green dot, where the Cartesian coordinate system is centered. The direction~$h$ is orthogonal to~$S^3$ and $\varphi^a$ are the remaining three orthogonal directions to~$h$.

We have seen above that the SM corresponds to the simple case of a flat four-dimensional scalar manifold. In the following we generalize our previous considerations by extending the SM by higher-dimensional operators and analyzing the geometric properties of the SMEFT. 
The scalar kinetic term in the SMEFT consists of all terms containing only Higgs doublets and exactly two derivatives acting on them. At dimension six in the Warsaw basis \cite{Grzadkowski:2010es} it reads
\begin{align}
\begin{split}
\L_\mathrm{SMEFT}^{H,\mathrm{kin}} 
&= \brackets{D_\mu H}^\dagger \brackets{D^\mu H} \\
&+ \frac{C_{HD}}{\Lambda^2} \brackets{H^\dagger D_\mu H}^\ast \brackets{H^\dagger D^\mu H} \\
&+ \frac{C_{H\Box}}{\Lambda^2} \brackets{H^\dagger H} \Box \brackets{H^\dagger H}
+ \ord{\Lambda^{-4}} \, .
\label{eq:geoSMEFT_L_kin1}
\end{split}
\end{align}
Using only $\SU{2}_L \times \mathrm{U}(1)_Y$ gauge invariance and the $\mathrm{SU}(2)$~Fierz identity in Eq.~\eqref{eq:SUN-Fierz} it is straight forward to show that at higher powers only two different and independent operator structures can appear at each mass dimension\footnote{See also \cite{Helset:2020yio} and references therein. Notice, however, that we use a different operator definition, and thus a different basis. The two bases differ only by a Fierz redefinition and thus reproduce the same metric.}
\begin{align}
    Q_{H,\mathrm{kin}}^{(8+2n)} &= \left( H^\dagger H \right)^{n+2} \brackets{D_\mu H}^\dagger \brackets{D^\mu H} \,,
    \label{eq:K-kin-1}
    \\
    Q_{H\!D}^{(8+2n)} &= \left( H^\dagger H \right)^{n+1} \brackets{H^\dagger D_\mu H}^\ast \brackets{H^\dagger D^\mu H} \,.
    \label{eq:K-kin-2}
\end{align}

Using Eq.~\eqref{eq:geoSMEFT_H_to_phi} we can express~$\L_\mathrm{SMEFT}^{H,\mathrm{kin}}$ (and these operators) in terms of the real scalar coordinates~$\phi^i$. The general expression for the kinetic term of the SMEFT in terms of the coordinates~$\phi^i$ is given by\footnote{Similar expressions have been given in \cite{Alonso:2016oah} for the SMEFT in the custodial limit. Notice, however, that the operator~$Q_{H\!D}$ breaks custodial symmetry. Therefore, the formulae presented here are more general.}
\begin{align}
\begin{split}
\L_\mathrm{SMEFT}^\mathrm{kin} 
= &\frac{1}{2} \left[A\brackets{\!\frac{\boldsymbol{\phi}\cdot\boldsymbol{\phi}}{\Lambda^2}\!} \brackets{D_\mu \boldsymbol{\phi}}\cdot\brackets{D^\mu \boldsymbol{\phi}} \right.
\\
&\left. +B\!\brackets{\!\frac{\boldsymbol{\phi}\cdot\boldsymbol{\phi}}{\Lambda^2}\!}\! \frac{(D_\mu \phi)^i \, \mathfrak{f}_{ij}(\phi) \, (D_\mu \phi)^j }{\Lambda^2} \right] ,
\end{split}
\label{eq:geoSMEFT_L_kin2}
\end{align}
where we have defined
\begin{align}
    \mathfrak{f}^{ij}(\phi) &= \begin{pmatrix}
	a & 0 & b & c \\
	0 & a & c & -b \\
	b & c & d & 0 \\
	c & -b & 0 & d
\end{pmatrix}
\,,
&
\begin{pmatrix}
    a\\b\\c\\d
\end{pmatrix}
&=
\begin{pmatrix}
    (\phi^1)^2+(\phi^2)^2
    \\
    \phi^1\phi^3 - \phi^2\phi^4
    \\
    \phi^1\phi^4 + \phi^2\phi^3
    \\
    (\phi^3)^2+(\phi^4)^2
\end{pmatrix}
.
\end{align}
This expression is, of course, only valid for a specific choice for the operator basis. Here, $A$~and~$B$ are defined through a power series expansion in their argument $z \equiv \brackets{\boldsymbol{\phi}\cdot\boldsymbol{\phi}}/\Lambda^2$, which simplifies to the usual EFT expansion in~$v_T^2 / \Lambda^2$ after spontaneous symmetry breaking. 
Since the above equation has to reduce to the SM case in the limit $\Lambda \to \infty$ we must have $A(0)=1$ and~$B(0)=0$. Comparing again to the general scalar kinetic term on a curved manifold in Eq.~\eqref{eq:geoSMEFT_SM_kinetic}, we find the SMEFT scalar field space metric
\begin{align}
g_{ij} (\phi) 
&= A\!\brackets{\!\frac{\boldsymbol{\phi}\cdot\boldsymbol{\phi}}{\Lambda^2}\!} \delta_{ij}
+ B\!\brackets{\!\frac{\boldsymbol{\phi}\cdot\boldsymbol{\phi}}{\Lambda^2}\!} \frac{\mathfrak{f}_{ij}(\phi)}{\Lambda^2} \, .
\label{eq:geoSMEFT_metric}
\end{align}
\newline
This metric describes, in general, a curved manifold~$\mathcal{M}_\mathrm{SMEFT}$ and only for~$B=0$ the manifold is flat. In the limit~$\Lambda \to \infty$ we find that the SMEFT metric reduces to the SM metric in Cartesian coordinates
\begin{align}
g_{ij}^\mathrm{SMEFT} (\phi) \quad \xrightarrow{~\Lambda\to\infty~} \quad \delta_{ij} = g_{ij}^\mathrm{SM} \, .
\end{align}

Therefore, the curvature of~$\mathcal{M}_\mathrm{SMEFT}$ is determined entirely by the EFT expansion parameter~$\smash{{v_T^2}/{\Lambda^2}}$. Furthermore, we can deduce all kinds of geometric quantities such as Christoffel symbols, Riemann curvature tensors,~etc. 

Taking Eq.~\eqref{eq:geoSMEFT_L_kin1} we find the scalar metric of the SMEFT up to dimension six \cite{Helset:2018fgq}
\begin{align}
g_{ij}(\phi) &= \delta_{ij} + \frac{C_{HD}}{2\Lambda^2} \mathfrak{f}_{ij}(\phi) -2 \frac{C_{H\Box}}{\Lambda^2} \phi_i \phi_j \,.
\label{eq:SMEFT-metric-d6}
\end{align}
Notice that the last term in the above expression does not match the general form of the metric in Eq.~\eqref{eq:geoSMEFT_metric}. This is because the operator~$Q_{H\Box}$ of the Warsaw basis does not agree with the definitions in Eqs.~\eqref{eq:K-kin-1}--\eqref{eq:K-kin-2}. It could, of course, be rewritten in that form by using integration by parts identities.
Apart from $\smash{Q_{H,\mathrm{kin}}^{(6)}}$, this would introduce further operators that have to be removed using field redefinitions.\footnote{The replacement reads $Q_{H\Box}=2Q_{H,\mathrm{kin}}^{(6)}+\ldots$, where the ellipsis denote terms that do not contribute to the metric after applying the appropriate redefinition of the Higgs field~$H$.} The latter, however, change the scalar field space metric. Thus, the last term in Eq.~\eqref{eq:SMEFT-metric-d6} could be removed in favor of a term proportional to~$\delta_{ij}$, but for consistency we decided to stick with the Warsaw basis at dimension six.
In this example we see that the explicit form of the metric is basis dependent. However, a geometric formulation of the SMEFT exists in every basis \cite{Helset:2020yio}.

We can now use Eqs.~\eqref{eq:K-kin-1}--\eqref{eq:K-kin-2} and our previous results to define the scalar field metric to all orders in the EFT power counting
\begin{align}
\begin{split}
g_{ij} 
&= \squarebrackets{1 + \sum_{n=0}^{\infty} \brackets{\frac{\boldsymbol{\phi}\cdot\boldsymbol{\phi}}{2}}^{\!n+2} \frac{C_{H,\mathrm{kin}}^{(8+2n)}}{\Lambda^{4+2n}}} \delta_{ij}  
\\
&\ +\frac{1}{2} \squarebrackets{C_{H\!D}^{(6)} + \sum_{n=0}^\infty \! \brackets{\frac{\boldsymbol{\phi}\cdot\boldsymbol{\phi}}{2}}^{\!n+1} \frac{C_{H\!D}^{(8+2n)}}{\Lambda^{4+2n}}} \! \mathfrak{f}_{ij}(\phi) 
\\
&\ -2 \frac{C_{H\Box}}{\Lambda^2} \phi_i \phi_j\, .
\end{split}
\end{align} 
Of course, this entails a choice of basis, nevertheless it is remarkable that we are able to define this geometric quantity to all orders in the EFT power counting.

The ideas discussed so far in this section apply to the Higgs two-point function leading to the scalar field space metric. Following \cite{Helset:2020yio}, we can generalize the concepts to higher $n$-point functions and other types of field connections by factorizing the operators in the SMEFT Lagrangian
\begin{align}
\L_\mathrm{SMEFT} &= \sum_{n} f_n\brackets{\mu,\alpha,\ldots} \, G_n\brackets{I,A,\ldots} \, .
\end{align}
The factors~$f_n$ are composite operator forms containing all non-scalar fields and all dependence on spacetime indices, i.e. Lorentz~$(\mu,\ldots)$, and spinor~$(\alpha,\ldots)$ indices. The~$f_n$ can only depend on the scalar field coordinates through derivatives acting on the scalars, e.g.~$\brackets{D_\mu H}$.
The factors~$G_n$, on the other hand, depend on the non-spacetime group indices~$(I,A,\ldots)$ and contain only scalar field coordinates and symmetry generators acting on them, i.e., expressions built only out of~$H^{(\dagger)}$ and~$\tau^I$. It is evident that after electroweak symmetry breaking the~$G_n$ collapse to a number and an appropriate power of Higgs~$h$ emissions, largely factoring out the expansion in~$v_T^2 / \Lambda^2$ from the remaining composite operator form~$f_n$, whereas the latter~$(f_n)$ contain the derivative expansion in~$p^2 / \Lambda^2$ and only retain a minimal dependence on the scalar coordinates and~$v_T$ mixing the two expansions \cite{Helset:2020yio}.

This allows us to define the scalar field metric
\begin{align}
g_{ij} \brackets{\phi} &= \left. \frac{g^{\mu\nu}}{D} \frac{\delta^2 \L_\mathrm{SMEFT}}{\delta\!\brackets{D^\mu \phi}^i \, \delta\!\brackets{D^\nu \phi}^j} \right|_{f_n\to 0} \, .
\end{align}
Similarly, we can now define all sorts of field-space connection, e.g., the Yukawa-type connection \cite{Helset:2020yio} we already encountered
\begin{align}
[\mathsf{Y}_{\psi}]_{pr} \brackets{\phi_i} &= \left. \frac{\delta \L_\mathrm{SMEFT}}{\delta ( \overline{\psi}{}^{L,i}_{p} \psi^R_{r} ) } \right|_{f_n\to 0} \, .
\end{align}
For this case we find $\smash{f_n} = \smash{\overline{\psi}{}^{L,i}_{p} \psi^R_{r}}$ containing the fermion bilinear and the factor $\smash{G_n}\sim \smash{\sum_k (H^\dagger H)^k H_i}$. 
Comparing to Fig.~\ref{fig:geoSMEFT} we see that $f_n$~corresponds to the fermion current, whereas the~$G_n$ corresponds to the emission of~$h$ and the vev~(marked by {\color{\myBlue}$\otimes$}). The operators contributing to~$G_n$ at all orders in this case are \cite{Helset:2020yio}
\begin{align}
	[Q_{\psi H}^{(6+2n)}]_{pr} &= \left(H^\dagger H\right)^{n+1} \left(\overline{\psi}{}^{L,i}_{p} \psi^R_{r} H_i\right)
\end{align}
leading to the all-order Yukawa connection
\begin{align}
\begin{split}
	[\mathsf{Y}_\psi]_{pr} \brackets{\phi} = &-H\brackets{\phi} [Y_\psi]_{pr} 
    \\
    &+ H\brackets{\phi} \sum_{n=0}^{\infty} \frac{[C_{\psi H}^{(6+2n)}]_{pr}}{\Lambda^{2+2n}} \brackets{\frac{\boldsymbol{\phi}\cdot\boldsymbol{\phi}}{2}}^{n+1}.
\end{split}
\end{align}
For more details and the definition of other field-space connections, as well as for more all order results, see \cite{Helset:2020yio}.
The key advantage of this formulation is the reduction of the number of relevant structures, especially when going beyond dimension six, and obtaining all-order results (or better: results independent of the operator power counting) 
for a series of relevant quantities, such as the physical fermion masses.

\subsubsection{Geometric formulation of the HEFT}
So far we have used Cartesian coordinates to describe the scalar field manifold~$\mathcal{M}$. We can equally well choose polar coordinates on~$\mathcal{M}$, since any measurable quantity does not depend on the choice of the coordinate system.
Following \cite{Alonso:2016oah}, we can use polar coordinates to write the real scalar fields as
\begin{align}
\boldsymbol{\phi} = \brackets{v+\hat{h}} \boldsymbol{n}(\pi) \quad \text{where} \quad \boldsymbol{n}(\pi)\in S^3 \sim \mathcal{G}\big/\mathcal{H}
\label{eq:geoSMEFT_polar_coordinates_parametriztion}
\end{align}
with the radial coordinate~$\hat{h}$ and the three angular coordinates~${\pi^a}/{v}$ associated to the Goldstone bosons of the broken generators. The three angular coordinates~$\pi^a$ form a four-dimensional unit vector~$\boldsymbol{n}(\pi^a) \in S^3$. The polar coordinate system is shown on the bottom of Fig.~\ref{fig:geoSMEFT_SM_manifold}. In the polar coordinates parametrization~\eqref{eq:geoSMEFT_polar_coordinates_parametriztion} there is no obvious relation among the physical Higgs field~$\hat{h}$ and the Goldstone bosons in~$\boldsymbol{n}$, contrary to the case of Cartesian coordinates in Eq.~\eqref{eq:geoSMEFT_Cartesian_coordinates} where such a relation is implicit as $h$~and~$\varphi^a$ transform together in the vector representation of~$\mathcal{G}=\mathrm{O}(4)$ as $\boldsymbol{\phi}\to O\boldsymbol{\phi}$ with~$O\in\mathcal{G}$.

The transformation properties of the coordinates under the chiral symmetry group~$\mathcal{G}$ are
\begin{align}
\hat{h} \xrightarrow{~\mathcal{G}~} \hat{h} \, , && \boldsymbol{n} \xrightarrow{~\mathcal{G}~} O\,\boldsymbol{n}\quad \text{with } O \in \mathcal{G} \, .
\label{eq:geoSMEFT_polar_transformation_laws}
\end{align}
The field~$\hat{h}$ is a singlet, whereas~$\boldsymbol{n}$ transforms linearly under~$\mathcal{G}$. However, due to the constraint $\boldsymbol{n}\cdot\boldsymbol{n}=1$, this four-component vector has only the three independent components~$\pi^a$ with~$a\in\{1,2,3\}$. Therefore, the~$\pi^a$ do not transform linearly under~$\mathcal{G}$, which is why this choice of coordinates is called the non-linear representation.\footnote{Contrary to Eq.~\eqref{eq:HEFT-non-linear-field-redefinition}, we use the vector notation with the field~$\boldsymbol{n}$ here rather than the matrix notation with the field~$U$. Nevertheless, the two formulation are completely equivalent.} 
Possible parametrizations are the square root parametrization $\boldsymbol{n} (\pi) = \smash{\left(\pi^1, \pi^2, \pi^3, \sqrt{v^2 - \boldsymbol{\pi}\cdot\boldsymbol{\pi}} \right)^\intercal \!\!\big/ v}$, and the exponential representation
\begin{align}
\begin{split}
    \boldsymbol{n} (\pi) 
    \!=\! \exp\!\! \left(\!\! \frac{1}{v}\!\! \left[\begin{matrix}
        0 & \!0 & \!0 & \!\pi^1
        \\
        0 & \!0 & \!0 & \!\pi^2
        \\
        0 & \!0 & \!0 & \!\pi^3
        \\
        -\pi^1 & \!-\pi^2 & \!-\pi^3 & \!0
    \end{matrix} \right]\!\right)\!\!\! 
    \begin{pmatrix}
        0 \\ 0 \\ 0 \\ 1
    \end{pmatrix} 
    \!\!=\!\!
    \begin{pmatrix}
        \!\sin\! \left(\!\frac{|\boldsymbol{\pi}|}{v}\!\right) \!\frac{\pi^1}{|\boldsymbol{\pi}|}\!\!\!
        \\
        \!\sin\! \left(\!\frac{|\boldsymbol{\pi}|}{v}\!\right) \!\frac{\pi^2}{|\boldsymbol{\pi}|}\!\!\!
        \\
        \!\sin\! \left(\!\frac{|\boldsymbol{\pi}|}{v}\!\right) \!\frac{\pi^3}{|\boldsymbol{\pi}|}\!\!\!
        \\
        \!\cos\! \left(\!\frac{|\boldsymbol{\pi}|}{v}\!\right)
    \end{pmatrix}\!,
\end{split}
\end{align}
where $|\boldsymbol{\pi}|=\sqrt{\boldsymbol{\pi}\cdot\boldsymbol{\pi}}$.
The latter corresponds to the standard coordinates of CCWZ \cite{Alonso:2016oah}. However, we will not pick any explicit parametrization here.

Using Eq.~\eqref{eq:geoSMEFT_polar_coordinates_parametriztion} to express the scalar part of the SM Lagrangian~\eqref{eq:geoSMEFT_SM_Lagrangian} in polar coordinates yields
\begin{align}
\begin{split}
\L 
= \frac{1}{2} & \brackets{v+\hat{h}}^2 \!\brackets{D_\mu \boldsymbol{n}} \! \cdot \! \brackets{D^\mu \boldsymbol{n}}
+ \frac{1}{2} \brackets{\partial_\mu \hat{h}} \! \brackets{\partial^\mu \hat{h}}
\\
&- \frac{\lambda}{8} \brackets{\hat{h}^2 + 2v\hat{h}}^2 \, ,
\end{split}
\label{eq:geoSMEFT_L_SM_polar}
\end{align}
where the Goldstone bosons of~$\boldsymbol{n}$ are only derivatively coupled and the potential is independent of the angular coordinates~$\pi^a$, contrary to the case of Cartesian coordinates in Eq.~\eqref{eq:geoSMEFT_SM_Lagrangian_2}. Instead of changing the coordinate system on the scalar field manifold~$\mathcal{M}$, we could equally well do a field redefinition. Using Eqs.~\eqref{eq:geoSMEFT_Cartesian_coordinates} and~\eqref{eq:geoSMEFT_polar_coordinates_parametriztion} we find $(v+h)^2+\boldsymbol{\varphi}\cdot\boldsymbol{\varphi} = (v+\hat{h})^2$ which yields
\begin{align}
	\hat{h} &= h + \frac{\boldsymbol{\varphi}\cdot\boldsymbol{\varphi}}{2v} - \frac{h}{2} \frac{\boldsymbol{\varphi}\cdot\boldsymbol{\varphi}}{v^2} + \mathcal{O}(v^{-3}) \,,
	\label{eq:geoSMEFT_higgs_field_relation}
\end{align}
where $h$ and $\hat{h}$ are the Higgs fields in Cartesian and polar coordinates, respectively.

As we have discussed before, in the Cartesian coordinate system, the Higgs field in Eq.~\eqref{eq:geoSMEFT_H_to_phi} or the corresponding real scalar fields~$\boldsymbol{\phi}$ in Eq.~\eqref{eq:geoSMEFT_Cartesian_coordinates} transform linearly under~$\mathcal{G}$ or the electroweak symmetry group. On the contrary, in the polar coordinate system the scalar fields~$\boldsymbol{\pi}$ do not transform linearly. However, physical observables must be independent of the choice of coordinates and, therefore, the SM~Lagrangians in Eq.~\eqref{eq:geoSMEFT_L_SM_polar} and Eq.~\eqref{eq:geoSMEFT_SM_Lagrangian_2} are equivalent as they only differ by a coordinate redefinition.
The question whether the Higgs transforms linearly or non-linearly under the electroweak symmetry group is thus depending on the choice of coordinate system and is therefore unphysical \cite{Alonso:2016oah}. The appropriate question is whether it is always possible to pick a coordinate system in which the Higgs field transforms linearly. As we have seen above this is true for the SM, but as we will discuss below this is not possible in general for all EFT extensions of the~SM. 

In fact, it is only possible if and only if the scalar field manifold~$\mathcal{M}$ has a $\mathcal{G}$~invariant fixed point \cite{Alonso:2016oah}. In a neighborhood of this fixed point it is then possible to pick a coordinate system in which the Higgs field transforms linearly under~$\mathrm{O}(4)$. For the SM, this fixed point is the origin~$\boldsymbol{\phi}=\boldsymbol{0}$ (black central dot) as can be seen in both parts of Fig.~\ref{fig:geoSMEFT_SM_manifold}.

We have seen in the Sec.~\ref{sec:geometric_formulation_SMEFT} that the SMEFT is the extension of the SM with higher-dimensional operators using Cartesian coordinates on~$\mathcal{M}$. The Higgs field~$H$ transforms linearly under~$\mathcal{G}$ or the electroweak gauge group and the SMEFT has a $\mathcal{G}=\mathrm{O}(4)$~fixed point at the origin~$\boldsymbol{\phi}=\boldsymbol{0}$. 
On the contrary, the HEFT is the EFT extension of the SM using polar coordinates on the scalar field manifold~$\mathcal{M}_\mathrm{HEFT}$. The corresponding scalar part of the HEFT Lagrangian is a generalization of Eq.~\eqref{eq:geoSMEFT_L_SM_polar} and given by
\begin{align}
\begin{split}
\L_\mathrm{HEFT} = &\frac{1}{2} v^2 F(\hat{h})^2 \brackets{D_\mu \boldsymbol{n}}\cdot\brackets{D^\mu \boldsymbol{n}}
\\
&+ \frac{1}{2} \brackets{\partial_\mu \hat{h}}\brackets{\partial^\mu \hat{h}} - V(\hat{h}) \, ,
\end{split}
\label{eq:geoSMEFT_L_HEFT}
\end{align}
where $V(\hat{h})$ is the scalar potential that only depends on the radial coordinate~$\hat{h}$, and $F(\hat{h})$ is a generic dimensionless function that is defined by a power expansion in~$\hat{h}/v$ with~$F(0)=1$ such that the radius of the vacuum manifold is fixed by~$v$ \cite{Alonso:2016oah}. The $\mathcal{G}$~transformation rules for the fields~$\hat{h}$ and~$\boldsymbol{n}$ are the same as in Eq.~\eqref{eq:geoSMEFT_polar_transformation_laws}. We can write
\begin{align}
F(\hat{h}) &= 1 + c_1 \brackets{\frac{\hat{h}}{v}} + c_2 \brackets{\frac{\hat{h}}{v}}^2 + \cdots
\, 
\end{align}
and, as already stated in Sec,~\ref{sect:HEFT1}, we recover 
the SM case $F_\mathrm{SM} (\hat{h}) = (1+\hat{h}/v)$ for $c_1=1$ and~$c_{n\geq 2}=0$. 

Defining the HEFT scalar field space metric as in Eq.~\eqref{eq:geoSMEFT_SM_kinetic} by choosing $\Phi=\smash{\big(\pi^1,\pi^2,\pi^3,\hat{h}\big)^\intercal}$, we find \cite{Alonso:2015fsp}
\begin{align}
g_{ij}^\mathrm{HEFT} (\phi) &= \begin{pmatrix}
	F(\hat{h}) \, g_{ab}(\pi) & 0 \\
	0 & 1
\end{pmatrix}_{\!\!ij} \,,
\end{align}
where $g_{ab}$~is the $\mathcal{H}=\mathrm{O}(3)$~invariant metric on the coset space~$S^3=\mathcal{G}/\mathcal{H}$ for the angular coordinates~$\boldsymbol{\pi}$. 

\begin{figure}
\centering
\includegraphics[]{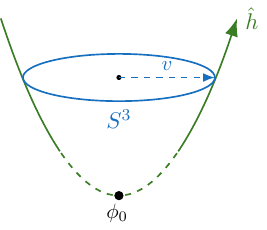}
\caption{The scalar field manifold~$\mathcal{M}_\mathrm{HEFT}$ of the HEFT is fibered with a three-sphere~$S^3$ of radius~$vF(\hat{h})$ for every~$\hat{h}$. An $\mathrm{O}(4)$~fixed point~$\phi_0$ does only exist if there is a value~$\smash{\hat{h}_\ast}$ such that the radius of the three-sphere vanishes~$\smash{F(\hat{h}_\ast)=0}$. If no fixed point~$\phi_0$ exists, the green dashed region does not exist and the manifold might either be smoothly connected without a fixed point or extend to infinity. The SMEFT corresponds to the theories with an $\mathrm{O}(4)$~fixed point at~$\phi_0=0$. For this type of theory it is possible to change from polar coordinates to Cartesian coordinates and vice versa in a neighborhood of the fixed point. Figure adapted from \cite{Alonso:2016oah}.
\label{fig:geoSMEFT_HEFT_manifold}
}
\end{figure}

The scalar field manifold~$\mathcal{M}_\mathrm{HEFT}$ for HEFT is shown in Fig.~\ref{fig:geoSMEFT_HEFT_manifold}. The manifold consists of~$\hat{h}$ (green arrow) and a sequence of three-spheres~$(S^3)$ of radius~$vF(\hat{h})$ fibered over any value of~$\hat{h}$ (the blue circle symbolizes the sphere for one particular value of~$\hat{h}$). From Eq.~\eqref{eq:geoSMEFT_polar_transformation_laws} we know that $\mathcal{G}$~acts on any point~$\boldsymbol{n}\in S^3$ by rotations on the surface of~$S^3$, i.e., rotations along the blue circle. Therefore, it is only possible to have a $\mathcal{G}=\mathrm{O}(4)$~invariant fixed point if the radius of the vacuum sphere is vanishing, meaning if there exists some~$\hat{h}_\ast$ for which we have~$F(\hat{h}_ \ast)=0$. In the SM this is the case for~$\hat{h}_\ast^\mathrm{SM}=-v$. However such a value~$\hat{h}_\ast$ does not exist in general as can be seen by the example~$F(\hat{h})=e^{\hat{h}/v} \smash{\cosh\big(1+\hat{h}/v\big)}$ which is non-vanishing for all~$\hat{h}$ \cite{Alonso:2016oah}. In this case the green dashed range of~$\hat{h}$ in Fig.~\ref{fig:geoSMEFT_HEFT_manifold} does not exist.

In summary, we have found that the most general EFT extension of the SM is the HEFT using polar coordinates on the scalar field manifold. In this framework the Goldstone bosons transform non-linearly under the electroweak symmetry group or the larger custodial symmetry group~$\mathcal{G}=\mathrm{O}(4)$. The subcategory of EFTs that have a $\mathcal{G}$~fixed point at the origin belong to the SMEFT class. For these theories it is possible to pick coordinates around this fixed point in which the Higgs field transforms linearly. Eventually, the SM is a subcategory of SMEFT with a flat scalar field manifold~$\mathcal{M}_\mathrm{SM}=\mathbb{R}^4$. We can therefore schematically write
\begin{align}
	\mathrm{SM} \subseteq \mathrm{SMEFT} \subseteq \mathrm{HEFT} \, .
  \label{eq:SMEFTHEFT}
\end{align}

Given the relation~\eqref{eq:SMEFTHEFT}, 
the last few years have seen an intense activity in 
identifying  concrete examples of UV models that cannot be well described by the 
SMEFT. 
As pointed out in \cite{Cohen:2020xca}, this can happen under two conditions:
i)~when (non-SM) particles which acquire mass via electroweak symmetry breaking are integrated out, introducing non-analytic dependence from the Higgs 
field in the corresponding EFT \cite{Falkowski:2019tft};
ii)~when additional sources of electroweak symmetry breaking are present besides a single scalar doublet.\footnote{An example was mentioned previously \cite{Manohar:2018aog}.} 
These two conditions signal that the $\mathrm{O}(4)$~fixed point 
in the scalar manifold is not the most convergent choice 
as origin for a Taylor expansion,
and that the cutoff of the effective theory is necessarily low, in fact below $4\pi v$~\cite{Banta:2021dek, Alonso:2021rac}. 
Depending on how these two conditions are realized (in terms of masses and couplings of the new states) the SMEFT represents a good or bad description of the underlying theory.
An instructive comparison of SMEFT and HEFT for a concrete UV model can be found in \cite{Buchalla:2016bse}.

On general grounds, a breakdown of the SMEFT description happens only if the non-SM states integrated out, which are connected to the mechanism of electroweak symmetry breaking (either as sources of the breaking or because they acquire mass via this breaking), are sufficiently close to the electroweak scale \cite{Banta:2021dek}.
It is fair to say that no indication of such states is present in current high-energy data.

At low energies, 
a pragmatic way to distinguish the two EFTs is by looking at transition amplitudes with identical electroweak and flavor structure, 
that differ only for the number of (massive) Higgs fields \cite{Isidori:2013cga,Isidori:2013cla,Brivio:2013pma}.
In the SMEFT, the linear realization implies a well-defined relation 
among all these processes at a given order in the EFT expansion. This relation can be broken at higher orders; however, the effect is expected to be small according to the power counting. On the contrary, in the HEFT the 
$F(\hat{h})$ function, and its analog for other electroweak structures, lead to a potential complete decoupling among processes with different number 
of Higgs fields. A particularly interesting study case is provided 
by non-universal corrections to the $Z\to f\bar f$ couplings \cite{Isidori:2013cga}.
Measurements at the $Z$~pole imply very small deviations from the SM, implying strong bounds on several operators in class~7 of Tab.~\ref{tab:Warsaw-basis} which control these effects. 
Within the SMEFT, this implies in turn tiny deviations from the SM
in the related processes $h\to Z f\bar f$ and $f \bar f \to Z h$.
A large deviation from the SM in the latter processes could occur naturally in the HEFT, while it would imply a breakdown of the SMEFT power counting.

\section{Low-Energy effective field theory}
\label{sect:LEFT}

\subsection{Introduction and overview}

The success of the SM rests to a large degree on tests in low-energy processes such as decays of Kaons, $D$-mesons, and even more on $B$~physics, because these processes can be calculated with rather high precision.\footnote{See \cite{Buchalla:1995vs} for an early review.} The tool for this is the so called low-energy effective field theory~(LEFT),\footnote{Sometimes this theory is also called the weak effective theory~(WET).} which is derived from the SM by integrating out the Higgs boson~($h$), the weak gauge bosons~($\mathcal{Z},\mathcal{W}$), as well as the top quark~($t_L, t_R$). Of course, this is a generalization of the original Fermi theory with the four-fermion interaction 
\begin{equation}
-\frac{4 G_F}{\sqrt{2}} \left( \overline{\psi}\gamma_\mu\psi)(\overline{\psi}\gamma^\mu\psi \right) \,,
\label{eq:4fermi}
\end{equation}
where the Fermi constant~$G_F$ is related to the vacuum expectation value~$v$ by $G_F = 1/(\sqrt{2} v^2)$. The method works so well because the relevant energies~$E$ are much smaller than $v$~or~$m_W$ (and of course than~$\Lambda$) and due to the asymptotic freedom of the strong interactions. Since the pioneering work in the mid 1970s, this theory has been developed to an astonishing degree of precision by including all kinds of strong and electromagnetic corrections. See, e.g., \cite{Buras:2020xsm} for a recent review. 

The LEFT is thus an $\mathrm{SU}(3)_c \times \mathrm{U}(1)_e$ invariant effective theory valid below the electroweak symmetry breaking scale containing five quark flavors $(u,d,s,c,b)$, three charged leptons $(e,\mu,\tau)$, three left-handed neutrinos~$\smash{(\nu_e,\nu_\mu,\nu_\tau)}$, the gluons, and the photon. The LEFT Lagrangian is the sum of the Lagrangians of QCD and QED of these particles and the mass terms of the fermions
\begin{align}
\mathcal{L}_\mathrm{SM}^\mathrm{broken} 
= &-\frac{1}{4}F_{\mu\nu}F^{\mu\nu} -\frac{1}{4}G_{\mu\nu}^A G^{A\,\mu\nu} -\theta_3\frac{g_3^2}{32\pi^2}G_{\mu\nu}^A \tilde{G}^{A\,\mu\nu}  
\nonumber\\
&+ \sum_{\psi=u,d,e,\nu_L} \sum_{X=L,R} \left( \overline{\psi}{}^X_{p} \, i\slashed{D} \,\psi^X_{p} \right)
\label{QFT}\\
&-  \left[ \sum_{\psi=u,d,e} [\mathcal{M}_\psi]_{pr} \, \left( \overline{\psi}{}^L_{p} \psi^R_{r} \right) + \mathrm{h.c.} \right]
\nonumber
\end{align}
and a series of higher-dimensional operators~$(\mathcal{Q})$, to be made precise later
\begin{align}
\mathcal{L}_\mathrm{LEFT} &= \mathcal{L}_\mathrm{SM}^\mathrm{broken} + \mathcal{L}_\mathrm{EFT} \,,
\\
\mathcal{L}_\mathrm{EFT} &=  \sum_{n=-1}^\infty \sum_i \frac{\mathcal{C}_i^{(n)}(\mu)}{v^n} \mathcal{Q}_i^{(n)} (\mu)
\label{leff}
\end{align}
arising from the interactions with the heavy particles that were integrated out; the best known being the four-fermion operator in Eq.~\eqref{eq:4fermi}. Here the flavor indices~$p,r$ run over the values~$1,2,3$ for~$\psi=d,e,\nu$ and over~$1,2$ for~$\psi=u$.
These operators are organized by their dimension, starting with terms of dimension three, and increasing powers of~$1/v$ (often expressed by the Fermi constant~$G_F$). Sometimes also $m_W={\mathcal{O}(1) \times v}$ is used as expansion parameter instead. In a SMEFT theory, there is an additional expansion in powers of $1/\Lambda = 1/v \times (v/\Lambda)$, and of course, more operators than in the~SM. The LEFT Wilson coefficients~$\mathcal{C}(\mu)$ multiplying the operators depend on the renormalization scale~$\mu$. As a rule, the renormalization scale should be chosen near
to the physically relevant energy, in order to avoid additional large corrections in matrix elements of the operators. On the other hand, the Wilson coefficients~$\mathcal{C}(v)$ from the matching to the underlying model, be it the SM or the SMEFT, are given at the weak scale,~$\mu \approx m_W$. 
The connection between the two scales is realized by the renormalization group and the running of the~$\mathcal{C}(\mu)$ described by the renormalization group equation
\begin{equation}
\dot{\mathcal{C}} = 16 \pi^2 \mu \frac{\dd}{\dd\mu} \mathcal{C} = \beta_\mathcal{C} \,,
\label{eq:RGE}
\end{equation}
where $\beta_\mathcal{C}$~is the beta-function of the coefficient~$\mathcal{C}$.
This implies that the Wilson coefficients can pick up large logarithmic correction of the form~$\log(m_b/m_W)$. Furthermore,  the running of the Wilson coefficients of lower-dimensional operators can be proportional the coefficients of higher-dimensional operators, due to the presence of light scales/masses in the theory. 

As mentioned, a lot of work has been done in developing the LEFT from the~SM. If the underlying theory is the SMEFT rather than the~SM, we need to match the Wilson coefficients of the LEFT to the coefficients in SMEFT. We have to do this matching in the broken phase of the SMEFT, in the same way we do for the~SM.
This implies that additional terms suppressed by appropriate powers of~$1/\Lambda$ must be added in the matching equations for the LEFT coefficients at the scale~$m_W$.

\subsection{Electroweak symmetry breaking in the SMEFT}
\label{sec:EWSB-in-SMEFT}
The Lagrangian for the SMEFT in the \textit{unbroken phase}, i.e., above the electroweak symmetry breaking~(EWSB) scale~$\sim v$, has been discussed in Sec.~\ref{smeft}. 
Here we will now consider EWSB in the SMEFT determining the Lagrangian in the \textit{broken phase}. We especially emphasize how EWSB is altered compared to the SM due to the presence of the additional higher-dimensional operators, which modify the definition of several SM parameters at tree level. For this, we will follow the discussions presented in \cite{Alonso:2013hga,Jenkins:2017jig}.

\subsubsection{The Higgs sector}
In the scalar sector both the Higgs kinetic term as well as the scalar potential are modified in the SMEFT and read
\begin{align}
    \mathcal{L}_{H} 
    &= (D_\mu H)^\dagger (D^\mu H) + m^2 H^\dagger H - \frac{\lambda}{2} (H^\dagger H)^2 
    \nonumber\\
    &+ \frac{C_{H\Box}}{\Lambda^2} (H^\dagger H) \Box (H^\dagger H) + \! \frac{C_{H\! D}}{\Lambda^2} (H^\dagger D_\mu H)^\ast (H^\dagger D_\mu H)
    \nonumber\\
    &+ \frac{C_H}{\Lambda^2} (H^\dagger H)^3 + \mathcal{O}(\Lambda^{-4}) \,.
    \label{eq:SMEFT-higgs-Lagrangian}
\end{align}
In unitary gauge we can write the Higgs doublet as
\begin{align}
H &= \frac{1}{\sqrt{2}} \begin{pmatrix}
    0
    \\
    [1+c_{H,\mathrm{kin}}] h + v_T
\end{pmatrix}\,, \quad \text{where}
\label{eq:Higgs-SMEFT-parametrization}
\\[0.2cm]
c_{H,\mathrm{kin}} &\equiv \left(\! C_{H\Box}-\frac{1}{4}C_{H\!D} \!\right) \frac{v^2}{\Lambda^2} \,, \quad 
v_T \equiv \left(\! 1+\frac{3\,C_{H}}{4\lambda} \frac{v^2}{\Lambda^2} \right)v \,.
\label{shift}
\end{align}
Here, $c_{H,\mathrm{kin}}$ guarantees a canonical normalization of the kinetic term of the physical real Higgs~$h$, and $v_T$~is the vacuum expectation value of the complex Higgs doublet~$H$ in the SMEFT, whereas $v=\sqrt{2 m^2/\lambda}$ is the vev of~$H$ in the SM.\footnote{Notice that we have $v_T^2=v^2+\mathcal{O}(v^2/\Lambda^2)$, and thus, when working at dimension six, we can always replace~$v_T$ by~$v$ when it multiplies a $d=6$~operator.} Substituting the above equations back into~\eqref{eq:SMEFT-higgs-Lagrangian} we find all self-interactions of the physical Higgs~$h$. For example, its mass term~$m_h$, defined by $\smash{\mathcal{L}_h \supset \frac{1}{2} m_h^2 h^2}$, reads
\begin{align}
    m_h^2 &= \lambda v_T^2 \left( 1-\frac{3 C_H}{\lambda} \frac{v^2}{\Lambda^2} + 2 c_{H,\mathrm{kin}} \right) \,.
\end{align}

Equally, all masses and couplings of the particles are modified. This concerns the fermions and their Yukawa couplings as well as the masses and couplings of the gauge bosons, which are modified by similar shifts.

\subsubsection{The Yukawa sector}
The fermion masses and Yukawa couplings in the broken phase 
\begin{align}
    \mathcal{L}_\mathrm{Yukawa}^\mathrm{broken}
    &= - [\mathcal{M}_\psi]_{pr} (\overline{\psi}{}^L_p \psi^R_r) - [\mathcal{Y}_{\psi h}]_{pr} (\overline{\psi}{}^L_p \psi^R_r) h + \mathrm{h.c.}
    \label{eq:LEFT-mass-Yukawa}
\end{align}
are determined by the parameters~$Y_\psi$ and~$C_{\psi H}$ through
\begin{align}
    [\mathcal{M}_\psi]_{pr} &= \frac{v_T}{\sqrt{2}} \left( [Y_\psi]_{pr} - \frac{1}{2} \frac{v^2}{\Lambda^2} [C_{\psi H}]_{pr} \right) \,,
    \\[0.2cm]
    [\mathcal{Y}_{\psi h}]_{pr} &= \frac{1}{\sqrt{2}} \left( (1+c_{H,\mathrm{kin}}) [Y_\psi]_{pr} - \frac{3}{2} \frac{v^2}{\Lambda^2} [C_{\psi H}]_{pr} \right)
\end{align}
for $\psi \in \{u,d,e\}$. Notice that contrary to the SM, the Yukawa matrices~$\mathcal{Y}_{\psi h}$ are no longer proportional to the mass matrices~$\mathcal{M}_\psi$. Therefore, both cannot be diagonalized simultaneously in general,\footnote{Both matrices also have different RG~equations.} and hence, when working in the mass basis, the Higgs boson~$h$ will have flavor violating couplings starting at~$\mathcal{O}(\Lambda^{-2})$ \cite{Alonso:2013hga}.

Similarly, the $d=5$~Weinberg operator~\eqref{eq:Weinberg-Operator} of the SMEFT yields a neutrino Majorana-mass matrix in the LEFT
\begin{align}
    \mathcal{L} &\supset -\frac{1}{2} [\mathcal{M}_\nu]_{pr} \left( \overline{\nu}{{}^L_{p}}^c \nu^L_{r} \right) + \mathrm{h.c.} \,,
\end{align}
where $\smash{[\mathcal{M}_\nu]_{pr}}=\smash{-[C_\mathrm{Weinberg}]_{pr} \, v_T^2 \big/ \Lambda_{\slashed{L}}}$. Here, $\Lambda_{\slashed{L}}$ denotes the new-physics scale of the Weinberg operator, where lepton number is violated by $\Delta L = 2$. As already mentioned, this scale is not necessarily related to the new-physics scale~$\Lambda$ of other operators and known to be very high $\Lambda_{\slashed{L}} \gtrsim 10^{13}\,\text{GeV}$, for an $\mathcal{O}(1)$ coefficient~$C_\mathrm{Weinberg}$, in order to explain the tiny neutrino masses of the order of~$\sim 1\,\text{eV}$. Notice that $\smash{[\mathcal{M}_\nu]_{pr}}$ is symmetric in the flavor indices~$p$ and~$r$, and that this is the only dimension-three operator present in the LEFT.

In general, the mass matrices~$\mathcal{M}_\psi$, for $\psi \in \{\nu,e,u,d\}$, are non-diagonal. To go to the mass basis we need to diagonalize them by unitary rotations~$U_{\psi_{L\!/\!R}}$ of the fermion fields $\psi_{L\!/\!R} \to U_{\psi_{L\!/\!R}} \psi_{L\!/\!R}$, such that 
\begin{align}
    U_{\psi_L}^\dagger \, \mathcal{M}_\psi \, U_{\psi_R} &\equiv \mathrm{diag}(m_{\psi_1},m_{\psi_2},m_{\psi_3}) \,.
\end{align}
Of course, in general we have $\smash{U_{d_L} \neq U_{u_L}}$ and $\smash{U_{e_L} \neq U_{\nu_L}}$ leading to the CKM and PMNS~matrix respectively
\begin{align}
    V_\mathrm{CKM} &= U_{u_L}^\dagger \, U_{d_L} \,,
    &
    V_\mathrm{PMNS} &=   U_{e_L}^\dagger \, U_{\nu_L}\, ,
\end{align}
which contribute to charged current interactions.
In the SM it is conventional to align the mass- and weak-eigenstate bases either in the up-sector~$(U_{u_L}=\mathds{1})$ or down-sector~$(U_{d_L}=\mathds{1})$. In the SM these two choices, and any other arbitrary alignment choice, are equivalent since the CKM~matrix is the only source of flavor violation in the SM, and it is determined experimentally. On the contrary, in the SMEFT there are potentially other sources of flavor violation due to the higher-dimensional operators. Therefore, the alignment of mass- and weak-eigenstates is crucial as different choices lead to different physics results (for a given set of Wilson coefficients). For example, for a four-fermion operator we find 
\begin{gather}
    [C]_{prst} \left( \overline{\psi}_{1,p} \Gamma \psi_{2,r} \right) \! \left( \overline{\psi}_{3,s} \Gamma \psi_{4,t} \right) 
    \nonumber\\
    \downarrow
    \\
    [C]_{prst} [U_1^\dagger]_{p^\prime \! p} [U_2]_{r r^\prime} [U_3^\dagger]_{s^\prime \! s} [U_4]_{t t^\prime} \! \left( \overline{\psi}_{1,p^\prime} \Gamma \psi_{2,r^\prime} \!\right) \!\! \left( \overline{\psi}_{3,s^\prime} \Gamma \psi_{4,t^\prime} \!\right)\! ,
    \nonumber
\end{gather}
where $\Gamma$~denotes some Dirac structure, possibly in combination with generators. We see that different alignment choices, i.e., different choices for the $U_n$~matrices, lead to different results for the operators in the mass basis.

\subsubsection{The gauge sector}
The kinetic terms for the gauge bosons in the broken phase receive additional contributions from the operators $Q_{HG}$, $Q_{HW}$, $Q_{HB}$, and~$Q_{HW\!B}$. To properly normalize the kinetic terms, we redefine the gauge fields and couplings \cite{Alonso:2013hga}
\begin{align}
    G_\mu^A &= \mathcal{G}_\mu^A \left(\! 1+ \frac{v_T^2}{\Lambda^2} C_{HG} \!\right) , 
    &
    \overline{g}_3 &= g_3 \left(\! 1+ \frac{v_T^2}{\Lambda^2} C_{HG} \!\right) ,
    \\
    W_\mu^I &= \mathcal{W}_\mu^I \left(\! 1+ \frac{v_T^2}{\Lambda^2} C_{HW} \!\right) ,
    &
    \overline{g}_2 &= g_2 \left(\! 1+ \frac{v_T^2}{\Lambda^2} C_{HW} \!\right) ,
    \\
    B_\mu &= \mathcal{B}_\mu \left(\! 1+ \frac{v_T^2}{\Lambda^2} C_{H\!B} \!\right) ,
    &
    \overline{g}_1 &= g_1 \left(\! 1+ \frac{v_T^2}{\Lambda^2} C_{H\!B} \!\right)
\end{align}
so that their products are left invariant (e.g. $\smash{g_2 W_\mu^I} = \smash{\overline{g}_2 \mathcal{W}_\mu^I}$).
This leads to canonically normalized kinetic terms for the gluons~$\smash{\mathcal{G}_\mu^A}$, but not for the weak gauge bosons $\smash{\mathcal{W}_\mu^I}$ and~$\smash{\mathcal{B}_\mu}$ due to the kinetic mixing induced by~$\smash{Q_{HW\!B}}$, which mixes the $\smash{\mathcal{W}_\mu^3}$~state with the $\smash{\mathcal{B}_\mu}$~state. For the kinetic and mass terms we find
\begin{align}
    \mathcal{L}_\mathrm{gauge}^\mathrm{broken} = 
    &-\frac{1}{2} \mathcal{W}_{\mu\nu}^{+} \mathcal{W}^{\mu\nu}_{-} -\frac{1}{4} \mathcal{W}_{\mu\nu}^{3} \mathcal{W}^{\mu\nu}_{3} +\frac{1}{4} \overline{g}_2^2 v_T^2 \mathcal{W}_\mu^{+} \mathcal{W}^{\mu}_{-}
    \nonumber\\
    &-\frac{1}{4} \mathcal{B}_{\mu\nu} \mathcal{B}^{\mu\nu} -\frac{1}{2} \frac{v_T^2}{\Lambda^2} C_{HW\!B} \mathcal{W}_{\mu\nu}^3 \mathcal{B}^{\mu\nu}
    \\
    &+\frac{1}{8}v_T^2 \left( 1 + \frac{1}{2} \frac{v_T^2}{\Lambda^2} C_{H\!D} \right) \left( \overline{g}_2 \mathcal{W}_\mu^3 - \overline{g}_1 \mathcal{B}_\mu \right)^2 \,,
    \nonumber
\end{align}
where $\smash{\mathcal{W}_\mu^\pm}=\smash{(\mathcal{W}_\mu^1 \mp i\mathcal{W}_\mu^2)/\sqrt{2}}$, and similarly for the field-strength tensors.
We can apply two rotations \cite{Grinstein:1991cd}
\begin{align}
    \begin{pmatrix}
        \mathcal{W}_\mu^3 
        \\ 
        \mathcal{B}_\mu
    \end{pmatrix}
    &=
    \begin{pmatrix}
        1 & \epsilon 
        \\
        \epsilon & 1
    \end{pmatrix}
    \begin{pmatrix}
        \overline{c}_\theta & \overline{s}_\theta
        \\
        -\overline{s}_\theta & \overline{c}_\theta
    \end{pmatrix}
    \begin{pmatrix}
        \mathcal{Z}_\mu 
        \\ 
        \mathcal{A}_\mu
    \end{pmatrix} \,,
\end{align}
where $\epsilon=- v_T^2 C_{HW\!B} \big/ (2\Lambda^2)$, to diagonalize the kinetic terms and go to the mass-eigenstate basis containing the photon~$\mathcal{A}_\mu$ and the $\mathcal{Z}_\mu$~boson.
The rotation angles are
\begin{align}
    \overline{s}_\theta &\equiv \sin\overline{\theta} = \frac{\overline{g}_1}{\sqrt{\overline{g}_1^2+\overline{g}_2^2}} \left[ 1 + \epsilon \, \frac{\overline{g}_2}{\overline{g}_1} \frac{\overline{g}_1^2-\overline{g}_2^2}{\overline{g}_1^2+\overline{g}_2^2} \right] \,,
    \\
    \overline{c}_\theta &\equiv \cos\overline{\theta} = \frac{\overline{g}_2}{\sqrt{\overline{g}_1^2+\overline{g}_2^2}} \left[ 1 - \epsilon \, \frac{\overline{g}_1}{\overline{g}_2} \frac{\overline{g}_1^2-\overline{g}_2^2}{\overline{g}_1^2+\overline{g}_2^2} \right] \,.
\end{align}
Of course, the photon~$\mathcal{A}_\mu$ remains massless due to $\mathrm{U}(1)_e$~gauge invariance, whereas the $\mathcal{W}^{\pm}_\mu$ and $\mathcal{Z}_\mu$~boson acquire the masses
\begin{align}
    m_W^2 &= \frac{\overline{g}_2^2 v_T^2}{4} \,,
    \\
    m_Z^2 &= \frac{v_T^2}{4}  \left[ \left( \overline{g}_1^2 + \overline{g}_2^2 \right)  \left( 1 + \frac{1}{2} \frac{v_T^2}{\Lambda^2} C_{H\!D} \right) - 4 \overline{g}_1 \overline{g}_{2} \epsilon \right].
\end{align}
We can then define the gauge couplings 
\begin{align}
    \overline{e} &= \overline{g}_2 \left( \overline{s}_\theta + \epsilon \, \overline{c}_\theta \right) \,,
    &
    \overline{g}_Z &= \frac{\overline{e}}{\overline{s}_\theta \overline{c}_\theta} \left( 1-\frac{\epsilon}{\overline{s}_\theta \overline{c}_\theta} \right)
\end{align}
and the covariant derivative of the broken phase
\begin{align}
\begin{split}
    D_\mu = \partial_\mu &- i\, \overline{g}_2 \left( \mathcal{W}_\mu^+ t^+ + \mathcal{W}_\mu^- t^- \right) 
    \\
    &- i\, \overline{g}_Z \left( T_3 - \overline{s}_\theta^2 Q \right) \mathcal{Z}_\mu - i\, \overline{e}\, Q \mathcal{A}_\mu \,,
\end{split}
\end{align}
where the electric charge is $Q=T_3+Y$ with the hypercharge~$Y$ and the third component of weak isospin~$T_3$. Furthermore, we defined $\smash{t^\pm}=\smash{(t^1 \pm i t^2)/\sqrt{2}}$, where $t^I$ are the $\mathrm{SU}(2)_L$~generators.
Also, the couplings of the $\mathcal{W}$ and $\mathcal{Z}$~boson to fermions are modified by operators of the class~$\psi^2 H^2 D$. 
For example, in the broken phase the operator $\smash{[Q_{Hl}^{(3)}]_{pr}}$ yields an interaction term of the form 
\begin{align}
    [Q_{Hl}^{(3)}]_{pr} 
    &\rightarrow 
    v_T^2 \bigg\{\frac{\overline{g}_Z}{2}\big[\!\left(\overline{\nu}{}^L_p \gamma^\mu \nu^L_r\right) - \left(\overline{e}{}^L_p \gamma^\mu e^L_r\right) \big] \mathcal{Z}_\mu
    \label{eq:QHl3}\\
    &+ \frac{\overline{g}_2}{\sqrt{2}} \big[ \! \left( \overline{\nu}{}^L_p \gamma^\mu e^L_r \right) \mathcal{W}_\mu^+ + \left( \overline{e}{}^L_p \gamma^\mu \nu^L_r \right) \mathcal{W}_\mu^- \big] \!\bigg\} + \mathcal{O}(h)
    \nonumber
\end{align}
additionally to the SM interactions.
For more details see, e.g., \cite{Jenkins:2017jig}.\footnote{Notice also that in the SMEFT the $\mathcal{W}$~boson can couple to right-handed fermions through the $Q_{Hud}$~operator.}

\subsection{Integrating out the weak-scale particles in SMEFT}
After having derived the SMEFT Lagrangian in the broken phase, we can construct the LEFT Lagrangian by removing the heaviest particles from the theory, i.e., the Higgs~$h$, $W/Z$-bosons, and the top quark~$t$. This procedure will be discussed in general in Sec.~\ref{sec:matching}, and here we only anticipate the determination of the Fermi interaction~\eqref{eq:4fermi} as an example.

\begin{figure}[t] 
    \centering
    \includegraphics[width=0.98\linewidth]{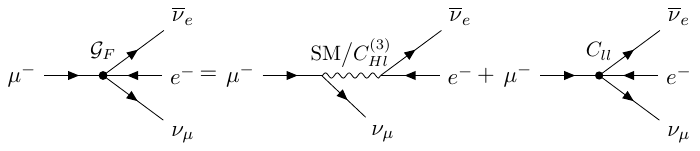}
    \caption{Tree-level diagrams contributing to the SMEFT-to-LEFT matching for the Fermi constant. The $\mathcal{W}$~boson propagators in the SMEFT diagrams are understood to be expanded in~$p^2/m_W^2$ yielding local contributions.}
    \label{fig:GFermi-matching}
\end{figure}

For illustration, consider the four-fermion interaction
\begin{align}
    -\frac{4 \mathcal{G}_F}{\sqrt{2}} \left( \overline{\nu}{}^L_{\mu} \gamma^\mu \mu^L \middle)\middle(\overline{e}{}^L \gamma_\mu \nu^L_{e} \right)
\end{align}
in the LEFT mediating the muon decay $\mu^- \to e^- + \nu_\mu + \overline{\nu}_e$, whose measurement allows to determine the value of the coupling constant~$\mathcal{G}_F$.
In the SMEFT this decay is mediated through the exchange of a $\mathcal{W}_\mu^-$~boson, either with its SM couplings or with the modified coupling due to~$\smash{Q_{Hl}^{(3)}}$ as shown in Eq.~\eqref{eq:QHl3}, or through the four-fermion SMEFT operator~$Q_{ll}$. The corresponding tree-level Feynman diagrams are shown in Fig.~\ref{fig:GFermi-matching}.

We can now compute the tree-level amplitudes in the LEFT and SMEFT, expanding the $\mathcal{W}$-propagators as $1/(p^2-m_W^2)=-1/m_W^2+\mathcal{O}(p^2 \big/ m_W^2)$, where $p$~is the momentum carried by the $\mathcal{W}$-boson, which in the range of validity of the LEFT is small $p^2 \ll m_W^2$. Equating our results we find \cite{Alonso:2013hga}
\begin{align}
\begin{split}
    -\frac{4\mathcal{G}_F}{\sqrt{2}}
    =- \frac{2}{v_T^2} &- \frac{2}{\Lambda^2}  \Big( [C_{Hl}^{(3)}]_{11} + [C_{Hl}^{(3)}]_{22} \Big) 
    \\
    &+ \frac{2}{\Lambda^2} \Re \big( [C_{ll}]_{1221} \big) \,,
\end{split}
\label{eq:4fermi-matching}
\end{align}
where we used $[C_{ll}]_{2112}=[C_{ll}^\ast]_{1221}$.
The above equation is called a matching condition and determines the LEFT coefficient~$\mathcal{G}_F$ in terms of the SMEFT parameters.
We see that the LEFT and SMEFT power expansions get mixed up in this case. In general LEFT operators of dimension~$d$ are suppressed by 
\begin{equation}
\frac{1}{\Lambda^a}\frac{1}{v^b}\,, \quad a+b=d-4\,, \quad a \geq 0 \,, 
\label{dimension}
\end{equation}
where~$a$ is always positive, since the SMEFT contains~$\Lambda$ only as multiplicative prefactor with negative powers, and~$b$ can be negative due to Higgs vev insertions. 

From Eq.~\eqref{eq:4fermi-matching} we see that what is actually extracted from experimental data on the muon decay is~$\mathcal{G}_F$ rather than the SM value~$\smash{G_F=(\sqrt{2}\,v^2)^{-1}}$. The modification of the tree-level relations of SM parameters like the above in the SMEFT requires extra care in extracting these from experiment. For more detail on the determination of the SM parameters and a discussion of appropriate input schemes for SMEFT computations see, e.g., \cite{Brivio:2021yjb,Brivio:2017bnu}.

The construction of a complete basis of the LEFT up to dimension six, and the derivation of all tree-level relations of the LEFT Wilson coefficients to the SMEFT Wilson coefficients is presented in \cite{Jenkins:2017jig}.
Subsequently, also LEFT operator bases at dimension seven \cite{Liao:2020zyx}, eight \cite{Murphy:2020cly,Li:2020tsi}, and nine \cite{Li:2020tsi} have been derived in the literature.
At the one-loop level, the matching requires the calculation of a large number of loop diagrams. This monumental program was taken up by~\textcite{Dekens:2019ept}. 
Obviously these are lengthy expressions and are most conveniently and usable given directly in digital form as in Appendix~G and the supplementary material of \cite{Dekens:2019ept}.\footnote{The SMEFT-to-LEFT one-loop matching is also implemented in codes such as \cmd{Wilson} \cite{Aebischer:2018bkb} and \cmd{DsixTools} \cite{Celis:2017hod,Fuentes-Martin:2020zaz}.}
Therein, the SMEFT coefficients appearing in the matching conditions are understood to be renormalized in the $\overline{\mathrm{MS}}$~scheme, and implicitly dependent on the matching scale~$\mu \sim m_W$.
When using these one-loop matching results it is important to strictly follow all the conventions used in their derivation also in all subsequent computation with the resulting LEFT Lagrangian to avoid any inconsistencies.
Eventually, we point out that the one-loop anomalous dimensions of all $d \leq 6$ LEFT coefficients have been derived in \cite{Jenkins:2017dyc} which complements the toolbox for LEFT computations.

\section{Beyond the Standard Model phenomenology with SMEFT}
\label{sect:practical}
In this section we showcase the practical application of effective field theory, and particularly of the SMEFT, in the search for new physics. Since the Standard Model predictions so far agree so well with experiment, we think that a detailed treatment is important. The bumpy history of some of the
precision results shows the importance of a systematic treatment. We will discuss the different EFTs involved, and how they are linked. We comment on existing work and consider in detail two explicit examples to illustrate all the steps needed in order to reliably constrain possible BSM scenarios.

\subsection{The SMEFT analysis workflow}
\begin{figure}[t]
    \centering
    \includegraphics[width= 0.95\linewidth]{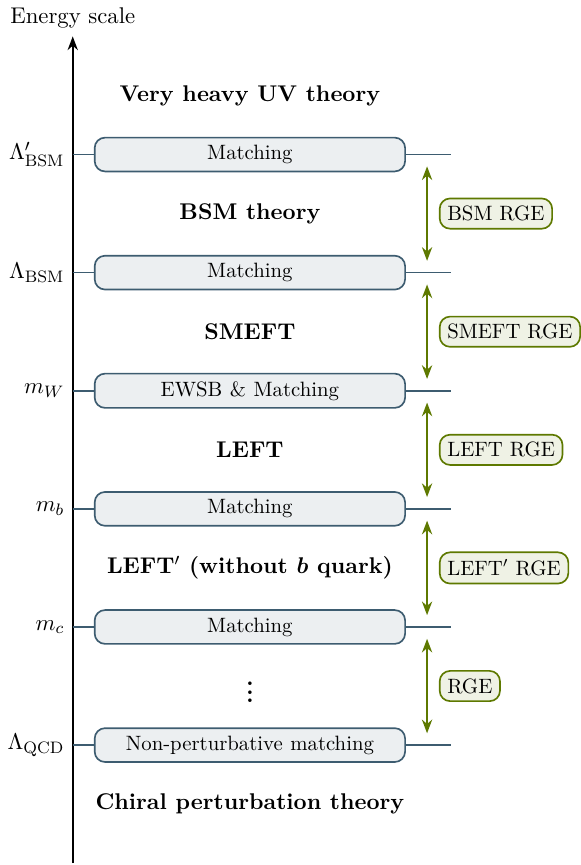}
    \caption{%
    Tower of effective field theories ranging from far UV scales to the low energies where experiments are performed. Depending on the observable, different EFTs might be appropriate. The various EFTs can be connected through matching and RG~evolution. Moreover, spontaneous symmetry breaking may occur at intermediate steps.}
    \label{fig:multiscale-eft}
\end{figure}

We start by discussing the general EFT workflow for new-physics searches, whereas detailed information on
the various steps involved will be discussed in the subsequent sections.
For a typical BSM analysis, we consider a tower of effective theories as illustrated in Fig.~\ref{fig:multiscale-eft}. Each of these EFTs provides an accurate description of nature at a given energy scale, only containing the relevant degrees of freedom at that energy and incorporating the effect of heavier states by higher-dimensional effective operators. To connect theories at different energy scales, matching computations are performed, which allow to integrate out the heavy particles from one theory to obtain the corresponding low-energy EFT. We already encountered a matching computation for the Fermi constant in Eq.~\eqref{eq:4fermi-matching}; further details on the procedure will be given in Sec.~\ref{sec:matching}.
Then, within each EFT, the corresponding renormalization group~(RG) equations can be used to evolve all couplings from the high energy scale, where the matching was performed, to the low energy scale of the heaviest particles remaining in the EFT (see Sec.~\ref{sec:rge}). These can then be integrated out in a next matching computation to obtain yet another EFT valid at even lower energies. This procedure has to be repeated until the desired energy scale --usually the energy range of experimental observables-- is reached. By using such a multi-step procedure of alternate matching and running, we can ensure a proper description of physics at all involved scales, as the RG~evolution allows to resum large logarithms that would appear if just a single matching would be performed at the lowest scale, or if we would even use only the full theory.
As one can easily imagine such a multi-scale computation can become highly involved  which requires automation of all the steps involved. There are already several computer tools available that automate some of these steps [for an overview see \cite{Proceedings:2019rnh,Aebischer:2023irs}]. However, complete automation is still a goal to be achieved in the future.

To be concrete, we can take a BSM theory containing some heavy particles with masses of order~$\mathcal{O}(\Lambda_\mathrm{BSM})$ that are not accessible at the energies of current experiments. 
We can then match such theory to the SMEFT where the effect of these heavy states is encoded in effective operators. 
In general, not all of the SMEFT operators will be generated by the matching, but only a certain subset. 
For example, a list of all operators that are generated at tree level in all possible BSM scenarios has been worked out in the UV/IR dictionary provided in \cite{deBlas:2017xtg}.
Consequently, the SMEFT RG~equations~\cite{Jenkins:2013zja,Jenkins:2013wua,Alonso:2013hga} are used to evolve the couplings from the matching scale~$\Lambda_\mathrm{BSM}$ down to the electroweak scale~$\sim m_W$. Through this RG~mixing further operators can be generated, that were absent in the matching. Then, at the electroweak scale spontaneous symmetry breaking takes place as discussed in Sec.~\ref{sec:EWSB-in-SMEFT}. After expanding the Higgs field around its vacuum expectation value, we end up with the SMEFT Lagrangian in the broken phase, invariant under the $\mathrm{SU}(3)_c \times \mathrm{U}(1)_e$ gauge symmetry, and containing only the physical Higgs~$h$ rather than the full Higgs doublet~$H$, which is only present in the unbroken phase above the electroweak scale. 
One can now integrate out the heaviest SM particles, i.e. the top quark~$t$, and the Higgs~$h$, $Z$,~and $W$~boson, to arrive at the~LEFT \cite{Jenkins:2017jig,Dekens:2019ept}.
Usually EWSB and the matching is performed at the single scale~$\sim m_W$. 
Since the masses of all particles that are integrated out here are very similar, no large logarithms arise even though we choose only a single matching scale.
Afterwards, the known LEFT~RG equations~\cite{Jenkins:2017dyc} can be used to evolve the theory down to the bottom-quark mass scale~$m_b$. Then, if necessary, the $b$~quark can be integrated out, and so on, until one reaches the energy scale of the experimental observables of interest. For example, for $B$~physics experiments it is enough to stop the procedure here, but for processes at even lower energies one might require also integrating out, e.g., the charm quark. See \cite{Buchalla:1995vs} for an excellent review of the EFTs at the $b$-scale.

Ultimately one could end up at the QCD confinement scale~$\Lambda_\mathrm{QCD}$, where one matches onto chiral perturbation theory~(ChPT). However, due to confinement, in this case the IR~degrees of freedom of~ChPT, i.e. the pions, are not the same as in the EFT above~$\Lambda_\mathrm{QCD}$, where we have the quarks. Because of the growth of the strong coupling constant, perturbativity is lost and the matching has to be performed non-perturbatively. The discussion of this process is, however, beyond the scope of this review and we refer to \cite{Bernard:2007zu} for further details.

The BSM theory we started with could well be just an EFT itself, originating from integrating out even heavier particles in some more fundamental theory valid at even higher scales~$\Lambda_\mathrm{BSM}^\prime$. 
In this EFT picture we can simply consider any theory as an effective description at a given energy containing only the relevant degrees of freedom at these energies and incorporating our agnosticism about the laws of nature valid at higher energies in terms of higher-dimensional effective operators. 
Relating the different EFTs through matching and running as discussed above, we can express the couplings in some UV theory through the couplings/Wilson coefficients of a low-energy EFT valid at the scale of experiments; therefore, allowing us to constrain these UV parameters from low-energy data.

Describing nature by a chain of EFTs is, by construction, always an approximation. However, it can be systematically improved by including higher orders in the EFT power counting, i.e., higher-dimensional operators, thereby allowing, in principle, to describe physical laws up to arbitrary precision. In practice, usually only the leading contributions given by the dimension-six operators and their interference with the~SM are relevant. 
Nevertheless, higher orders, such as dimension-six squared or $d\geq 8$~operators, can be relevant in the case where they introduce new interactions that are not generated by the leading order \cite{Hays:2018zze,Corbett:2021eux} or when they exhibit some energy enhancement as in the case of high-$p_T$ Drell-Yan tails that we will discuss in Sec.~\ref{sec:Drell-Yan}. 
For a discussion of higher-order operators, and the EFT validity see also Sec.~\ref{sec:d8-operators} and~\ref{subsec:SMEFT_Constraints}.

Any SMEFT computation always involves a double expansion in the EFT power-counting order and the loop order, and the truncation of both has to be chosen individually according to the desired precision and the process of interest.\footnote{Contrary to that, in the HEFT the two expansions are linked together and cannot be truncated individually.} Going to the one-loop level in the matching computation might be especially required in cases where the operators of interest are not generated at the tree level, i.e., for all loop generated operators (see Sec.~\ref{sec:tree-loop-generated-operators}). In principle,  consistency in the expansion parameters should be obeyed. For instance, a one-loop matching computation would require two-loop running to obtain scheme independent results. However, in practice nearly always only the one-loop RG equations are considered as the full two-loop SMEFT RG equations are not yet known and only partial results are available. Also, usually one-loop running suffices for the required precision and one-loop matching computations are only required for operators that cannot be generated at the tree level.

Depending on the question we ask, the EFT analysis starts at the top or bottom of the tower of EFTs shown in Fig.~\ref{fig:multiscale-eft}, which we already denoted as top-down or bottom-up studies.
In the top-down approach one starts with a given BSM theory and matches it to the SMEFT and consecutively to the LEFT and so on, to determine the implications of this given theory at low energies. The ultimate goal is usually to constrain specific parameters of the BSM theory from suitable low-energy measurements. The strong advantage of using EFTs in this approach is the simplification of the initial problem and the resummation of large logarithms.

In the bottom-up approach, on the other hand, the idea is trying to be agnostic about the UV completion of the SM. In this case, one can try to use many low-energy data sets to constrain a large number of SMEFT Wilson coefficients. In principle, the goal of this approach is to perform a global fit to determine all SMEFT parameters. In practice, due to the large number of free parameters, such a fit is (currently) unfeasible. These types of analyses only consider a certain subset of parameters
based either on general dynamical hypotheses and/or symmetry assumptions, such as the flavor symmetries discussed in Sec.~\ref{sec:GlobalSymmetries}.
Eventually, constraints provided in this manner should still be used to constrain different BSM scenarios that just have to be matched to the SMEFT, rather than having to perform the full analysis for every model individually.

Sometimes there is also no clear distinction between the two approaches. 
A~key point worth stressing is that performing an EFT analysis is especially useful if there is a signal for new physics. Without such a signal, one can only put constraints on the huge parameter space of the SMEFT, gaining limited knowledge about the underlying structure of new physics. This is particularly problematic as not all the SMEFT parameters are equally relevant for experimental observations in different BSM theories. 
In absence of a new-physics signal, the constraints are more efficiently expressed as direct bounds on possible deviations from the SM for a given set of observables. 
In this respect, an interesting approach is that of {\em pseudo-observables}~\cite{Bardin:1999gt,Passarino:2010qk}
i.e.~the definition of a suitable set of  on-shell
amplitudes unambiguously connected to measurable quantities,  
able to characterize in general terms deviations from the SM of short-distance 
origin. This approach, originally introduced to describe electroweak precision tests at the $Z$~pole \cite{Bardin:1999gt}, and later extended also to Higgs physics \cite{David:2015waa,Ghezzi:2015vva, Gonzalez-Alonso:2014eva, Greljo:2015sla,Passarino:2010qk}, can be viewed as an intermediate, consistent, and economical step between measurements and their possible EFT interpretation. 
Of course, also in absence of deviations from the SM the EFT interpretation of data provides a useful guiding principle in model building, but the strong power of the EFT approach in predicting new effects and testing the validity of a given BSM hypothesis cannot be exploited.

In the following we will discuss all the steps involved in a new-physics SMEFT analysis. We also analyze in detail one example of a top-down and bottom-up analysis each, for illustration.

\subsection{Matching BSM models to the SMEFT}
\label{sec:matching}
The matching of a BSM theory to the SMEFT is quite involved and there are several variants. To expose the important points, we will consider the matching procedure in the context of EFTs in general, as similar computations are required in all the matching steps illustrated in Fig.~\ref{fig:multiscale-eft}.
By matching a given UV theory to its corresponding low-energy EFT, we fix the value of all the Wilson coefficients of the EFT such that both theories reproduce the same physics in the low-energy limit. Therefore, the Wilson coefficients have to be determined as functions of the UV parameters, and constraints on the EFT coefficients can be directly translated into bounds on the BSM parameters. 
There are two different approaches that allow us to ensure that two theories describe the same physics at low energies, known as \textit{off-shell} and \textit{on-shell} matching. 

The most restrictive requirement we can enforce is that the effective action~$\Gamma$ of both theories, taken as a function of the light fields~$\phi$ only, agrees 
\begin{align}
\Gamma_\mathrm{UV}[\phi]=\Gamma_\mathrm{EFT}[\phi] \,.
\label{eq:matching-condition-general}
\end{align}
This ensures that all off-shell amplitudes with light external particles agree in both theories at low energies. Therefore, this method is commonly known as \textit{off-shell} matching. The matching condition~\eqref{eq:matching-condition-general} then determines the value of all Wilson coefficient in terms of the UV parameters. 
One way to determine the effective action is by calculating all relevant Feynman diagrams with external light fields. Contrary to the usual computation of the effective action, where we need to consider all one-particle-irreducible~(1PI) diagrams, it is required in this case to compute all one-light-particle-irreducible~(1LPI) Feynman diagrams, i.e., the diagrams that cannot be split in two by cutting any light internal line. 
This is because we consider the effective action as a function of light fields only. 
More on this ``diagrammatic matching procedure''  is provided in Sec.~\ref{sec:diagrammatic-matching}. 
An alternative approach to calculating the effective action is through its path integral representation, which we discuss afterwards in Sec.~\ref{sec:functional-matching}.

Requiring off-shell amplitudes to agree is actually more restrictive than necessary. It suffices to ensure that all physical observables computed in either theory agree, which is equivalent to equating the $S$-matrices of UV and EFT for all scattering processes with only light particles in the external states:
\begin{align}
    \langle\phi| S_\mathrm{UV} |\phi\rangle = \langle\phi| S_\mathrm{EFT} |\phi\rangle \,.
\end{align}
This amounts to equating on-shell amplitudes and is therefore known as \textit{on-shell} matching. We can again compute the $S$-matrix diagrammatically, but contrary to the off-shell matching computation we now need to consider all contributing diagrams and not only the 1LPI ones. This can significantly increase the number of diagrams that need to be included in the computation. Also the matching of the reducible diagrams is computationally more challenging, which is why in practice mostly off-shell matching is used.

The main advantage of on-shell matching is that it suffices to consider EFT operators from a minimal basis for the matching computation.
This is not the case for off-shell matching, since for off-shell amplitudes further kinematical structures are allowed, and therefore additional operators need to be included. When discussing the construction of EFT bases in Sec.~\ref{sec:redundant-operators}, we used field redefinitions to reduce the operator list to a minimal basis. Recall that to do so, we argued that the LSZ-formula guarantees that physical observables remain unchanged under field redefinitions. For off-shell matching, however, we do not compute physical observables. Therefore, we are not allowed to use the LSZ-formula and have to use a larger set of operators, that is a basis up to field redefinitions. Such an operator set is commonly referred to as a Green's basis.
We will discuss the diagrammatic off-shell matching procedure with a concrete example and briefly mention the differences to the on-shell computation.
Afterwards, we provide a short introduction to functional matching.

Before we continue with the different matching prescriptions, a comment on the calculation of the loop integrals is in order.
Because loop integrals in the matching can depend on light~$m$ and heavy~$M$ scales/masses,
certain regions of the internal momentum~$k$ of the loop are enhanced. This is encapsulated in the method of \textit{expansion by regions}~\cite{Beneke:1997zp,Jantzen:2011nz} which can be applied in all of the previously mentioned matching techniques.
We can expand each integrand in the region where~$k$ is hard $(k \sim M \gg m)$, and where it is soft $(k\sim m \ll M)$. 
Then performing both integrals separately over the full $D$-dimensional space and summing their results yields the same outcome as computing the full integral and expanding afterwards in powers of~$m/M$. 
Applying the method of regions offers the advantage of separating the UV from the IR physics. 
By construction, the full theory and its corresponding EFT describe the same low-energy dynamics, i.e., IR physics. 
This means that the soft region of the full theory integrals must be equal to the soft region of the corresponding EFT diagrams. 
Thus, we only have to compute the hard region of all integrals to determine the matching conditions, since these incorporate all of the short distance dynamics that must be captured by the Wilson coefficients. 
Next, we notice that the hard region of EFT integrals only yields scaleless integrals, which vanish in dimensional regularization, since the EFT only depends analytically on~$M$. Therefore, we only need to consider tree-level EFT diagrams with insertions of one-loop coefficients and full theory diagrams with at least one heavy propagator in the loop to obtain the full matching conditions. See Appendix~\ref{app:method-of-regions} and \cite{Manohar:2018aog} for more details.

\subsubsection{Diagrammatic matching} \label{sec:diagrammatic-matching}
In order to demonstrate the diagrammatic matching procedure, we work with a concrete example. Consider the extension of the SM by a heavy colored scalar~$S_1$ transforming as~$(\boldsymbol{\bar{3}},\boldsymbol{1})_{1/3}$ under the SM gauge group. The $S_1$~leptoquark couples to both quarks and leptons (thus its name), and its BSM Lagrangian is
\begin{align}
\mathcal{L}_{S_1} \! = \mathcal{L}_\mathrm{SM} &+ (D_\mu S_1)^\dagger (D^\mu S_1) - M_{S}^2 S_1^\dagger S_1
\label{eq:S1}\\
&-\left[ \lambda_{pr}^L (\overline{q}^c_p \varepsilon \ell_r) S_1 + \lambda_{pr}^R (\overline{u}^c_p e_r) S_1 + \mathrm{h.c.} \right] \,, \nonumber
\end{align}
where we neglect any direct coupling of the $S_1$ to the Higgs doublet~$H$. 
For the off-shell matching we need to consider a Green's basis for the SMEFT, i.e., a basis of operators up to field redefinitions. Such basis is given in \cite{Gherardi:2020det}, where also the full matching computation for the given model is presented. Here, we will only reproduce partial results of this derivation to illustrate the procedure.

\begin{figure}[t]
    \centering
    \includegraphics[width=0.80\linewidth]{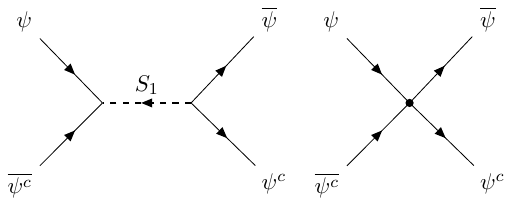}
    \caption{Tree-level Feynman diagrams for the $S_1$~model (left) and the SMEFT (right), that are relevant to the matching of the four-fermion operators. The UV diagram has to be expanded in powers of $1/M_{S}$ before equating it with the EFT diagram.}
    \label{fig:S1-tree-matching}
\end{figure}

First, we realize that for tree-level matching only interaction terms of the Lagrangian~\eqref{eq:S1} with at most one heavy field can contribute. 
Operators with more heavy fields can only contribute at loop level. 
It is then obvious that only four-fermion operators are generated in the EFT at tree level. 
The resulting EFT Lagrangian is
\begin{align}
    \mathcal{L}_\mathrm{EFT} &= \mathcal{L}_\mathrm{SM} + \sum_X \frac{[C_X^{(R)}]_{prst}}{\Lambda^2} [R_X]_{prst} \,,
\end{align}
where the sum runs over the ``redundant'' operators
\begin{subequations}
\begin{align}
    [R_{q^c l}]_{prst} &= (\overline{q}^c_{i p} \ell_{j r}) (\overline{\ell}{}_{s}^j {q^c_{t}}^i) \,,
    \\
    [R_{q^c l}^\prime]_{prst} &= (\overline{q}^c_{i p} \ell_{j r}) (\overline{\ell}{}_{s}^i {q^c_{t}}^j) \,,
    \\
    [R_{e^c u}]_{prst} &= (\overline{e}^c_p u_r) (\overline{u}_s e_t^c) \,,
    \\
    [R_{u^c e l q^c}]_{prst} &= (\overline{u}^c_p e_r) \varepsilon_{ij} (\overline{\ell}{}_{s}^i {q_{t}^c}^j)
\end{align}%
\label{eq:operators_S1}%
\end{subequations}%
defined following the conventions in~\cite{Fuentes-Martin:2022vvu}.
The Feynman diagrams in the UV and the EFT relevant to the matching are shown on the left- and right-hand side of Fig.~\ref{fig:S1-tree-matching}, respectively. Inserting the appropriate interaction terms from the UV and EFT Lagrangian we can compute the corresponding amplitudes. Afterwards, we expand the UV amplitudes as power series in~$1/M_{S}$. Here we work to mass-dimension six, thus we need to truncate the results at~$\mathcal{O}(M_{S}^{-2})$. We can then equate the amplitudes of both EFT and UV to find the matching conditions
\begin{subequations}%
\begin{align}
    [C_{q^c l}^{(R)}]_{prst} &= -[C_{q^c l}^{\prime\,(R)}]_{prst} = \lambda^{L}_{p r} \lambda^{L\ast}_{t s} \,,
    \\
    [C_{e^c u}^{(R)}]_{prst} &= \lambda^{R}_{r p} \lambda^{R\ast}_{s t} \,,
    \\
    [C_{u^c e l q^c}^{(R)}]_{prst} &= -\lambda^{R}_{p r} \lambda^{L\ast}_{t s} \,,
\end{align}%
\end{subequations}%
where we also identify the new-physics scale~$\Lambda$ with the mass of the $S_1$~state: $\Lambda = M_S$.

For convenience, we want to rewrite our result in the Warsaw basis. The operators in Eq.~\eqref{eq:operators_S1} are related to the operators $\smash{Q_{lq}^{(1,3)}}$, $\smash{Q_{eu}}$, and $\smash{Q_{lequ}^{(1,3)}}$ of the Warsaw basis through the Fierz transformations~\eqref{eq:Fierz-ids}. The matching conditions in the Warsaw basis read
\begin{subequations}
\begin{align}
    [C_{lq}^{(1)}]_{prst} &= \frac{1}{2} [C_{q^c l}^{(R)}]_{trps} + \frac{1}{4} [C_{q^c l}^{\prime\,(R)}]_{trps} \,,
    \\
    [C_{lq}^{(3)}]_{prst} &= \frac{1}{4} [C_{q^c l}^{\prime\,(R)}]_{trps} \,,
    \\
    [C_{eu}]_{prst} &= \frac{1}{2} [C_{e^c u}^{(R)}]_{rtsp} \,,
    \\
    [C_{lequ}^{(1)}]_{prst} &= -4 [C_{lequ}^{(3)}]_{prst} = -\frac{1}{2} [C_{u^c e l q^c}^{(R)}]_{trps} \,,
\end{align}%
\label{eq:S1-Fierz}%
\end{subequations}%
where evanescent operators can be ignored since we work at tree level. In the present case there are no integration by parts relations or field redefinitions required to reduce the matching result to the Warsaw basis.

\begin{figure}[t]
    \centering
    \includegraphics[height=1.9cm]{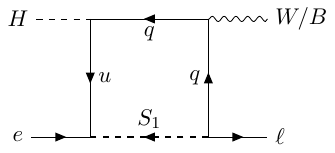}
    \includegraphics[height=1.9cm]{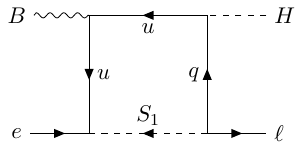}
    \\[0.1cm]
    \includegraphics[height=1.9cm]{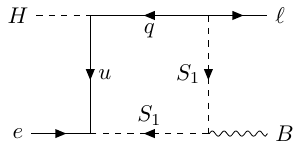}
    \\[0.1cm]
    \includegraphics[height=1.9cm]{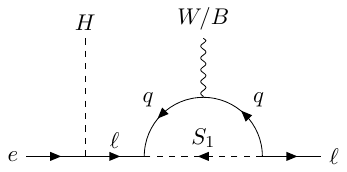}
    \includegraphics[height=1.9cm]{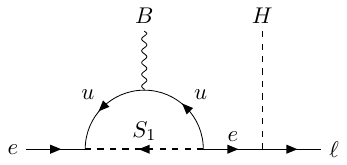}
    \\[0.1cm]
    \includegraphics[height=1.9cm]{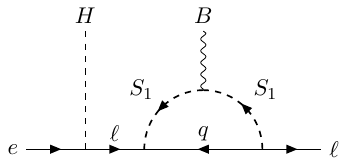}
    \includegraphics[height=1.9cm]{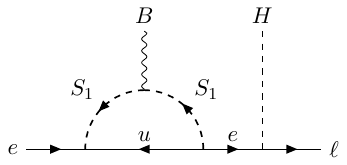}
    \\[0.2cm]
    \includegraphics[height=2cm]{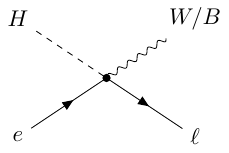}
    \caption{One-loop diagrams relevant to the on-shell matching of the leptonic dipole operators for the $S_1$~model. The first four rows show the diagrams of the UV theory, whereas the last row contains the only EFT diagram when using the method of regions.}
    \label{fig:S1-loop}
\end{figure}

Next, we want to perform the one-loop matching. Since the entire computations is rather lengthy and the full results are shown in~\cite{Gherardi:2020det}, we focus here only on the contributions to the leptonic dipole operators 
\begin{align}
    [Q_{eB}]_{pr} &= (\overline{\ell}_p \sigma^{\mu\nu} e_r) H B_{\mu\nu} \,,
    \label{eq:QeB}
    \\
    [Q_{eW}]_{pr} &= (\overline{\ell}_p \sigma^{\mu\nu} e_r) \tau^I H W_{\mu\nu}^I \,,
    \label{eq:QeW}
\end{align}
which can only be generated at loop level. 
We choose to use on-shell matching, which allows us to single out the dipole matching contributions, for which the relevant diagrams are shown in Fig.~\ref{fig:S1-loop}. 
The first four rows display the diagrams of the UV theory, whereas the diagram in the last row is the only EFT diagram. Recall that, since we employ the method of regions, we only have to consider EFT tree diagrams with one-loop coefficients, but no loop diagrams. 
After expanding in the hard loop-momentum region and performing the Dirac algebra, including the application of the spinor equations of motions and the Gordon identity, the diagrams in the third and fourth row provide a local contribution to the leptonic dipole operators.
If we had chosen off-shell matching instead, we had to consider further topologies that do not directly match onto the dipole, but do contribute to it only after applying field redefinitions to reduce the Green's basis to the Warsaw basis. 
However, it is not straight forward to identify which topologies to consider, which is why we choose to work out this explicit contribution on-shell.\footnote{In the case of off-shell matching we could also neglect the diagrams in the third and fourth row of Fig.~\ref{fig:S1-loop}, since we only have to consider 1LPI diagrams. Their contribution would be shifted to the additional operators in the Green's basis that reduce to the dipole by applying field redefinitions, such that the final results of both methods agree.}

Computing and equating the amplitudes corresponding to the diagrams shown in Fig.~\ref{fig:S1-loop}, where for the UV amplitudes we only keep the terms with the Lorentz structure matching that of the dipole, since the remaining terms will match onto other operators of the Warsaw basis that we are not interested in, we find at~$\mathcal{O}(M_{S}^{-2})$
\begin{align}
    \label{eq:S1-CeB}
    [C_{eB}]_{pr}
    &= \frac{1}{16\pi^2} \frac{g_1}{8} \Bigg\{ -[Y_e]_{pt} \lambda_{st}^{R\ast} \lambda^R_{sr}  \\
    &+ \lambda^{L\ast}_{sp} [Y_u^\ast]_{st} \lambda^R_{tr} \left[\frac{19}{2}+5\log\!\left(\!\frac{\mu_m^2}{M_{S}^2}\!\right)\!\right] \!\!\Bigg\} + [\Delta_{eB}]_{pr}  ,
    \nonumber\\[0.1cm]
    \label{eq:S1-CeW}
    [C_{eW}]_{pr}
    &= \frac{1}{16\pi^2} \frac{g_2}{8} \Bigg\{ \lambda_{sp}^{L\ast} \lambda^L_{st} [Y_e]_{tr} \\
    &- 3 \lambda^{L\ast}_{sp} [Y_u^\ast]_{st} \lambda^R_{tr} \left[\frac{3}{2}+\log\!\left(\!\frac{\mu_m^2}{M_{S}^2}\!\right)\!\right] \!\!\Bigg\} + [\Delta_{eW}]_{pr} , \nonumber
\end{align}
where $\mu_m$ is the matching scale, which we can conveniently choose as $\mu_m = M_{S}$ to eliminate all logarithms in the matching conditions.\footnote{Other choices for~$\mu_m$ are possible, but it should not be chosen too far away from the mass threshold to avoid large logarithms and a worsening of the perturbative expansion.}

However, since we work at the one-loop level now, we can no longer use the Fierz identities applied in the tree-level matching in Eq.~\eqref{eq:S1-Fierz}. 
As stated before, these are intrinsically four-dimensional identities that must not be applied in combination with a computation in dimensional regularization, since they lead to evanescent operators. 
Instead, we apply the corrected \textit{one-loop Fierz transformations} as discussed in Sec.~\ref{sec:evanescent} and \cite{Fuentes-Martin:2022vvu}, which effectively project the evanescent operators onto the physical four-dimensional Warsaw basis, allowing us to ignore evanescent contributions afterwards. 
We focus on the dipole operator, whose additional contributions arising due to the evanescent operators generated by applying the $D=4$~Fierz identities in Eq.~\eqref{eq:S1-Fierz} are labeled by $\Delta_{eB/eW}$ in Eqs.~\eqref{eq:S1-CeB}--\eqref{eq:S1-CeW}. The corresponding shift of the action was derived in Sec.~\ref{sec:evanescent} and is given in Eq.~\eqref{eq:evanescent-dipole-shift} with which we find
\begin{align}
    [\Delta_{eB}]_{pr} &= \!- \frac{1}{16\pi^2} \frac{5}{8} g_1 [Y_u^\ast]_{ts} (1-\xi_\mathrm{rp}) [C_{u^c e l q^c}^{(R)}]_{srpt},
    \\
    [\Delta_{eW}]_{pr} &= \frac{1}{16\pi^2} \frac{3}{8} g_2 [Y_u^\ast]_{ts} (1-\xi_\mathrm{rp}) [C_{u^c e l q^c}^{(R)}]_{srpt},
\end{align}
where $\xi_\mathrm{rp}$ is the parameter denoting the reading point ambiguity when using NDR to evaluate the loop integrals (see Sec.~\ref{sec:evanescent} and Appendix~\ref{app:gamma5} for more details). 
For convenience, we decide to read all EFT loop integrals ending with the EFT operator as suggested by \textcite{Fuentes-Martin:2022vvu}.
As mentioned before, we then have to follow this prescription for all subsequent computations within the EFT to obtain consistent results
independent of the choice for the reading point.
The given prescription yields $\xi_\mathrm{rp}=1$ (which also agrees with the results obtained in the 't~Hooft--Veltman scheme) and thus the evanescent contributions to the dipole happens to vanish.
Nevertheless, there are additional non-vanishing and unambiguous evanescent contributions to further operators, that we do not consider here. 
For more details on the reading point ambiguity see Appendix~\ref{app:gamma5} and \cite{Fuentes-Martin:2022vvu}.

\subsubsection{Functional matching}\label{sec:functional-matching}
We now recap the functional formalism  
worked out in \cite{Dittmaier:1995ee,Henning:2014wua,delAguila:2016zcb,Henning:2016lyp,Fuentes-Martin:2016uol,Zhang:2016pja,Cohen:2020fcu},
and employ it for a specific matching computation.
We make use of the background-field method \cite{Abbott:1980hw,Abbott:1983zw,Denner:1994xt,Denner:1996wn} by separating all fields~$\eta \to \hat\eta + \eta$ in a background field configuration~$\hat\eta$, which satisfies the classical equations of motion, and a pure quantum component~$\eta$. 
In Feynman diagrams $\hat\eta$ then corresponds to tree-level lines, whereas $\eta$ corresponds to lines in loops.
Expanding the action to one-loop accuracy we find
\begin{align}
	S[\hat\eta + \eta] &=S[\hat\eta] + \frac{\sigma_{\eta_j}}{2} \bar{\eta}_i \left. \frac{\delta^2 S}{\delta\bar\eta_i \delta\eta_j} \right|_{\eta=\hat\eta} \eta_j + \mathcal{O}(\eta^3) \,,
\end{align}
where $\sigma_{\eta_j}= 1$ if $\eta_j$ is bosonic and $\sigma_{\eta_j}=-1$ if it is Grassmann due to anti-commuting $\eta_j$ to the right-hand side of the above equation.
The linear term vanishes due to the equations of motion, and higher-order terms only contribute at two-loop order and beyond. We identify the term quadratic in the quantum fields as the \textit{fluctuation operator}
\begin{align}
    \Omega_{ij}[\hat\eta] &= \left. \sigma_{\eta_j} \frac{\delta^2 S}{\delta\bar\eta_i \, \delta\eta_j} \right|_{\eta=\hat\eta} \,.
    \label{eq:fluctuation_operator}
\end{align}
The effective action of the theory is then given by
\begin{align}
    \exp \! \big( i\Gamma[\hat\eta] \big) \! &= \!\! \int\!\!\mathcal{D}\eta \, \exp\!\left( \! i S[\hat\eta] + \frac{i}{2} \bar{\eta}_i \, \Omega_{ij}[\hat\eta] \, \eta_j + \mathcal{O}(\eta^3) \!\right) \,.
\end{align}
Thus, we find the tree-level effective action $\Gamma^{(0)}[\hat\eta]=S^{(0)}[\hat\eta]$ and the one-loop effective action
\begin{align}
\begin{split}
    \Gamma^{(1)}[\hat\eta] &= S^{(1)}[\hat{\eta}] - i\log \left(\mathrm{SDet}\,\Omega^{(0)}[\hat\eta]\right)^{-1/2} 
    \\
    &= S^{(1)}[\hat{\eta}] + \frac{i}{2} \mathrm{STr} \log \Omega^{(0)}[\hat\eta] \,,
\end{split}
\end{align}
where $S^{(1)}[\hat{\eta}]$ contains all local one-loop contributions; that is, in renormalizable theories $S^{(1)}[\hat{\eta}]$ contains only the counterterms required to renormalize the theory.
In EFTs the one-loop induced Wilson coefficients are included in~$S^{(1)}[\hat{\eta}]$, too.
We furthermore introduce the superdeterminant~($\mathrm{SDet}$) and the supertrace~($\mathrm{STr}$), which are generalizations of the determinant and trace to operators with mixed spin. The supertrace is a trace over all internal degrees of freedom and therefore involves an integration over all loop momenta
\begin{align}
    \mathrm{STr} \, \log \Omega[\hat\eta] &= \pm \int \frac{\mathrm{d}^D k}{(2\pi)^D} \langle k | \mathrm{tr} \, \log \Omega[\hat\eta] | k \rangle \,,
\end{align}
where $\mathrm{tr}$ denotes the regular trace over all internal degrees of freedom apart from momentum, and the sign depends on the spin of the considered field, with $+~(-)$ for bosonic (fermionic) states.
These operator traces can be evaluated using the so-called \textit{covariant derivative expansion} \cite{Gaillard:1985uh,Chan:1986jq,Cheyette:1987qz}. However, a~discussion of the supertrace evaluation is beyond the scope of this review and we refer to~\cite{Henning:2014wua,Cohen:2020fcu,Fuentes-Martin:2020udw} for further details.

Since the path integral formulation allows to compute the effective action, we can also use it to calculate the off-shell matching condition in Eq.~\eqref{eq:matching-condition-general}. At tree level we find
\begin{align}
    S^{(0)}_\mathrm{EFT}[\hat\eta_L] &= S^{(0)}_\mathrm{UV}[\hat\eta_L,\hat\eta_H] \,,
\end{align}
where we separated the fields into light~$\hat\eta_L$ and heavy~$\hat\eta_H$. The heavy background fields are understood as the solution to their equation of motion as a power series in~$1/M$, where $M$~is their mass, so that they can be entirely expressed in terms of the light fields $\hat\eta_H=\hat\eta_H[\hat\eta_L]$. 

Taking the Lagrangian~\eqref{eq:S1} of the $S_1$~leptoquark example, we find the equation of motion for~$S_1$
\begin{align}
    D^2 S_1 + M_{S}^2 S_1 - \lambda^{L\ast}_{pr} (\overline{\ell}_r \varepsilon q_p^c) + \lambda^{R\ast}_{pr} (\overline{e}_r u_p^c) &= 0 \,.
\end{align}
Its power series solution is given by
\begin{align}
    S_1 &= \frac{1}{M_{S}^2} \left[ \lambda^{L\ast}_{pr} (\overline{\ell}_r \varepsilon q_p^c) - \lambda^{R\ast}_{pr} (\overline{e}_r u_p^c) \right] + \mathcal{O}(M_{S}^{-4}) \,.
\end{align}
Substituting this solution back into the Lagrangian~\eqref{eq:S1} yields the same matching condition as in the diagrammatic computation shown in Eq.~\eqref{eq:operators_S1}. 

For the one-loop matching it is convenient to split the fluctuation operator into a kinetic and an interaction term
\begin{align}
    \Omega_{ij} &\equiv \delta_{ij} \Delta_i^{-1} - X_{ij} ,
    \ \text{with} \ 
    \Delta_i^{-1} = \left\{ \begin{matrix} -(D^2 + M_i^2) \\ i\slashed{D} - M_i \\ g^{\mu\nu} (D^2 + M_i^2) \end{matrix} \right. ,
\end{align}
for scalars, fermions, and vector bosons, respectively.
For simplicity we use the Feynman gauge for the quantum fluctuations of the gauge fields. 
This does not imply any particular choice for the gauge of the background fields, which remain in the general $R_\xi$~gauge \cite{Henning:2014wua}. For more details on gauge fixing the SMEFT in the background field method see \cite{Helset:2018fgq}. 
The interaction terms~$X_{ij}$ are implicitly defined by the above equation. This allows to write the one-loop effective action of the UV theory as
\begin{align}
    \Gamma_\mathrm{UV}^{(1)} &= \frac{i}{2} \mathrm{STr} \log \Delta^{-1} + \frac{i}{2} \mathrm{STr} \log (1-\Delta X) \,.
\end{align}
We can again apply the method of regions splitting $\Gamma_\mathrm{UV}^{(1)}$ into a hard and soft part, which are computed by expanding the loop integrands in the soft or hard momentum region, respectively. By construction we have $\Gamma_\mathrm{EFT}^{(1)}|_\mathrm{soft} = \Gamma_\mathrm{UV}^{(1)}|_\mathrm{soft}$, which ensures that both theories describe the same long distance dynamics. Therefore, we find the one-loop EFT Lagrangian to be given by $\int \mathrm{d}^ D x\,\mathcal{L}_\mathrm{EFT}^{(1)} = \Gamma_\mathrm{UV}^{(1)}|_\mathrm{hard}$, and thus
\begin{align}
    \int \!\!\mathrm{d}^D x\,\mathcal{L}_\mathrm{EFT}^{(1)} 
    &= \!\frac{i}{2} \mathrm{STr} \log \Delta^{-1} \bigg|_\mathrm{hard} \!\!\!\!\! + \!\frac{i}{2} \sum_{n=0}^\infty \frac{1}{n} \mathrm{STr} (\Delta X)^n \bigg|_\mathrm{hard}
    \label{eq:matching-master-formula}
\end{align}
where we expanded the logarithm in the latter term. This is the master formula for functional one-loop matching, expressing the EFT Lagrangian in terms of \textit{log-type} and \textit{power-type} supertraces. These can be evaluated using the covariant derivative expansion as discussed in \cite{Cohen:2020fcu,Fuentes-Martin:2020udw}. 
The main advantage of the functional formalism is that Eq.~\eqref{eq:matching-master-formula} directly yields all generated EFT operators and, unlike the diagrammatic approach, no a~priori knowledge of an operator basis is required. However, the Lagrangian obtained by Eq.~\eqref{eq:matching-master-formula} is in a non-minimal form, and redundant operator need to be removed to recover the EFT in a minimal basis.

A~computation for the $S_1$ example discussed before using functional methods is rather tedious and thus not discussed here, but further details can be found in \cite{Fuentes-Martin:2020udw,Dedes:2021abc}.\footnote{For the application of the functional matching formalism to other simple BSM theories see, e.g., \cite{Zhang:2021jdf,Li:2022ipc,Liao:2022cwh,Dittmaier:2021fls,Du:2022vso}.} 
However, it is a purely algebraic problem that can be solved by a computer. 
The Mathematica package \cmd{Matchete} \cite{Fuentes-Martin:2022jrf} is the first tool that fully automatizes the functional one-loop matching.\footnote{Earlier codes such as \cmd{STrEAM} \cite{Cohen:2020qvb} and \cmd{SuperTracer} \cite{Fuentes-Martin:2020udw} allow only to compute the supertraces, but do not perform the full matching computation. In particular they do not perform operator reductions on the resulting EFT Lagrangian. See also \cmd{MatchingTools} \cite{Criado:2017khh} for a pure tree-level matching implementation.}
Previously, the diagrammatic one-loop matching technique was already automated in the \cmd{MatchMakerEFT} \cite{Carmona:2021xtq} tool.
This greatly simplifies phenomenological BSM analyses and gives the possibility of validating matching results with different methods.
Another tool for one-loop matching is \cmd{CoDEx} \cite{DasBakshi:2018vni} using the universal one-loop effective action~(UOLEA) \cite{Drozd:2015rsp,Ellis:2016enq,Ellis:2017jns,Ellis:2020ivx,Kramer:2019fwz}, which is also based on the path integral approach explained above. A~more detailed discussion of the UOLEA technique is, however, beyond the scope of this review.

\subsection{Renormalization group evolution}
\label{sec:rge}
The Wilson coefficients of the SMEFT Lagrangian obtained from the matching are related to the UV parameters at the matching scale~$\mu_m$, usually taken at the mass threshold $\mu_m \sim M$. 
Next, we have to evolve the coefficients down to the electroweak scale~$(\sim \! m_W)$
using the SMEFT RG~equations. 
These have been computed at one loop for the dimension-six operators of the Warsaw basis shown in Tab.~\ref{tab:Warsaw-basis} in \cite{Jenkins:2013zja,Jenkins:2013wua,Alonso:2013hga}.
The RG equations of the baryon- and lepton-number violating operators listed in Tab.~\ref{Tab:Warsaw-basis_BV} have been derived in \cite{Alonso:2014zka} also including operators with right-handed neutrinos. The RG~equations for the dimension-five and -seven operators have been derived in \cite{Babu:1993qv,Davidson:2018zuo,Liao:2016hru,Liao:2019tep}, whereas for dimension eight only partial results are yet available \cite{Chala:2021pll,DasBakshi:2022mwk}. 
Results for specific sectors of 
the two-loop anomalous dimension matrix 
have been derived in \cite{Aebischer:2022anv,Bern:2020ikv}.
For some recent phenomenological analyses of the SMEFT RG~mixing effects see, e.g., \cite{Chala:2021juk,Kumar:2021yod,Isidori:2021gqe,Aoude:2022aro}. 
A~careful analysis of the flavor structure of the 2499-by-2499 anomalous-dimension matrix of the SMEFT is presented in \cite{Machado:2022ozb}.

An important feature of the RG~evolution is the mixing of different operator classes. 
In particular, an operator that is not generated by the matching can obtain a non-vanishing coefficient through the running. 
This leads to non-trivial relations among different operator types, that need to be carefully considered in a phenomenological analysis.

As an example, we consider the RG~evolution of the leptonic dipole operators in Eq.~\eqref{eq:QeB} and~\eqref{eq:QeW}, and the Yukawa interactions, that are described by
\begin{align}
    \mu\frac{\mathrm{d}}{\mathrm{d}\mu} [C_{X}]_{pr} &= \frac{1}{16\pi^2} [\beta_X]_{pr}
\end{align}
with the beta-functions given by
\begin{subequations}
\begin{align}
    [\beta_{eB}]_{pr} 
    &=  3 |y_t|^2 [C_{eB}]_{pr} - 10 g_1 y_t^\ast [C_{lequ}^{(3)}]_{pr33} \,, 
    \label{eq:CeB-RGE}
    \\
    [\beta_{eW}]_{pr} 
    &= 3 |y_t|^2 [C_{eW}]_{pr} + 6 g_2 y_t^\ast [C_{lequ}^{(3)}]_{pr33} \,,
    \label{eq:CeW-RGE}
    \\
    [\beta_{Y_e}]_{pr}
    &= 3 \lambda \frac{v^2}{\Lambda^2} \left( [C_{eH}]_{pr} - y_t^\ast [C_{lequ}^{(1)}]_{pr33} \right) \approx 0 \,,
    \label{eq:Y-RGE}
    \\
    [\beta_{eH}]_{pr}
    &= 9 |y_t|^2 [C_{eH}]_{pr} + 12 y_t^\ast |y_t|^2 [C_{lequ}^{(1)}]_{pr33} \,,
    \label{eq:CeH-RGE}
\end{align}%
\label{eq:SMEFT-beta-functions}%
\end{subequations}%
where for simplicity we only keep numerically relevant terms, i.e., top Yukawa~$(y_t)$ enhanced terms that are not multiplied by~$\lambda$.
Thus, we can write the Wilson coefficients at a low scale~$\mu_l$, in terms of the coefficients at the matching scale~$\mu_m$ with one-loop accuracy as
\begin{align}
    [C_{X}]_{pr} (\mu_l) &= [C_{X}]_{pr} (\mu_m) + \frac{1}{16\pi^2} \log\left(\frac{\mu_l}{\mu_m}\right) [\beta_{X}]_{pr} \,.
    \label{eq:SMEFT-one-loop-running}
\end{align}

The RG~evolution of the Warsaw basis operators is also automated in computer programs such as \cmd{DSixTools} \cite{Celis:2017hod,Fuentes-Martin:2020zaz} and \cmd{Wilson} \cite{Aebischer:2018bkb}, making a phenomenological analysis using the full 2499-by-2499 anomalous-dimension matrix of the $d=6$~SMEFT feasible.

\subsection{Low-energy constraints in the LEFT}
\label{sec:g-2_SMEFT}

Having discussed the matching of the BSM model defined in Eq.~\eqref{eq:S1}
onto the dipole operators $Q_{eB}$ and $Q_{eW}$, we now relate these to the photon dipole operator
\begin{align}
    [\mathcal{Q}_{e\gamma}]_{pr} = \frac{v}{\sqrt{2}} \overline{e}_p^L \sigma^{\mu\nu} e_r^R F_{\mu\nu} \,.
    \label{eq:photon-dipole}
\end{align} 
This allows us to illustrate how the low-energy constraints on this effective operators can be used 
for constraining the high-energy couplings of the $S_1$~field.

To this end, we write the SMEFT Lagrangian in the broken phase\footnote{Notice that for convenience we use here a different definition for the Yukawa and mass matrices compared to Eq.~\eqref{eq:LEFT-mass-Yukawa} . Moreover, for the dipole operators we directly apply the SMEFT instead of the LEFT power counting.}
\begin{align}
    \Delta\mathcal{L}^\mathrm{broken} 
    = &-[\mathcal{Y}_e]_{pr} \frac{v}{\sqrt 2} (\bar e_{p}^L e_{r}^R )
- [\mathcal{Y}_{he}]_{pr} \frac{h}{\sqrt{2}} (\bar e_{p}^L e_{r}^R )  
\nonumber\\
&+ \frac{[\mathcal{C}_{e\gamma}]_{pr}}{\Lambda^2} \frac{v}{\sqrt 2} (\bar e_{p}^L \sigma^{\mu\nu} e_{r}^R) F_{\mu\nu}
\\
&+ \frac{[\mathcal{C}_{eZ}]_{pr}}{\Lambda^2} \frac{v}{\sqrt 2} (\bar e_{p}^L \sigma^{\mu\nu} e_{r}^R) Z_{\mu\nu} + \ldots
\nonumber
\end{align}
Here, we also included the mass term, the Yukawa, and the $Z$-boson dipole, where the latter two are phenomenologically not relevant for the present analysis.

Assuming that new physics is not affecting the electroweak symmetry breaking pattern, 
i.e.~assuming the relations between quantities in the broken and unbroken phase are the same as in the SM (e.g. $\overline{g}_1=g_1$, $\overline{s}_\theta=s_\theta$, $v_T=v$~\ldots),
we can use the results presented in Sec.~\ref{sect:LEFT}
to relate the coefficients of the broken phase Lagrangian to the ones of the unbroken phase by
\begin{align}
\begin{pmatrix}
	 [\mathcal{C}_{e\gamma}]_{pr} \\[0.1cm]
     [\mathcal{C}_{eZ}]_{pr}
\end{pmatrix}
&=
\begin{pmatrix}
	c_\theta & -s_\theta \\[0.1cm]
	-s_\theta & -c_\theta
\end{pmatrix}
\begin{pmatrix}
	[C_{eB}]_{pr} \\[0.1cm]
	[C_{eW}]_{pr}
\end{pmatrix} \,, 
\label{eq:dipole-rotation}
\\[0.2cm]
\begin{pmatrix}
	[\mathcal{Y}_e]_{pr} \\[0.1cm]
	[\mathcal{Y}_{he}]_{pr}
\end{pmatrix} 
&=
\begin{pmatrix}
	1 & -\frac{1}{2} \\[0.1cm]
	1 & -\frac{3}{2}
\end{pmatrix}
\begin{pmatrix}
	[Y_e]_{pr} \\[0.1cm]
	\frac{v^2}{\Lambda^2}[C_{eH}]_{pr}
\end{pmatrix} \, ,
\label{eq:Yukawa-rotation}
\end{align}
where
\begin{align}
    c_\theta &= \frac{g_2}{\sqrt{g_1^2 + g_2^2}} = \frac{e}{g_1} \,, \quad  s_\theta = \frac{g_1}{\sqrt{g_1^2 + g_2^2}} = \frac{e}{g_2} \,.
\end{align}

We can now combine our results for the relations to the broken phase, shown in Eqs.~\eqref{eq:dipole-rotation} and~\eqref{eq:Yukawa-rotation}, with the RG~evolution equations above the electroweak scale in Eqs.~\eqref{eq:SMEFT-beta-functions} and~\eqref{eq:SMEFT-one-loop-running} to express the electromagnetic dipole and the mass Yukawa at the electroweak scale~$\mu_w$ in terms of the SMEFT Wilson coefficients at the new-physics/matching scale~$\mu_m\!\sim\!\Lambda$:
\begin{align}
\begin{split}
    [\mathcal{C}_{e\gamma}]_{pr} (\mu_w) 
    &= 
    \brackets{1 - 3 \hat{L} y_t^2 } [\mathcal{C}_{e\gamma}]_{pr} (\mu_m) 
    \\
    &\qquad + 16 \hat{L} y_t e \, [C_{lequ}^{(3)}]_{pr33} (\mu_m) \,,
\end{split}
\label{eq:physical_dipole_low}
\\[0.1cm]
    [\mathcal{Y}_e]_{pr} (\mu_w) 
    &= 
    \squarebrackets{Y_{e}}_{pr}(\mu_m) - \frac{v^2}{2 \Lambda^2} [C_{eH}]_{pr}(\mu_m) 
    \label{eq:physical_Yukawa_low}\\
    &\qquad + 6 \frac{v^2}{\Lambda^2} \hat{L} \left[ y_t^3 [C_{lequ}^{(1)}]_{pr33} +  \frac34 y_t^2 [C_{eH}]_{pr} \right]_{\mu_m} \!\!\!\! ,
    \nonumber
\end{align}
where we assume the Yukawa couplings to be real, and we define $\hat{L}\equiv (1/16\pi^2) \log(\mu_m/\mu_w)$.
We find that the semileptonic triplet operator~$\smash{Q_{lequ}^{\scriptscriptstyle (3)}}$ can generate the electromagnetic dipole~$\mathcal{Q}_{e\gamma}$ at the low scale, whereas the semileptonic singlet operator~$\smash{Q_{lequ}^{\scriptscriptstyle (1)}}$ as well as~$Q_{eH}$ run into the mass terms~$\mathcal{Y}_e$.

We can now investigate the RG evolution below the electroweak scale, which is given by \cite{Jenkins:2017dyc}
\begin{align}
    \mu \frac{\dd}{\dd\mu} [\mathcal{C}_{e\gamma}]_{pr} &= \frac{1}{16\pi^2} \frac{170}{9} e^2 [\mathcal{C}_{e\gamma}]_{pr} \,,
    \label{eq:Cegamma-RGE}
    \\
    \mu \frac{\dd}{\dd\mu} [\mathcal{Y}_{e}]_{pr} &= -\frac{1}{16\pi^2} 6e^2 [\mathcal{Y}_{e}]_{pr} \,,
    \label{eq:Ye-RGE}
\end{align}
where we consider all other operators to be turned off,\footnote{For the SMEFT, the only numerically relevant contributions in the running are due to the $y_t$~enhanced terms. In the LEFT, however, the top quark is integrated out and top loops cannot contribute, thus no such RG effects are present below the electroweak scale.}
and thus we only have the self renormalization of the dipole and the mass term, which leave the flavor structure unchanged. 
Notice also that the LEFT dipole operator in Eq.~\eqref{eq:photon-dipole} is a dimension-five operator, thus, in principle, we had to consider double insertions of this operator for the RG~evolution. 
However, from the matching conditions~\eqref{eq:S1-CeB} and~\eqref{eq:S1-CeW} we know that such contribution is of order~$\mathcal{O}(M_{S}^{-4})$ in the SMEFT power counting and can thus be neglected.
Equation~\eqref{eq:Cegamma-RGE} then allows to evolve the photon dipole to the low-energy scales of experimental measurements, which for muons is~$\mu_l\sim m_\mu$. Notice that in the present case it is not required to integrate out any other particles, such as the $b$~quark, since these do not affect the RG~evolution in good approximation due to their small Yukawa couplings.

Experimental measurements (usually) constrain couplings in the mass basis, whereas our Wilson coefficients are given in the generic flavor basis of the UV theory. Thus, rotating the fermion fields to the mass basis is the last missing piece of our analysis. To do this, we need to diagonalize the mass matrix $[\mathcal{Y}_e]_{pr}$ which is determined in terms of the SMEFT operators in Eq.~\eqref{eq:Yukawa-rotation}. Assume the mass term is diagonalized $\smash{(\mathrm{diag.} = U_L \, \mathcal{Y}_e \, U_{R}^\dagger)}$ when rotating the lepton fields~by
\begin{align}
    e_L^\prime = U_L \, e_L \,, \qquad e_R^\prime = U_R \, e_R \,,
\end{align}
where $U_{L,R}$ are unitary matrices and $e_{L,R}^\prime$ denote the mass-basis fields. Then the mass-basis dipole~$\mathcal{C}_{e\gamma}^\prime$ is given~by
\begin{align}
    \mathcal{C}_{e\gamma}^\prime &= U_L \, \mathcal{C}_{e\gamma} \, U_R^\dagger \,.
    \label{eq:Cegamma-mass-basis-rotation}
\end{align}

The most sensitive probe of this operator is the lepton flavor violating transition $\mu \to e \gamma$; however, also the anomalous magnetic moment of the muon $(g-2)_\mu$ is interesting, especially given the tension of the recent FNAL measurement \cite{Muong-2:2021ojo} with the SM prediction by \textcite{Aoyama:2020ynm}, summarized in Eq.~\eqref{eq:gm2exp}. 
For mere illustrative purposes, we take the latter result as reference input of our analysis, despite the recent doubts on its validity mentioned in Sec.~\ref{sect:gm2intro}.
Taking into account also the upper bound on the branching ratio $\mathcal{B}(\mu^+ \to e^+ \gamma)$ determined by the MEG experiment \cite{MEG:2016leq},
we can then write
\begin{align}
\begin{split}
    \mathcal{B}(\mu^+ \to e^+ \gamma) &= \frac{m_\mu^3 v^2}{8\pi \Gamma_\mu} \frac{\big|[\mathcal{C}_{e\gamma}^\prime]_{12}\big|^2 + \big|[\mathcal{C}_{e\gamma}^\prime]_{21}\big|^2}{\Lambda^4}
    \\
    &< 4.2 \times 10^{-13} \quad \text{(90\% CL)} \,,
\end{split}
    \\[0.2cm]
    \Delta a_\mu  
    &\equiv a_\mu^\mathrm{Exp} - a_\mu^\mathrm{SM} 
    = -\frac{4m_\mu}{e} \frac{v}{\sqrt{2}} \frac{\mathrm{Re}[\mathcal{C}_{e\gamma}^\prime]_{22}}{\Lambda^2}
    \nonumber\\
    &= (251 \pm 59) \times 10^{-11} \,,
\end{align}
which leads to
\begin{align}
    \left|\frac{[\mathcal{C}_{e\gamma}^\prime]_{12(21)}}{\Lambda^2}\right| &\lesssim 2.1 \times 10^{-10} \,\mathrm{TeV}^{-2} \,,
    \label{eq:Cegamma12-constraint}
    \\[0.1cm]
    \frac{\mathrm{Re} [\mathcal{C}_{e\gamma}^\prime]_{22}}{\Lambda^2} &\simeq -1.0 \times 10^{-5} \,\mathrm{TeV}^{-2} \,.
    \label{eq:Cegamma22-constraint}
\end{align}

We can now combine all our results: the low-energy constraints in Eqs.~\eqref{eq:Cegamma12-constraint}--\eqref{eq:Cegamma22-constraint}, the rotation to the mass basis~\eqref{eq:Cegamma-mass-basis-rotation}, the LEFT RG equations~\eqref{eq:Cegamma-RGE}--\eqref{eq:Ye-RGE}, the EWSB relations~\eqref{eq:dipole-rotation}--\eqref{eq:Yukawa-rotation}, the SMEFT running~\eqref{eq:SMEFT-beta-functions}, and the matching conditions~\eqref{eq:S1-CeB}--\eqref{eq:S1-CeW}, where the last three results have already been combined in Eq.~\eqref{eq:physical_dipole_low} and~\eqref{eq:physical_Yukawa_low}.\footnote{Notice that we have chosen $\xi_\mathrm{rp}=1$ for convenience, which fixes the NDR reading point that has to be used in all consecutive EFT calculations.}

\begin{figure}[t]
    \centering
    \includegraphics[width=0.8\linewidth]{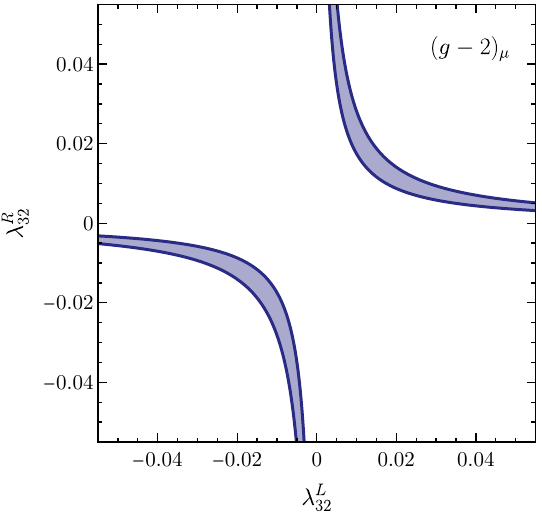}
    \\[0.2cm]
    \includegraphics[width=0.8\linewidth]{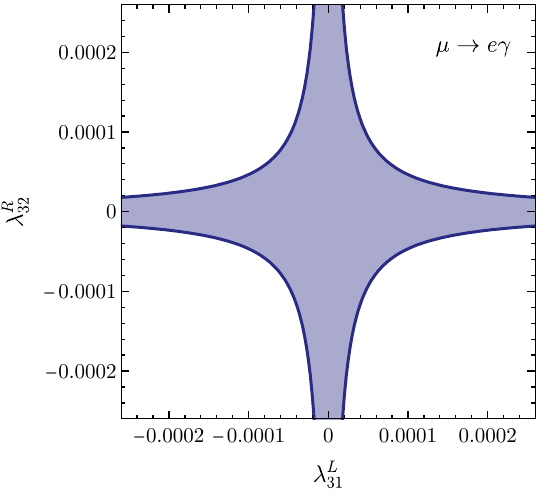}
    \caption{Constraints on the $S_1$~leptoquark couplings derived from the measurements of the $\mu \to e \gamma$ transition (upper plot), and from the $(g-2)_\mu$ measurement (lower plot). The leptoquark mass is chosen as $M_{S}=2\,\text{TeV}$, and only top-Yukawa enhanced contributions are considered in the numerical analysis. See text for more details.}
    \label{fig:S1-constraints}
\end{figure}

For simplicity we also consider $[C_{eH}]_{pr}=0$, which holds at tree level in the considered $S_1$~model. 
We also assume $Y_e$ to be diagonal such that the mass matrix is already diagonal and we can set~$U_{L,R}=\mathds{1}$. 
Notice that this is a strong assumption on a marginal operator appearing in the~UV, and in general we have to consider rotation matrices~$U_{L,R} \neq \mathds{1}$. 
The resulting constraints on the $S_1$~couplings, assuming these are real quantities, are shown in Fig.~\ref{fig:S1-constraints}, where we set the leptoquark mass to~$M_{S}=2\,\text{TeV}$. In the upper plot, the constraints derived from the $\Delta a_\mu$~measurement are shown, whereas the lower plot shows the constraints from the $\mu \to e \gamma$ decay, where we set $\lambda^R_{31}=0$ for simplicity. Also couplings to quarks other than the top are neglected as these are not $y_t$~enhanced. 

As can be seen, the scales of the two figures are very different, signaling that underlying models 
able to explain the $(g-2)_\mu$ anomaly, while being consistent with $\mu \to e \gamma$, require a 
peculiar flavor-alignment mechanism.
A~more detailed phenomenological analysis of the given model and a discussion of the implied flavor structure can be found in \cite{Isidori:2021gqe}, see also \cite{Aebischer:2021uvt}.

There are also tools automating large parts of such analysis. For example the \cmd{flavio}~\cite{Straub:2018kue} package has a large set of low-energy measurements implemented that can be used to constrain Wilson coefficients. Also the SMEFT to LEFT matching as well as the RG~evolution in both ETFs is available in the code [trough the \cmd{Wilson} package \cite{Aebischer:2018bkb}; see also \texttt{DsixTools} \cite{Celis:2017hod}]. A~global likelihood based on the data available in \cmd{flavio} can be constructed with the \cmd{smelli} package \cite{Aebischer:2018iyb}, which can simplify analyses.

\subsection{\texorpdfstring{SMEFT at high-$p_T$}{SMEFT at high-pT} and global fits}
While the SMEFT (in combination with the LEFT) is very practical to relate low-energy measurements to UV parameters, it can also be used to analyze measurements from higher energies in a model independent way. This makes it a powerful tool for combined analyses of multiple data sets from various types of processes at different energy scales. This is in particular advantageous in light of the plethora of measurements of different processes performed at LHC and~LEP. We can use the SMEFT for phenomenological analyses of all these observables in 
Higgs \cite{Ellis:2014dva,Corbett:2015ksa,Corbett:2012ja}, 
Di-boson \cite{Grojean:2018dqj,Gomez-Ambrosio:2018pnl,Butter:2016cvz,Biekoetter:2018ypq}, 
and top physics \cite{Hartland:2019bjb,Aoude:2022deh,Aoude:2022aro,Brivio:2019ius}, 
as well as for electroweak precision studies \cite{Falkowski:2019hvp,Han:2004az,Breso-Pla:2021qoe,Efrati:2015eaa,Falkowski:2014tna,Almeida:2021asy}, 
and Drell-Yan tails \cite{Allwicher:2022gkm,Greljo:2022jac}. 
Global fits considering multiple of the above data sets have been performed, e.g., in \cite{Ellis:2018gqa,Ellis:2020unq,Ethier:2021bye,daSilvaAlmeida:2018iqo}, see also \cite{Dawson:2020oco}.
Such combined analyses of different types of data are necessary since in any reasonable new-physics model multiple SMEFT operators are generated when integrating out the heavy particles \cite{Jiang:2016czg}. 
These operators can contribute to different processes that can be probed at various energies. 
Also RG~mixing can generate further operators contributing to even more processes. 
Therefore, to carefully evaluate the plausibility of a given BSM theory, it is not enough to look at only a single measurement, but we have to perform a global SMEFT fit.

One of the main challenges for these fits are the large number of free parameters in the SMEFT. 
Thus, one has to apply some simplifying assumptions to reduce the degrees of freedom in a fit. 
For example, one can decide only to look at a specific set of operators that is particularly relevant for a given set of observables (e.g.~those involving only top and bottom quarks, and electroweak gauge bosons).
Moreover, one can apply some flavor symmetry assumptions as discussed in Sec.~\ref{sect:Flavor}. 
As shown in Tab.~\ref{tab:U3new}, the latter allow to significantly lower the number of parameters that we have to fit, while still allowing to describe the SM flavor structure in good approximation.

\begin{figure*}[t]
\begin{center}
\includegraphics[width=1.0\linewidth]{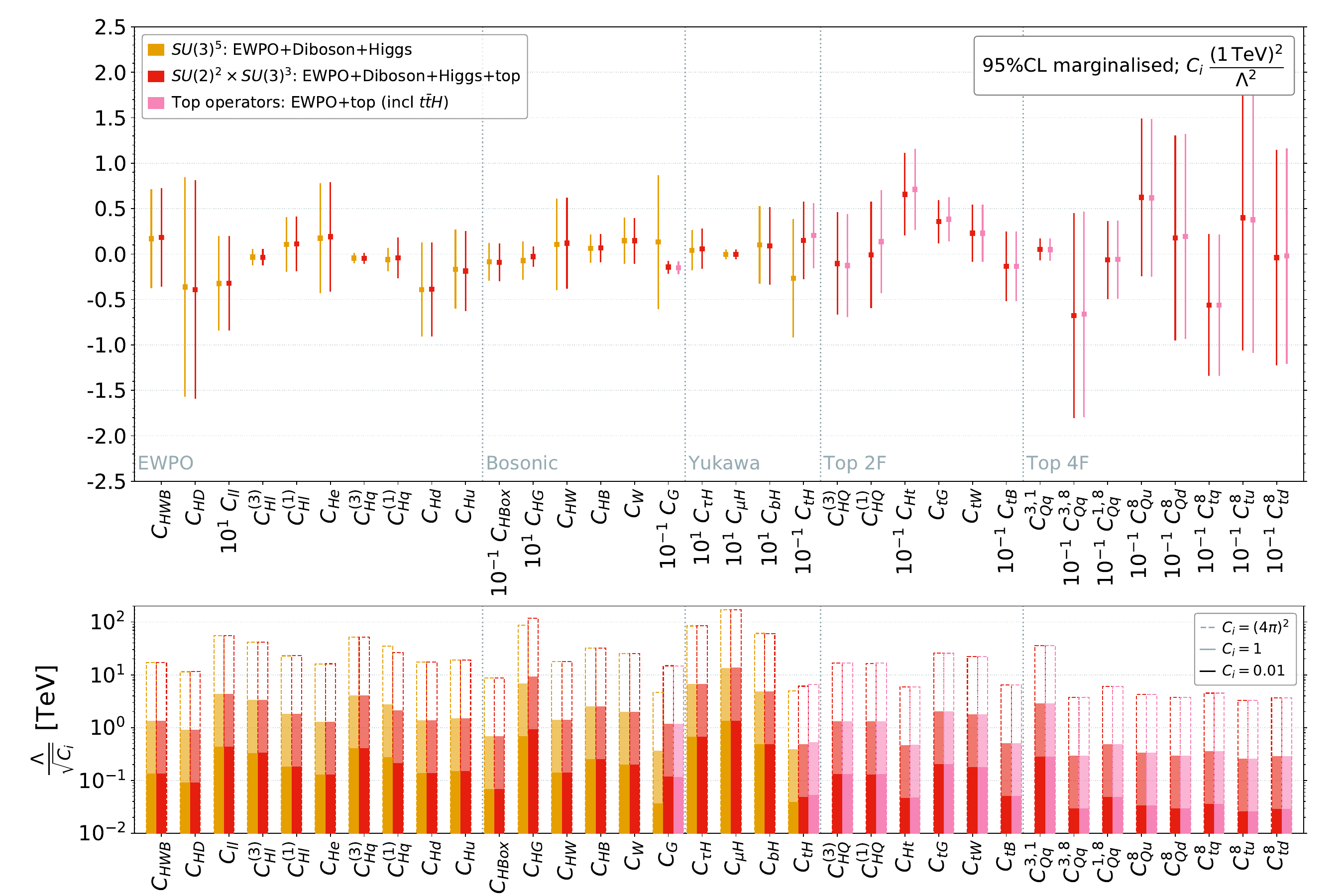}
\caption{Bounds on SMEFT effective coefficients as obtained by~\textcite{Ellis:2020unq}.
The top panel indicates the bounds on the 
coefficients assuming 
a reference effective scale of 1~TeV.
The corresponding bounds on the effective scales, for different reference 
hypotheses for the Wilson coefficients,
are shown in the bottom panel. 
The light yellow points are obtained in the 
$U(3)^5$ symmetric limit. The remaining points are obtained 
employing the  $U(2)^3\times  U(2)_u \times U(2)_q$ flavor symmetry,
which allow us to treat separately top-physics observables.} 
\label{fig:Globalfits}
\end{center}
\end{figure*}

On the one hand, if experimental data show deviations from the SM predictions, global fits are essential to determine the best-fit values of all relevant Wilson coefficients in order to be simultaneously compatible with multiple possibly correlated measurements.
On the other hand, if no clear signal for new physics is present in the data, global fits only allow us to put upper bounds on the coefficients. In general, the constraints obtained  
depend on the assumptions entering the fit. Since a truly global fit with all 2499~parameters of the $d=6$~SMEFT is unfeasible, a selection of certain operators, e.g., by choosing a specific flavor symmetry, has to be made. Therefore, one should keep in mind that 
the results of the  simplified fit cannot necessarily be applied to generic BSM scenarios.

As an illustration of the present 
state of the art of global fits, in Fig.~\ref{fig:Globalfits} we report the results of one of the most updated and extensive global analysis of SMEFT coefficients~\cite{Ellis:2020unq}.
The results are obtained considering all the relevant 
operators constrained by 
electroweak precision observables, 
di-boson processes, and top-physics measurements from the LHC.
The flavor symmetries 
$\mathrm{U}(3)^5$ or $\mathrm{U}(2)^3\times \mathrm{U}(2)^2$
are employed  (see  Sec.~\ref{sect:Flavor}).
The results for each Wilson coefficient are obtained marginalizing over the remaining ones.
Despite not fully generic, the number
of independent coefficients varied at the same time is quite impressive. 
One the most important message emerging from this analysis is that, under
motivated flavor-symmetry assumptions, present data are compatible with an effective cutoff scale for the SMEFT in the few-TeV domain.

\subsubsection{Drell-Yan tails}
\label{sec:Drell-Yan}
\begin{figure}[t]
    \centering
    \includegraphics[width=0.95\linewidth]{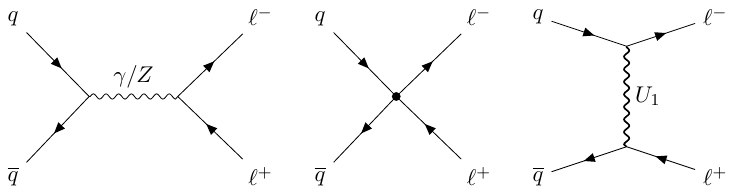}
    \caption{Tree-level Feynman diagrams contributing to Drell-Yan in the SM~(left), the SMEFT~(center), and the $U_1$~leptoquark model~(right).}
    \label{fig:Feynman-diagrams-Drell-Yan}
\end{figure}

In this section we analyze in detail the specific case of the Drell-Yan process $p p \to \ell^+ \ell^-$, which represents a good example of a high-energy transition  constraining SMEFT  Wilson coefficients. 
In the SM this process is mediated by the photon and the $Z$-boson, whereas in the SMEFT the dominant contributions are given by four-fermion operators~$(\psi^4)$, dipole operators~$(\psi^2 X H)$, and operators modifying the $Z$-boson couplings~$(\psi^2 H^2 D)$. 
The relevant tree-level Feynman diagrams are shown in Fig.~\ref{fig:Feynman-diagrams-Drell-Yan}. 
The plot on the left-hand side shows the SM contribution and the center plot the contribution by $\psi^4$~contact interactions. 
The diagram for the $(\psi^2 X H)$ and $(\psi^2 H^2 D)$ operators is similar to the SM diagram with the SM interaction vertices replaced by the respective SMEFT interactions. 
The dominant contribution depends on the energy range we are investigating. 
The operators modifying the $Z$-boson couplings can be best probed at the $Z$-pole, i.e., for invariant masses of the dilepton system of around~$m_{\ell\ell} \sim m_Z$. 
At higher energies the four-fermion contact interaction yield the dominant contribution since their amplitude is energy enhanced compared to the~SM. 
This is what allows us to probe effects due to the exchange of resonances 
with a mass even above the center-of-mass energy of the collider, as pointed out by \textcite{Greljo:2017vvb}.

In the following, we will focus on the high-$p_T$ constraints on $\psi^4$~operators involving mainly third-generation fermions, which have received considerable interest in the recent literature \cite{Allwicher:2022gkm,Allwicher:2022mcg,Greljo:2022jac,Angelescu:2020uug,Fuentes-Martin:2020lea,Faroughy:2016osc,Dawson:2018dxp,Greljo:2018tzh,Endo:2021lhi,Marzocca:2020ueu,Jaffredo:2021ymt,Boughezal:2023nhe,Boughezal:2021tih,Boughezal:2022nof,Alioli:2020kez}. 
We will also consider measurements of low-energy meson decays that are mediated by the same effective operators. 
Therefore, we can utilize the SMEFT framework to combine these complementary high-$p_T$ and low-energy constraints to asses the validity of a given BSM scenario.
The first analysis of this type, focused of light-generation four-fermion operators, has been presented in \cite{Cirigliano:2012ab}.

In this example, we consider the $U_1$~vector leptoquark contributing to Drell-Yan (see the diagram on the right-hand side of Fig.~\ref{fig:Feynman-diagrams-Drell-Yan}) and to charged-current semileptonic $B$-meson decays with the underlying $b \to c \tau \nu$ transition. We will follow the discussion laid out in \cite{Aebischer:2022oqe}.
The example is particularly interesting due to deviations currently observed in these low-energy decays, known as the $B$-anomalies, mentioned already in Sec.~\ref{sec:LFUV-intro}. We are especially interested in the lepton-flavor-universality ratios~$R_{D^{(\ast)}}$ defined in Eq.~\eqref{eq:RD-ratio} currently showing a $3.1\,\sigma$ discrepancy with the SM expectation \cite{HFLAV:2022pwe}.\footnote{Notice that while the fate of this anomaly, as for any anomaly, is unclear, the discussion presented here still remains an illustrative example of a SMEFT analysis.}

Consider the $U_1$~Lagrangian
\begin{align}
    \mathcal{L}_{U_1}
    &= \mathcal{L}_\mathrm{SM} -\frac{1}{2} U_{\mu\nu}^\dagger U^{\mu\nu} + M_{U}^2 U_\mu^\dagger U^\mu + \left( U_\mu J^\mu + \mathrm{h.c.} \right) \,,
    \\
    J^\mu &= \frac{g_U}{\sqrt{2}} \left[ \beta^L_{pr} \left( \overline{q}_p \gamma^\mu \ell_r \middle)  + \beta^R_{pr}  \middle(\overline{d}{}_p \gamma^\mu e_r \right) \right] \,.
\end{align}
We now integrate out the $U_1$ at tree-level using its equation of motion $\smash{U_\mu = -J_\mu^\dagger \big/ M_{U}^2 + \mathcal{O}(M_{U}^{-4})}$, we find 
\begin{align}
    \mathcal{L}_\mathrm{EFT} &= \mathcal{L}_\mathrm{SM} - \frac{1}{M_{U}^2} J_\mu^\dagger J^\mu \,.
\end{align}
Then, using the Fierz identities in Eqs.~\eqref{eq:Fierz-ids} and~\eqref{eq:SUN-Fierz}
we find the EFT Lagrangian in the Warsaw basis 
\begin{align}
\begin{split}
    \mathcal{L}_\mathrm{W}
    &= \mathcal{L}_\mathrm{SM} -\frac{g_U^2}{2 M_{U}^2} \bigg\{ \frac{1}{2} \beta^L_{pr} \beta^{L\ast}_{st} \left( [Q_{lq}^{(1)}]_{trps} + [Q_{lq}^{(3)}]_{trps}  \right)
    \\
    & + \beta^R_{pr} \beta^{R\ast}_{st} [Q_{ed}]_{trps} - \left( 2 \beta^R_{pr} \beta^{L\ast}_{st} [Q_{ledq}]_{trps} + \mathrm{h.c.} \right) \! \bigg\} \,.
\end{split}
\label{eq:U1-Warsaw}
\end{align}
Notice that since we restrict our analysis to the tree level, we do not have to consider evanescent contributions here.

This Lagrangian provides the appropriate description for interactions at energies above the electroweak scale but below~$M_{U}$. Thus, we can use it to describe the tails of Drell-Yan distributions where we consider events with~$200\,\text{GeV} \lesssim m_{\ell\ell} \lesssim M_{U}$. For a discussion of the EFT validity in the case where the EFT cutoff scale~$M_{U}$ is not sufficiently high, see the end of this section and \cite{Allwicher:2022gkm}.

The event yield~$\mathcal{N}$ in a given bin of the measured $m_{\ell\ell}$~distribution can then be schematically written as
\begin{align}
    \mathcal{N} &= \mathcal{L}_\mathrm{int} \, (\mathcal{A}\times\epsilon) \int_{m_{\ell\ell,\,\mathrm{min}}^2}^{m_{\ell\ell,\,\mathrm{max}}^2} \mathrm{d}s \, \frac{\mathrm{d}\sigma}{\mathrm{d}s} \,,
\end{align}
where $\mathcal{L}_\mathrm{int}$ is the integrated luminosity and $(\mathcal{A}\times\epsilon)$ parametrizes the acceptance and efficiency of the detector and has to be extracted using Monte Carlo simulations. The cross section~$\sigma$ is computed as a function of the Wilson coefficients or new-physics couplings, thus allowing to constrain these. For more details see \cite{Allwicher:2022gkm}. The event yields can also be automatically extracted using codes like \cmd{HighPT} \cite{Allwicher:2022mcg} or \cmd{flavio} \cite{Greljo:2022jac}.

The operators in Eq.~\eqref{eq:U1-Warsaw} contribute also to low-energy processes, of course. In particular, $\smash{[Q_{lq}^{(3)}]_{3323}}$ and $\smash{[Q_{ledq}]_{3332}}$ can contribute to the $b \to c \tau \nu$ transitions that we are interested in.
The relevant low-energy Lagrangian can be written as
\begin{align}
\begin{split}
    \mathcal{L}_{b \to c}
    = -\frac{4 \, G_F}{\sqrt{2}} V_{23} \Big[ &\left( 1 + \mathcal{C}_{LL}^c \right) \left( \overline{c}_L \gamma^\mu b_L \middle) \middle( \overline{\tau}_L \gamma_\mu \nu_L \right)  
    \\
    &- 2\,\mathcal{C}_{LR}^c \left( \overline{c}_L b_R \middle) \middle( \overline{\tau}_R \nu_L \right) \Big]
\end{split}
\end{align}
where $G_F$ is Fermi's constant and $V_{23}=V_{cb}$ is a CKM matrix element.
The coefficients are related to the Warsaw basis Wilson coefficients by
\begin{align}
    \mathcal{C}_{LL}^c
    &= -\frac{1}{\sqrt{2} G_F} \frac{1}{M_{U}^2} \sum_{k=1}^3 \frac{[C_{lq}^{(3)}]_{33k3} V_{2k}}{V_{23}} \,,
    \\
    \mathcal{C}_{LR}^c
    &= \frac{1}{4\sqrt{2} G_F} \frac{1}{M_{U}^2} \sum_{k=1}^3 \frac{[C_{ledq}^\ast]_{333k} V_{2k}}{V_{23}} \,,
\end{align}
where we assume that the flavor basis of the new physics is given by the down-quark and charged-lepton mass basis so that we can write
\begin{align}
    q_p = \! \begin{pmatrix}
        V_{rp}^\ast u_r^L \\[0.05cm] d_p^L
    \end{pmatrix}\!,
    \, 
    u_p = u_p^R ,
    \ 
    d_p = d_p^R ,
    \ 
    \ell_p = \! \begin{pmatrix}
        \nu_p^L \\[0.05cm] e_p^L
    \end{pmatrix}\!,
    \ 
    e_p = e_p^R . 
\end{align}

Following \cite{Cornella:2021sby}, we can express the LFU ratios~$R_{D^{(\ast)}}$ in terms of these parameters as 
\begin{align}
    \frac{R_D}{R_D^\mathrm{SM}} &= \left| 1 \! + \mathcal{C}_{LL}^c \right|^2 \! - 3.0 \mathrm{Re} \big[\! \left( 1 \! + \mathcal{C}_{LL}^c \right) \mathcal{C}_{LR}^{c\ast} \big] + 4.12 \left| \mathcal{C}_{LR}^c \right|^2  \!,
    \\
    \frac{R_{D^\ast}}{R_{D^\ast}^\mathrm{SM}} &= \left| 1 \! + \mathcal{C}_{LL}^c \right|^2 \! - 0.24 \mathrm{Re} \big[\! \left( 1 \! + \mathcal{C}_{LL}^c \right) \mathcal{C}_{LR}^{c\ast} \big] + 0.16 \left| \mathcal{C}_{LR}^c \right|^2  \!\!.
\end{align}
As numerical input we use the world average for the experimental measurements and SM predictions for these observables as provided by the HFLAV collaboration in \cite{HFLAV:winter2023,HFLAV:2022pwe}, respectively:
\begin{align}
    R_{D} &= 0.356 \pm 0.029 \,,
    &
    R_{D}^\mathrm{SM} &= 0.298(4) \,,
    \\
    R_{D^\ast} &= 0.284 \pm 0.013 \,,
    &
    R_{D^\ast}^\mathrm{SM} &= 0.254(5) \,.
\end{align}

The LEFT beta-functions of the coefficients are given by \cite{Jenkins:2017dyc}
\begin{align}
    \beta_{\mathcal{C}_{LL}^c} &= -4 e^2 \mathcal{C}_{LL}^c \,, 
    & 
    \beta_{\mathcal{C}_{LR}^c} &= \left( \frac{4}{3} e^2 - 8 g_3^2 \right) \mathcal{C}_{LR}^c \,.
\end{align}
We use the LEFT RG~equations\footnote{The dominant contribution is due to the strong coupling constant~$\alpha_s=g_3^2/4\pi$, which runs as $\alpha_s(\mu)=\frac{4\pi}{\beta_0 \ln(\mu^2/\Lambda_\mathrm{QCD}^2)}$ at one loop, with the one-loop QCD beta-function~$\beta_0$.} to directly run the low-energy coefficients from the scale~$\mu \sim m_b$ up to $\mu = 1\,\text{TeV}$, which is the appropriate scale for measurements of the high-$p_T$ Drell-Yan tails at LHC. There we directly match to the SMEFT and neglect the SMEFT running in good approximation since it only yields a small logarithmic contribution.

\begin{figure}[t]
    \centering
    \includegraphics[width=0.85\linewidth]{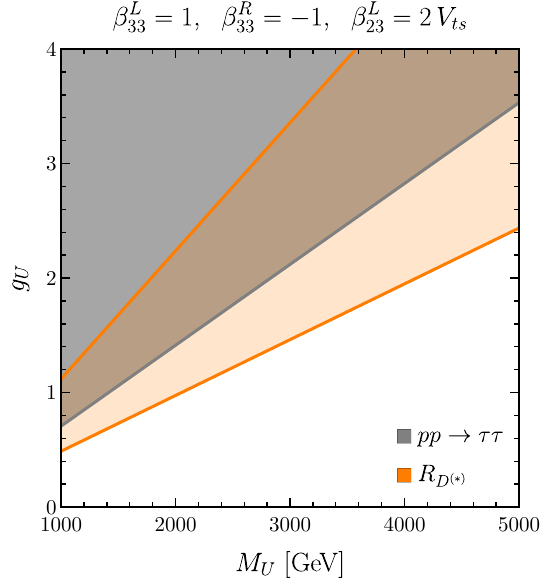}
    \caption{Constraints on the $U_1$~model in the coupling~$g_U$ versus mass~$M_{U}$ plane. Shown in light orange is the region preferred by the low-energy fit of the $R_{D^{(\ast)}}$~anomalies, and in dark gray we show the parameter space excluded by the ATLAS search \cite{ATLAS:2020zms} for new physics in $pp \to \tau\tau$ scatterings. Both constraints are given at $95\%$~CL.}
    \label{fig:U1-fit}
\end{figure}

To perform the combined fit of the high-$p_T$ Drell-Yan data and the low-energy measurements of~$R_{D^{(\ast)}}$, we assume that all couplings except for $\beta_{33}^{L/R}$ and $\beta_{23}^{L}$ vanish, i.e., the~$U_1$ couples dominantly to the third generation. Furthermore, we choose to set $\beta_{33}^L=-\beta_{33}^R=1$ and $\beta_{23}^L=2V_{ts}$,
adopting the hypothesis of a minimal breaking of the flavor symmetry  \cite{Aebischer:2022oqe}.
The combined constraints on the $U_1$~model in the coupling versus mass plane are shown in Fig.~\ref{fig:U1-fit}. We used the \cmd{HighPT} package \cite{Allwicher:2022mcg} to derive the constraints from the Drell-Yan search for new physics in $pp \to \tau\tau$ scattering by the ATLAS collaboration \cite{ATLAS:2020zms}. The $95\%$~CL region preferred by our low-energy constraint discussed above is shown in light orange, whereas the region excluded at $95\%$ by LHC is shown in dark gray. In combination, only a fraction of parameter space is left viable, thus showing the complementarity of the low- and high-energy constraints.\footnote{Interestingly enough, CMS data currently indicates a $3\sigma$ excess of events in $pp \to \tau\bar\tau$, well compatible 
with a possible $U_1$ contribution in this parameter region~\cite{CMS:2022zks}. }
For more details on this analysis see \cite{Aebischer:2022oqe,Cornella:2021sby}.

In the case of very low masses of the leptoquark $(M_U \sim 1\,\text{TeV})$ one might question the validity of the EFT approach to Drell-Yan measurements, since the kinematical distributions contain events with corresponding center-of-mass energies~$\sqrt{s}$ of the same order. 
Therefore the EFT expansion in~$s/M_U^2$ can converge poorly or even break down. To improve the convergence one can include higher-dimensional operators.  
We can either fit them as additional free parameters, marginalize over them,\footnote{Notice that when marginalizing over $d=8$ operators no correlation among the $d=6$ and $d=8$ operators is assumed, which is not true in concrete BSM scenarios. In particular the interference of $d=6$ and $d=8$ operators with the SM amplitude are allowed to have opposite sign, leading to cancellations.} or we can match them to the parameters of a given UV model, such as the $U_1$~leptoquark, depending on the scenario we are considering.
If we are too close to the mass threshold of the heavy BSM states there might be no way to analyze the high-energy data apart from using a concrete UV model. However, in this case the model independence of the EFT approach might be less important as the signal for a concrete new-physics model should be stronger.
A~short discussion of the EFT validity in Drell-Yan tails can be found in \cite{Allwicher:2022gkm}. For more details see also Sec.~\ref{sect:dim8} and \cite{Brivio:2022pyi}.

\section{Conclusion}
The Standard Model has set a natural and successful framework for the qualitative and quantitative understanding of the elementary particles and their interactions. It has been possible to calculate its predictions with enormous precision, allowing comparison with a similar progress on the experimental side. 
On the other hand, as already stated in the introduction, there is a number of observational and theoretical issues with the SM, such as neutrino masses, baryon asymmetry, a natural bridge to gravity, and the instability of the Higgs quadratic term. 
This is why it is widely believed, and we share this point of view, that the SM is the remnant of a more complete theory with new degrees of freedom showing up at some higher energy scale. By this statement we do not imply there cannot be also other light states beyond the SM ones, but rather that the SM fields are embedded into a more complete QFT with heavy fields in the UV, addressing many of the currently open issues.

The outstanding agreement between experiment and theory, that in various cases reach the sub-percent level, suggests that the energy scale where new heavy particles will appear, and the SM will manifestly become an incomplete description of nature, is well above the electroweak scale. This fact does not prevent the observation of effects related to the new degrees of freedom in current and near-future experiments. However, these effects will be indirect manifestation of new physics, and their interpretation in terms of hypothetical new dynamics require a suitable effective theory approach.

In this article we review the EFT approach to physics beyond the SM, focusing in particular on the linear realization of the mechanism of electroweak symmetry breaking, 
i.e., the SMEFT. Given all measurements of the 125\,GeV scalar particle discovered at the LHC are consistent with the properties expected for the SM Higgs boson, the SMEFT emerges as most natural EFT approach to physics beyond the~SM. In Sec.~\ref{sect:SMEFT} we extensively reviewed the construction of the basis of effective operators, the power counting, and various other technical aspects of this EFT. In Sec.~\ref{sec:HEFT} we also illustrated the more general approach represented by the~HEFT, or the possibility of a non-linear realization of the mechanism of electroweak symmetry breaking. An option that, despite being not favored by current data, cannot be excluded at present. 

An important role in effective field theories is played by exact and approximate symmetries
emerging in the low-energy limit of the theory, the so-called accidental symmetries.
We extensively reviewed this aspect in Sec.~\ref{sec:GlobalSymmetries}, focusing in particular on flavor symmetries, which represent the vast majority of possible global symmetries in the SMEFT. As we argued, in the absence of flavor symmetries the SMEFT approach is not particularly useful: severe bounds from flavor-violating observables would imply a very high scale of new physics, rendering the whole construction not particularly appealing. 
On the other hand, with the help of motivated hypotheses about a symmetry and symmetry-breaking, 
resulting from general dynamical hypothesis in the~UV, it is possible to consistently reduce the bounds on the new-physics scales and provide an {\em a~posteriori} justification for the observed mass hierarchies. 
In this theoretically motivated limit, we can both 
reduce the number of free parameters of the SMEFT, and
combine information from flavor-changing and flavor-violating processes.

In Sec.~\ref{sect:LEFT} and, especially, in Sec.~\ref{sect:practical} we show in detail the techniques used to put the SMEFT at work in analyzing data and possibly extracting information about physics beyond the~SM. These involve a large array of theoretical concepts and methods developed over the last decades, which we bring together here. From the use of low-energy effective theories valid below the electroweak scale, to the running of the SMEFT, and finally the 
matching to explicit beyond-the-SM theories. We reviewed various technical aspects of this workflow, both in a bottom-up perspective as well as in top-down approaches. 
We expect that the remarkable progress of these calculations will continue over the immediate future.

The SMEFT is already a mature subject and many studies exist in the literature, 
including excellent reviews such as the one by \textcite{Brivio:2017vri}.
However, most of the existing studies are focused mainly on the use of this tool in setting bounds on possible new-physics scenarios. 

In this review we emphasized the advantage of using the SMEFT 
in case of a “positive” signal of new physics.  While new-physics bounds can efficiently be set, in many cases, directly at the level of the observables, the full power of the EFT approach manifests itself in presence of a new-physics signal. In this case the SMEFT, being a consistent QFT, allows us to connect a signal in one observable to those in other processes and possibly recognize the underlying origin of the new dynamics.
We illustrated this chain via two specific examples in Sec.~\ref{sect:practical}, inspired by `anomalies' (i.e.~deviations from the SM predictions) present in current data: the $(g-2)_\mu$ anomaly and the deviation from lepton-flavor universality in $b\to c\ell\nu$ decays. While none of these effects is statistically compelling, we have analyzed them in detail since they provide a very clear and rather general illustration of the power of the EFT approach. 

This leads us to the important question of how to design a strategy for future experiments and where to focus theoretical work. 
A~general analysis of all experimental results, aiming at a global fit to all the 2499 SMEFT dimension-six coefficients, is neither a viable nor a particularly useful option. 
It is hardly feasible because of the huge dimension of the parameter space, while also being not especially illuminating, given that in realistic models only a subset of the operators play a relevant phenomenological role. 
We believe that a more purposeful strategy is to work out the main features of representative classes of models as UV~conditions on the SMEFT, correspondingly identify the relevant subsets of operators, and then proceed in the comparison with experiments.
As discussed in Sec.~\ref{sect:practical}, the new generation of automated tools for the matching, RG~evolution, and computation of experimental observables in the involved EFTs make such an approach feasible.
An important role in the data--theory comparison is also played by formulating hypotheses on flavor symmetries and corresponding symmetry-breaking terms.
These symmetries not only reduce the number of relevant free parameters, but they also allow us to consider more compelling new-physics scenarios in the few-TeV energy range which can be probed directly by current and near-future experiments, as shown in Fig.~\ref{fig:flavorbounds}.

Concerning experimental work, a fruitful direction is to investigate possible differences between  HEFT and SMEFT. As discussed at the end of Sec.~\ref{sect:HEFT1}, progress has been made in constructing UV~models which cannot be described by the SMEFT. Correspondingly, some experimental signatures that would signal a breakdown of the SMEFT description have also been identified. 
We believe that a~comprehensive strategy for how to distinguish the two effective theories could lead to meaningful results in the near future.

The applicability of the SMEFT rests on the validity of the effective theory approach. This itself relies on the hypothesis of having identified all degrees of freedom and symmetries
relevant at low energies. 
In this respect, the wide class of SM extensions with light new degrees of freedom, such as axions or axion-like particles, is not entirely covered by the SMEFT as described here. 
In such models we can imagine that the BSM physics produces two low-energy sectors, one of which is the SM and another one in the world of light particles (such as axions). 
These two sectors are necessarily weakly coupled to each other and generate low-energy axion physics. 
In this sense, the SMEFT is part (probably the major part) of a larger effective theory. 
The inclusion of additional light particles is conceptually simple once the symmetry properties of the new fields are specified, see e.g. \cite{Bauer:2020jbp,Agrawal:2021dbo,Galda:2021hbr}.

More generally, EFT approaches are based on the concept of scale separation, a key paradigm which guided the progress in particle physics for several decades by now. 
The absence of TeV-scale new physics, as expected by na\"ive EFT considerations, has stimulated theorists to consider alternatives to this paradigm. 
See for instance \cite{Giudice:2017pzm}.
While this is certainly an interesting possibility, we believe that our knowledge of TeV-scale physics is still far from being complete. 
The possibility of new physics just around the corner of the current energy and precision frontiers remains an extremely motivating option, and the SMEFT represents the most suitable tool to analyze it.

\begin{acknowledgments}
We are grateful to Ilaria Brivio, Wilfried Buchm\"uller, and Peter Stoffer, for very useful comments on the manuscript. 
DW thanks the High Energy Physics group (CHEP) at the Indian institute of science for hospitality.
This project has received funding from the European Research Council~(ERC) under the European Union's Horizon~2020 research and innovation programme under grant agreement 833280~(FLAY), and by the Swiss National Science Foundation~(SNF) under contract~200020\_204428.
Figures~\ref{fig:pulls2} and~\ref{fig:Globalfits} are reprinted here under the conditions of the licences 
\href{https://creativecommons.org/licenses/by-nc-nd/4.0/}{CC BY-NC-ND 4.0}
and \href{https://creativecommons.org/licenses/by/4.0/}{CC BY 4.0},  respectively.

\end{acknowledgments}

\appendix
\section{Dimensional regularization in the SMEFT}
\label{app:dim-reg}
As in any QFT, divergences can occur in the computation of one-loop diagrams in EFTs, which need to be regulated. 
Afterwards, the theory can be renormalized, and physical predictions can be derived.
By far the most commonly used regularization scheme in SMEFT computations is dimensional regularization, which we also use throughout this review, where we work in $D=4-2\epsilon$ spacetime dimensions. 
The most common renormalization scheme for SMEFT computations is the modified minimal subtraction~$\smash{(\overline{\mathrm{MS}})}$ scheme, which we also use in most of this work. 
The only exception is when we deal with evanescent operators, where we choose to work in an evanescent free version of~$\smash{\overline{\mathrm{MS}}}$ (see Sec.~\ref{sec:evanescent}). 
To be precise, we work in a modified version of $\smash{\overline{\mathrm{MS}}}$ that contains additional finite counterterms that compensate the effects of evanescent operators to physical observables so that these can be neglected in all computations.

In the following, we discuss two topics related to the use of dimensional regularization. First, we discuss the method of regions that is often used in EFT computations, e.g., when computing one-loop matching conditions. Afterwards, we comment on the issue of chiral fermions in dimensional regularization, i.e., the generalization of the $\gamma^5$~matrix to $D$~dimensions.

\subsection{The method of regions}
\label{app:method-of-regions}
\begin{figure*}[t]
    \centering
    \includegraphics[width=0.9\linewidth]{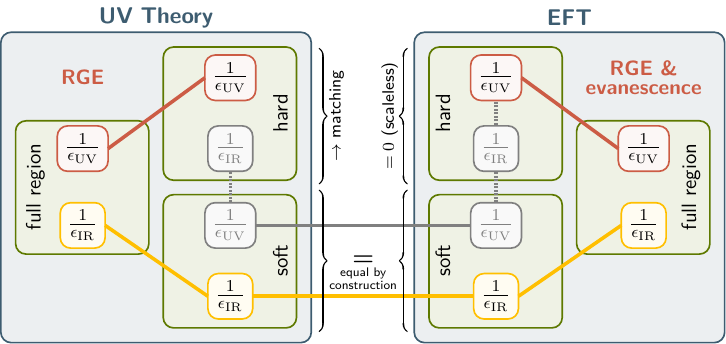}
    \caption{Schematic illustration of the method of regions applied for EFT matching. The separation of the full integration region into a hard and soft region (green frames) is shown for the UV~theory and the corresponding~EFT. For each region the UV~(IR) poles are highlighted in red (yellow) at the top (bottom) of each box. The UV~divergences require counterterms and allow to extract the RGE and the contributions of evanescent operators. The artificial divergences introduced by the method of region are shown in gray and cancel between the soft and the hard region as indicated  by the dashed lines connecting them. Divergences that are equal are connected by a solid line, whereas divergences that have the same magnitude but opposite sign are linked by dashed lines. The soft region of both theories are equal by construction, and the hard region of the EFT only contains scaleless integrals and thus vanishes in dimensional regularization.}
    \label{fig:method-of-regions}
\end{figure*}

The method of \textit{expansion by regions} \cite{Beneke:1997zp,Jantzen:2011nz} simplifies the calculation of multi-scale loop integrals in the presence of a power counting. Loop integrals can depend on several different scales (e.g. masses or external momenta), each defining an integration region. For each scale we can expand the loop integrand in the quantities that are small in the respective region and then perform the resulting integral over the entire $D$-dimensional space. The method of regions states that doing this for all regions and summing the results yields the same answer as performing the full original integral and expanding afterwards.

As an example, consider the loop integral
\begin{align}
    \mathcal{I} &= \int \frac{\dd^D k}{(2\pi)^D} \, \frac{1}{k^2-M^2} \, \frac{1}{k^2-m^2}
    \\
    &= \!\frac{i}{16\pi^2} \!\!\left[ \frac{1}{\epsilon} \!+\! \log\!\left(\!\frac{\mu^2}{M^2}\!\right)\! +\! 1 \!+ \!\frac{m^2}{M^2} \log\!\left(\!\frac{m^2}{M^2}\!\right) \!\right] \!+ \mathcal{O}(M^{-4})
    \nonumber
\end{align}
which entails two regions called soft~$(k \sim m)$ and hard~$(k \sim M)$.
Expanding the propagators in the soft $(k^2 \sim m^2 \ll M^2)$ and hard $(k^2 \sim M^2 \gg m^2)$ region before the integration
\begin{align}
    \frac{1}{k^2-M^2} &= -\frac{1}{M^2} \left[ 1 + \frac{k^2}{M^2} + \mathcal{O}\left(\frac{k^4}{M^4}\right) \right] \,,
    \\
    \frac{1}{k^2-m^2} &= \frac{1}{k^2} \left[ 1 + \frac{m^2}{k^2} + \mathcal{O}\left(\frac{m^4}{k^4}\right) \right] 
\end{align}
we find the corresponding integrals in each region
\begin{align}
    \mathcal{I}\big|_\mathrm{soft} &= -\frac{1}{M^2} \int \frac{\dd^D k}{(2\pi)^D} \, \left[ \frac{1}{k^2-m^2} + \ldots \right]
    \\
    &= - \frac{i}{16\pi^2} \frac{m^2}{M^2} \left[ \frac{1}{\epsilon} + \log\left(\frac{\mu^2}{m^2}\right) + 1 \right] + \mathcal{O}(M^{-4}) \,,
    \nonumber\\[0.1cm]
    \mathcal{I}\big|_\mathrm{hard} &= \int \frac{\dd^D k}{(2\pi)^D} \, \frac{1}{k^2} \, \frac{1}{k^2-M^2} \left[ 1 + \frac{m^2}{k^2} + \ldots \right]
    \\
    &= \! \frac{i}{16\pi^2} \!\! \left[ \frac{1}{\epsilon} + \log\!\left(\!\frac{\mu^2}{M^2}\!\right)\! + 1 \right]\!\! \left(\! 1 +\! \frac{m^2}{M^2} \!\right) \!+ \mathcal{O}(M^{-4}) \,.
    \nonumber
\end{align}
Thus, we find working at the order~$\mathcal{O}(M^{-2})$
\begin{align}
    \mathcal{I} &= \mathcal{I}\big|_\mathrm{hard} + \mathcal{I}\big|_\mathrm{soft} + \mathcal{O}(M^{-4}) \,,
\end{align}
as dictated by the method of regions.

As discussed in Sec.~\ref{sec:matching}, the method of regions provides a very powerful tool for EFT matching computations. These are, of course, multi-scale problems and applying this method allows for a separation of the hard UV dynamics from the soft IR behavior. In these computations we have to determine Green's functions in the UV~theory and the corresponding~EFT. In both theories we can split these into a hard and soft region. Since we require both theories to describe the same IR physics the soft regions --which contain exactly the low-energy dynamics-- of both theories must agree. Thus, the one-loop matching conditions for the EFT Wilson coefficients are solely determined by the hard regions encoding the UV~dynamics. However, since the EFT does by definition not contain any UV scales, its hard scale loop integrals must be scaleless and thus vanish exactly in dimensional regularization. This holds for all integrals apart from
\begin{align}
    \int \frac{\dd^D k}{(2\pi)^D} \, \frac{1}{k^4} &= \frac{i}{16\pi^2} \left( \frac{1}{\epsilon_{\scriptscriptstyle\mathrm{UV}}} - \frac{1}{\epsilon_{\scriptscriptstyle\mathrm{IR}}} \right) = 0 \,,
    \label{eq:scaleless-integral}
\end{align}
which vanishes only since we identify $\epsilon_{\scriptscriptstyle\mathrm{UV}} = \epsilon_{\scriptscriptstyle\mathrm{IR}}$ by analytic continuation in dimensional regularization.
Therefore, it is enough to only consider the hard region of the Green's functions of the UV theory which contains all information required to determine the EFT Wilson coefficients. The entire procedure of applying the method of regions is schematically shown in Fig.~\ref{fig:method-of-regions}.

This illustration also highlights the connection of the different UV and IR~divergences encountered in the computation. The UV~poles (red) and IR~poles (yellow) of a theory must match the corresponding poles in hard and soft region, respectively. However, applying the method of regions introduces additional artificial divergences (gray) in both regions. But since the sum of both regions must yield back the full solution, these must cancel between soft and hard region. We recall that also the UV and IR~poles of the hard EFT region must cancel due to Eq.~\eqref{eq:scaleless-integral}. In Fig.~\ref{fig:method-of-regions} cancelling divergences are connected by dashed lines, whereas equal poles are linked by solid lines.
When performing a computation we use the renormalized versions of these theories, i.e., we introduce counterterms cancelling the UV~poles. As mentioned before, for a matching computation we only need to compute the hard region of the UV~theory. The $\epsilon_{\scriptscriptstyle\mathrm{UV}}$~poles of this region are cancelled by the appropriate counterterms, and from Fig.~\ref{fig:method-of-regions} we see that the artificial IR~poles provide exactly the right counterterms to cancel the UV~poles of the resulting EFT. Thus the EFT is automatically renormalized. 

Eventually, notice that the method of regions is also useful to extract only the UV~divergences of a theory, since these are entirely encoded in its hard region. Therefore, it simplifies the extraction of the RG equations of a theory, and also the computation of the physical effect of evanescent operators (see Sec.~\ref{sec:evanescent}), since both are entirely determined by~$\epsilon_{\scriptscriptstyle\mathrm{UV}}$.

\subsection{Treatment of \texorpdfstring{$\gamma_5$}{gamma5} in \texorpdfstring{$D$}{D}~dimensions}
\label{app:gamma5}
When working in $D$~dimensions the Dirac algebra is infinite dimensional for non-integer~$D$, as mentioned already in Sec.~\ref{sec:evanescent}. While the usual Dirac matrices are defined by interpolation of the $D$~dimensional Dirac basis~$\gamma^\mu$ for $\mu \in \{0,\ldots,D=2n\}$ with an integer~$n\geq2$, the $\gamma_5$~matrix is not easily generalizable to $D \neq 4$~dimensions.
This is due to the intrinsically four-dimensional relation 
$\gamma_5 = -\frac{i}{4!} \varepsilon_{\mu\nu\rho\sigma} \gamma^\mu \gamma^\nu \gamma^\rho \gamma^\sigma$
linking it to the Levi-Civita tensor that can only be defined for~$D=4$.
Thus, any regularization and renormalization scheme must provide a prescription for treating $\gamma_5$ in dimensional regularization.

Throughout this review, we employed the (semi-) na\"ive dimensional regularization~(NDR) scheme, which assumes that the four-dimensional anti-commutation relations \cite{Korner:1991sx,Kreimer:1989ke,Nicolai:1980km}
\begin{align}
    \{\gamma^\mu,\gamma^\nu\} &= 2 g^{\mu\nu} \,,
    &
    \{\gamma^\mu,\gamma_5\} &= 0 \,,
    &
    \gamma_5^2 &= \mathds{1}
\end{align}
hold also away from~$D=4$. This is inconsistent with the cyclicity of the trace and $\mathrm{tr} \left( \gamma^\mu \gamma^\nu \gamma^\rho \gamma^\sigma \gamma_5 \right) \neq 0$. To reproduce the correct four-dimensional limit we formally substitute
\begin{align}
    \mathrm{tr} \left( \gamma^\mu \gamma^\nu \gamma^\rho \gamma^\sigma \gamma_5 \right) &= -4i\varepsilon^{\mu\nu\rho\sigma} \,,
    \label{eq:NDR-trace}
\end{align}
with $\varepsilon^{0123}=+1$. This breaks the cyclicity of traces with six or more $\gamma^\mu$-matrices and an odd number of~$\gamma_5$, thus introducing a reading point ambiguity. That means these traces depend on which $\gamma$-matrix is put first/last in the trace. 
For example, when computing the Feynman diagrams in Fig.~\ref{fig:evanescent-dipole} with insertions of the operator $Q_{lequ}^{(3)}$ we find, depending on where we start reading the closed fermion loop, the two Dirac traces
\begin{align}
    \mathrm{tr}_1 &\equiv \mathrm{tr}\left( \gamma^\alpha \gamma^\rho \gamma^\sigma \gamma_\alpha \gamma^\mu \gamma^\nu \gamma_5 \right) = 4i (4-D) \varepsilon^{\mu\nu\rho\sigma} \,,
    \\
    \mathrm{tr}_2 &\equiv \mathrm{tr}\left( \gamma^\rho \gamma^\sigma \gamma_\alpha \gamma^\mu \gamma^\nu \gamma_5 \gamma^\alpha \right) = -4i (4-D) \varepsilon^{\mu\nu\rho\sigma} \,,
\end{align}
which can be shown using Eq.~\eqref{eq:NDR-trace} and $\gamma^\alpha \gamma^\mu \gamma^\nu \gamma_\alpha = 4 g^{\mu\nu} \mathds{1} - (4-D) \gamma^\mu \gamma^\nu$.
We thus find
\begin{align}
     \mathrm{tr}_1 - \mathrm{tr}_2 = \mathcal{O}(\epsilon) \neq 0
\end{align}
in contradiction to the cyclicity of the trace. 
In EFT analyses using the NDR scheme, we must therefore carefully apply a consistent reading point prescription throughout all computations to obtain consistent results \cite{Fuentes-Martin:2020udw,Fuentes-Martin:2022vvu,Carmona:2021xtq}.

If one wants to avoid the ambiguities related to the reading point of Dirac traces one can resort to the 't~Hooft--Veltman~(HV) scheme \cite{tHooft:1972tcz,Breitenlohner:1977hr}, which is the only $\gamma_5$--scheme that is proven to be self-consistent to all orders.
In this scheme we define
\begin{align}
    \{\gamma^\mu,\gamma_5\} &= 0 & &\text{for } \mu \in \{0,1,2,4\} \,,
    \\
     [\gamma^\mu,\gamma_5] &= 0 & &\text{otherwise} \,.
\end{align}
While being the only known self-consistent scheme, HV comes with the subtlety that it breaks chiral symmetry and thus the Ward identities, which need to be restored by finite renormalizations. Also the HV scheme is computationally more expensive than NDR due to the splitting of the Dirac algebra in a four and a $D-4$~dimensional part. We therefore stick to the NDR scheme throughout this review, which is sufficient for the topics discussed here.

For a more detailed discussion of regularization schemes in $D$~dimensions and the problems of extending $\gamma_5$ to $D$~dimensions see \cite{Gnendiger:2017pys,Jegerlehner:2000dz} and references therein.

\bibliography{references}

\end{document}